\documentclass[journal,a4paper]{IEEEtran}

\usepackage{balance}
\usepackage[T1]{fontenc}
\usepackage{amsmath}
\usepackage{amsfonts}
\usepackage{mleftright}
\usepackage{amssymb}
\usepackage{amsbsy}
\usepackage{graphicx}
\usepackage{times}
\usepackage{default}
\usepackage{cite}
\usepackage{enumerate}
\usepackage{enumitem}
\usepackage[utf8]{inputenc}
\usepackage{hyperref}
\usepackage{mathtools}
\usepackage{xspace}
\usepackage{dsfont} %doublestroke for indicator function
\usepackage[justification=centering,font={small}]{caption}
\usepackage{tikz, pgfplots}
\usetikzlibrary{matrix,patterns,calc,decorations.fractals,arrows,external,pgfplots.colormaps}
\usepgfplotslibrary{external}
\allowdisplaybreaks[3]

\definecolor{tblblue}{rgb}{0.93,0.93,1.0}
\definecolor{tblred}{rgb}{1,0.93,0.93}
\definecolor{darkblue}{rgb}{0,0,0.7} 
\definecolor{darkgreen}{RGB}{20,120,43} 
\definecolor{darkred}{rgb}{0.8,0,0} 
\definecolor{lightblue}{RGB}{101,124,191}
\definecolor{skyblue}{RGB}{135,206,235}
\definecolor{gold}{RGB}{204,168,66}
\definecolor{strongblue}{RGB}{60,146,228}
\definecolor{lightgray}{gray}{0.5}
\definecolor{verylightgray}{RGB}{101,124,191}
\definecolor{mistyrose}{RGB}{238,213,210}
\definecolor{firebrick3}{RGB}{205,38,38}

\def\av{\boldsymbol{a}}\def\bv{\boldsymbol{b}}
\def\ev{\boldsymbol{e}}

\def\pv{\boldsymbol{p}}
\def\qv{\boldsymbol{q}}\def\sv{\boldsymbol{s}}
\def\uv{\boldsymbol{u}}\def\vv{\boldsymbol{v}}\def\wv{\boldsymbol{w}}\def\xv{\boldsymbol{x}}
\def\yv{\boldsymbol{y}}\def\zv{\boldsymbol{z}}\def\0v{{\boldsymbol{0}}}\def\1v{\boldsymbol{1}}

\def\am{\boldsymbol{A}}\def\bm{\boldsymbol{B}}
\def\fm{\boldsymbol{F}}\def\hm{\boldsymbol{H}}

\def\um{\boldsymbol{U}}

\def\arm{\boldsymbol{\mathsf{A}}}

\def\arv{\boldsymbol{\mathsf{a}}}

\def\wrv{\boldsymbol{\mathsf{w}}}\def\xrv{\boldsymbol{\mathsf{x}}}
\def\yrv{\boldsymbol{\mathsf{y}}}\def\zrv{\boldsymbol{\mathsf{z}}}

\def\as{\mathcal{A}}\def\bs{\mathcal{B}}

\def\ss{\mathcal{S}}\def\ts{\mathcal{T}}
\def\us{\mathcal{U}}\def\vs{\mathcal{V}}\def\ws{\mathcal{W}}

\def\IN{\mathbb{N}}

\def\IR{\mathbb{R}}

\def\IrX{\mathbb{\mathsf{X}}}
\def\IrY{\mathbb{\mathsf{Y}}}
\def\IrZ{\mathbb{\mathsf{Z}}}

\def\pr{\mathbb{P}}

\def\rank{\operatorname{rank}}

\def\leb{\operatorname{Leb}}

\def\dim{\operatorname{dim}}

\def\MIL{\underline{\dim}_\text{B}}
\def\MIU{\overline{\dim}_\text{B}}
\def\MILU{{\dim}_\text{B}}

\def\ba#1\ea{\begin{align*}#1\end{align*}}	%Shortcut for align-environment
\def\ban#1\ean{\begin{align}#1\end{align}}	%Shortcut for align-environment, numbered
\def\bac#1\eac{\vspace{\abovedisplayskip}{\par\centering$\begin{aligned}#1\end{aligned}$\par}\addvspace{\belowdisplayskip}}	% the same, centered

\makeatletter
\newcommand{\vast}{\bBigg@{2.5}}
\newcommand{\Vast}{\bBigg@{4}}
\makeatother

\newcommand{\lefto}{\mathopen{}\left}

\newtheorem{theorem}{Theorem}

  \newtheorem{WVV}{Theorem (\hspace*{-2truemm}\cite[Thm.~18, 2)]{WV10})}

\newtheorem{lemma}{Lemma}
\newtheorem{definition}{Definition}
\newtheorem{proposition}{Proposition}
\newtheorem{corollary}{Corollary}
\newtheorem{remark}{Remark}%[section]

\mathtoolsset{showonlyrefs=true}

\title{Almost Lossless Analog Signal Separation\\ and Probabilistic Uncertainty Relations}

\author{
\IEEEauthorblockN{David Stotz$^\dagger$, Erwin Riegler$^\star$,  Eirikur Agustsson$^\star$, and  Helmut B\"olcskei$^\star$}\\
\IEEEauthorblockA{$^\dagger$Kantonsschule Schaffhausen, 8200 Schaffhausen, Switzerland, Email: david-stotz@web.de}\\
\IEEEauthorblockA{$^\star$Dept. of IT \& EE, ETH Zurich, 8092 Zurich, Switzerland, Email:  \{eriegler, boelcskei\}@nari.ee.ethz.ch}, aeirikur@vision.ee.ethz.ch\\
\thanks{The material in this paper was presented in part  at the IEEE Int.\ Symp.\ on Inf.\ Theory, Istanbul, Turkey, Jul.~2013  \cite{SRB13}. An Online Addendum \cite{SRAB15extended} to this paper is available at  \url{http://www.nari.ee.ethz.ch/commth/research/downloads/sigsep_addendum.pdf}.}
}

\begin{document}
\maketitle
\begin{abstract}
We propose an information-theoretic framework for analog signal separation. Specifically, we consider the problem of recovering two analog signals, modeled as general random vectors, from the noiseless sum of linear measurements of the signals. Our framework is inspired by the groundbreaking work of Wu and Verd\'u (2010) on  analog compression and encompasses, inter alia,  inpainting, declipping,  super-resolution, the recovery of signals corrupted by impulse noise, and the separation of (e.g., audio or video) signals into two distinct components. The main results we report are  general achievability bounds for the compression rate, i.e., the number of measurements relative to the dimension of the ambient space the signals live in, under either measurability or  H\"older continuity imposed on the separator. Furthermore, we find a matching converse for sources of mixed  discrete-continuous distribution. For measurable separators our  proofs are based on  a new probabilistic uncertainty relation which shows that the intersection of generic subspaces with general sets of sufficiently small Minkowski dimension is empty.  H\"older continuous separators are dealt with by introducing the concept of regularized probabilistic uncertainty relations.
The probabilistic uncertainty relations we develop are inspired by embedding results in dynamical systems theory  due to  Sauer et al.\  (1991) and---conceptually---parallel classical Donoho-Stark and Elad-Bruckstein uncertainty principles at the heart of compressed sensing theory. Operationally, the new uncertainty relations take the theory of sparse signal separation beyond traditional sparsity---as measured in terms of the number of non-zero entries---to the more general notion of low description complexity as quantified by Minkowski dimension. Finally, our approach also allows to significantly strengthen key results in Wu and Verd\'u (2010). 
\end{abstract}


\section{Introduction}\label{sec:intro}
We consider the following signal separation problem: Recover  the vectors $\yv$ and $\zv$ from the noiseless observation%\\[-.9cm] 
\begin{align}\label{eq:problem1} 
\wv=\am\yv+\bm\zv ,
\end{align}  
where $\am$ and $\bm$ are measurement matrices. 
Numerous signal processing problems can be cast in the form  \eqref{eq:problem1}, e.g.,  inpainting, declipping,  super-resolution, the recovery of signals corrupted by impulse noise, and the separation of (e.g., audio or video) signals into two distinct components.
% signals corrupted by impulse noise or narrowband interference such as electric hum, the separation of (audio or video) signals into two distinct components, image inpainting, and super-resolution. 
 For a detailed exposition of the specifics of \eqref{eq:problem1} for each of these applications and for corresponding references, we refer the interested reader to \cite[Sec.~1]{SKP12}.

%This problem is encountered in numerous signal processing applications  \cite[Sec.~1]{SKP12}. For example, the analog-to-digital conversion often causes clipping of the amplitudes when they exceed a certain threshold. Formally, we can write the output by $\wv=\yv+\ev$, where $g$ denotes the entry-wise clipping map and $\ev=g(\yv)-\yv$ represents the excess. Another application of the signal separation problem is the recovery of signals that are corrupted by impulse noise, e.g., caused by erroneous readings from memory, or by narrowband interference, e.g., electric hum caused by  deficient audio devices. Finally, also super-resolution and inpainting can be cast as a signal separation problem, since in both cases the observation $\wv$ consists only of a subset of the entries of the compete signal and therefore $\wv$ can be written as a difference of the desired complete signal and the missing entries.

%As detailed in \cite[Sec.~1]{SKP12} this   problem has numerous applications such as inpainting, super-resolution, and the recovery of clipped signals and of signals that are corrupted by impulse noise or narrowband interference. 
 
The  sparse signal recovery literature \cite{SKP12, Li12, WM10, DMM09, DK10, Tro08, CT12, PBS13,  DH01, DS89} provides separation guarantees under sparsity constraints on the vectors $\yv$ and $\zv$. Specifically, the sparsity  
thresholds in, e.g., \cite{SKP12,DK10,DH01,DS89} are functions of the coherence parameters \cite{SKP12} of the matrices $\am$ and $\bm$ and hold for \emph{all}  $\yv$ and $\zv$, but suffer from the ``square-root bottleneck'' \cite{Tro08}, which states that the number of measurements, i.e., the number of entries of $\wv$, has to scale quadratically in the total number of non-zero entries in $\yv$ and $\zv$.
For random signals $\yv$ and $\zv$, the probabilistic results in \cite{Li12, WM10, PBS13} overcome the square-root bottleneck, but hold ``only'' with overwhelming probability.  
For $\bm$  equal to the identity matrix and $\am$ a random orthogonal matrix it is shown in \cite{CT12}  that the probability of failure of an $\ell_1$-based separation algorithm decays exponentially in the dimension of the ambient space, provided that $\yv$ and $\zv$ satisfy certain convex cone  conditions. 
%\cite{SKP12,DK10,DS89, DH01,DMM09, WM10,Li12, PBS13,Tro08,CT12} 

% on the decay of the  volume. 
\par\smallskip

 \textit{\;\,\; Contributions.}\,
%\paragraph*{Contributions}
The goal of this paper is to develop an information-theoretic framework for signal separation. 
Specifically, inspired by the groundbreaking work of Wu and  Verd\'u on  analog compression \cite{WV10}, we consider the problem of recovering $\yrv$ and $\zrv$  from 
$\wrv=\am\yrv+\bm\zrv$, where  $\yrv$ and $\zrv$ are random, possibly dependent,  and of general distributions, i.e.,  mixtures of discrete, absolutely continuous, and singular distributions. Our results are asymptotic in the sense that the probability of error can be made arbitrarily small by increasing the dimensions of $\wrv$, $\yrv$, and $\zrv$. 
%The goal of this paper is to develop an information-theoretic framework for the problem of recovering $\yv$ and $\zv$ from $\wv=\am\yv+\bm\zv$. Specifically, inspired by the groundbreaking work by Wu and  Verd\'u on  analog compression \cite{WV10}, we derive asymptotic (in the dimensions of $\wv$, $\yv$, and $\zv$)
In practical signal separation problems of this form one often encounters a specific structure for one of the matrices (assumed to be $\bm$ here, without loss of generality (w.l.o.g.)); for example, the matrix could represent a certain dictionary under which a class of signals is sparse. We will therefore be interested in statements that  hold for a given $\bm$. % for almost all (a.a.)  matrices $\am$, instead of only for a.a.\ concatenated matrices  ${[\am\:\:  \bm]}$. 
Our separation guarantees will, indeed, be seen to apply to deterministic  $\bm$ and  a.a.\footnote{Throughout the paper a.a. stands for ``Lebesgue almost all''.}  $\am$ (with the set of exceptions for $\am$ depending on the specific choice of $\bm$). Moreover, they  do not provide worst-case guarantees like the coherence-based results in, e.g., \cite{SKP12,DK10,DH01,DS89}, but are rather in terms of probability of separation error with respect to (w.r.t.) the constituents $\yrv$ and $\zrv$,
and as such do not depend on coherence parameters. 
%do not ,
%Since we assume $\yv$ and $\zv$ to be random, the statements
% are in terms of probability of separation error with respect to (w.r.t.) the constituents $\yv$ and $\zv$, and hence do not provide worst-case guarantees like the coherence-based results in, e.g., \cite{SKP12,DK10,DH01,DS89}. 
Specifically, we study the asymptotic setting $\ell, n \to \infty$ where the random vectors $\yrv\in\IR^{n-\ell}$ and $\zrv\in\IR^{\ell}$ are sections of random processes; for each $n$, we let $\ell=\lfloor \lambda n\rfloor$ and $k=\lfloor R n\rfloor$ for parameters $\lambda , R\in [0,1]$. % and measurement matrices $\am\in\IR^{k\times (n-\ell)}$ and $\bm\in \IR^{k\times \ell}$, with $k\geqslant \ell$.  
We refer to  $R$ as the compression rate as it equals (approximately) the ratio between the number of measurements, $k$, and the total number of entries, $n$, in $\yrv$ and $\zrv$.
Our first main result, Theorem~\ref{th2}, shows that for each (deterministic) full-rank matrix $\bm\in \IR^{k\times \ell}$, with $k\geqslant \ell$, and  a.a.\ matrices $\am\in\IR^{k\times (n-\ell)}$, there exists a measurable\footnote{Throughout the paper, the term measurable refers to Borel measurability.} separator recovering $\yrv$ and $\zrv$ from $\wrv$  with arbitrarily small probability of  error, provided that $n$ is sufficiently large and the compression rate $R$ is larger than 
the description complexity of the concatenated random source vector $\xrv:=[\yrv^{\operatorname{T}} \:\: \zrv^{\operatorname{T}}]^{\operatorname{T}}$ as quantified by its  Minkowski dimension compression rate ${R}_{\text{B}}(\varepsilon)$ (see\ Definition~\ref{defminkowskirate}). In practice, when recovery is to be performed from noisy, quantized, or otherwise perturbed versions of the measurement $\wrv$, it is desirable to impose continuity/smoothness constraints on the separator. The second main result of this paper, reported in  Theorem~\ref{thm:hoeldersepneu}, shows that for each (deterministic) full-rank matrix  $\bm\in \IR^{k\times \ell}$,  with $k\geqslant \ell$, and  a.a.\ matrices $\am\in\IR^{k\times (n-\ell)}$, there exists  a $\beta$-H\"older continuous separator achieving  error probability $\varepsilon$ provided that $R>R_{\text{B}}(\varepsilon)$ and $\beta < 1-\frac{{R}_{\text{B}}(\varepsilon)}{R}$. 
We hasten to add that we do not specify explicit separators that achieve our thresholds, rather we prove  existence results absent computational considerations. In contrast, many of the recovery thresholds available in the literature pertain to $\ell_1$-norm-based recovery algorithms, see, e.g.,  \cite{SKP12, Li12, WM10, DK10, Tro08, CT12, PBS13,  DH01, DS89}. 
In the case of mixed discrete-continuous source distributions a converse matching the general---w.r.t. the nature of the source distributions---achievability statements in Theorems~\ref{th2} and \ref{thm:hoeldersepneu} can be obtained. This establishes the Minkowski dimension compression rate ${R}_{\text{B}}(\varepsilon)$ as the critical rate for successful separation when the source distributions are mixed discrete-continuous.

In principle one could rewrite \eqref{eq:problem1} 
 in the form 
 $\wv = {[\am\:\: \bm]} [\yv^{\operatorname{T}} \:\: \zv^{\operatorname{T}}]^{\operatorname{T}}$ 
% $\wv = {[\am\:\: \bm]} \mleft[ \begin{matrix} \yv \\ \zv  \end{matrix} \mright] $ 
 and consider applying the results in \cite{WV10} with $\hm = {[\am\:\: \bm]}$. 
However, the theory developed in \cite{WV10} leads to statements that apply to a.a.\ matrices $\hm$, whereas here, for reasons mentioned above, we seek statements that apply for a \emph{given} matrix $\bm$, 
  and fixing $\bm$ results in $\hm={[\am\:\: \bm]}$-matrices supported on a set of Lebesgue measure zero. A direct application of the results in \cite{WV10} to the signal separation problem 
is hence not possible; we therefore develop a new proof methodology and new mathematical tools.  % (cf. Remark \ref{rem.rev}).  
% We therefore  have to  develop a new proof methodology and new mathematical tools.
%Since the proof techniques  in \cite{WV10} for the analog compression framework %,  related to our results Theorem~\ref{th2} and Theorem~\ref{thm:hoeldersepneu}, in the context of almost lossless analog compression
 %can not be adapted to our setting, %where the effective measurement matrix is the concatenation of a random matrix  and a fixed matrix, 
%we develop a new  methodology and tools, 
%which  we consider the third main contribution of the paper. 
The foundation of our  approach   stems from dynamical systems theory \cite{SYC91}. Specifically, we establish  a new technique for showing that the intersection of generic subspaces (of finite-dimensional Euclidean spaces) and arbitrary sets of sufficiently small Minkowski dimension is empty. This leads to statements that have the flavor of  a probabilistic uncertainty relation akin to the classical (deterministic) Donoho-Stark \cite{DS89} and  Elad-Bruckstein \cite{EB02} uncertainty relations underlying much of compressed sensing theory.
% A novel  concentration of measure result, developed in Lemma~\ref{lemma2}, turns out to be an essential ingredient of this technique. 
Our result on H\"older continuous separators is based on a regularized probabilistic uncertainty relation, a concept which does not seem to have a counterpart in classical compressed sensing theory.
%which parallels the original probabilistic uncertainty relation. 
Finally, we note that applying our mathematical machinery to the analog compression framework in \cite{WV10} leads to a simplification of the proof of \cite[Thm.~18, 1)]{WV10} and  to 
 significant strengthening of \cite[Thm.~18, 2)]{WV10}, as detailed in Section~\ref{sec:simple}. 
%Specifically, this strenthening consists of a converse result that applies to a.a. measurement matrices, as opposed  
 %of \cite[Thm.~18, 1)]{WV10}.

%For $\yv$ and $\zv$ mixed discrete-continuously distributed with mixing parameters $\rho_1$ and  $\rho_2$, respectively, we show that the Minkowski dimension compression rate can be evaluated explicitly to 
%\begin{align}\label{eq:threshold}
%(1-\lambda)\rho_1+\lambda\rho_2.
%\end{align}
%What is more, this threshold is tight in the sense that there is a converse if the compression rate $R$ is smaller than \eqref{eq:threshold}.

\textit{\;\,\; Notation.}\,
For a relation $\# \in\{<,>,\leqslant,\geqslant,=,\neq,\in, \notin\}$, we write 
$f(n)\ \overset{\textbf{.}}{\#}\ g(n)$ if there exists an $N\in \IN$ such that $f(n)\ \# \ g(n)$ holds for all $n\geqslant N$. 
$\leb^n$ stands for the $n$-dimensional Lebesgue measure and $\bs^{\otimes n}$ refers to the Borel $\sigma$-algebra on $\IR^n$. Matrices are denoted by capital boldface and vectors by lowercase boldface letters.  We let $\| \cdot \|$ be  the $\ell_2$-norm on $\IR^n$ and set  $\|\am\| := \sup_{\|\xv\|=1} \|\am \xv \|$.
The $n\times n$ identity matrix is  $\boldsymbol{I}_n$ and $\fm_n$ stands for the $n$-dimensional discrete Fourier transform (DFT) matrix.
%For $\xv\in \mathbb R^n$ and $\ts\subseteq \{1,\ldots , n\}$, we let $\xv_{\ts}$ denote the $|\ts|$-dimensional subvector that consists of the components of $\xv$ corresponding to the indices in $\ts$. 
Sets are represented by calligraphic letters. For $\as, \bs\subseteq \mathbb R^n$, we set $\as\ominus\bs :=\{\av-\bv : \av\in \as,\bv\in\bs\}$.  
%$\overline{\as}$ denotes the closure of $\as\subseteq\IR^n$. 
$B^n(\xv,\delta)$ is the open $\ell_2$-ball of radius $\delta$ centered at $\xv\in\IR^n$, and   its  volume is  given by $\alpha(n,\delta)=\leb^n(B^n(\xv,\delta))$.  %The closure of a set   $\us\subseteq\mathbb R^n$ is  denoted by $\overline{\us}$.
 We use sans-serif letters, e.g.\ $\xrv$, for random quantities and roman letters, e.g. $\xv$, for deterministic quantities. For a random variable $\IrX$ and a random vector $\xrv$, $\mu_{\IrX}$ and $\mu_{\xrv}$ denote the respective distributions,  integration w.r.t. these distributions is indicated by $\mu_{\IrX}(\mathrm dx)$ and $\mu_{\xrv}(\mathrm d \xv)$. For Borel sets $\mathcal A$, we let $\mathds{1}_\mathcal A(x)$ be the indicator function on $\mathcal A$. %We write $\1v_{\IrX\in\as}$ for the characteristic function associated with the event $\IrX \in \as$.  
Constants which depend exclusively  on parameters $\alpha_1,...,\alpha_n$ are written as $c(\alpha_1,...,\alpha_n)$ or $C(\alpha_1,...,\alpha_n)$, where  the constants may take on different  values in different appearances.

 %a repeated appearance of $C(\alpha_1,...,\alpha_n)$ does not necessarily refer the same  expression. 

\textit{\;\,\; Outline of the paper.}\,
%\paragraph*{Outline of the paper}
In Section~\ref{sec:mainresults}, we first introduce our information-theoretic framework for the signal separation problem and then state 
 the achievability result for measurable separators, Theorem~\ref{th2},  followed by  the achievability result for H\"older continuous separators, Theorem~\ref{thm:hoeldersepneu}.
%general achievability results of the paper in  Theorems~\ref{th2} and~\ref{thm:hoeldersepneu}.
%our general achievability result with either measurability or with Hölder continuity imposed on the separator.
 Section~\ref{sec:probuncrel} contains
 %  the third contribution, namely
   the probabilistic uncertainty relation  the proof of Theorem~\ref{th2} is based on.
In Section~\ref{proofTh1}, we present the proof of Theorem~\ref{th2}.
%the general achievability result, Theorem~\ref{th2}, for measurable separators. 
Section~\ref{sec:regprobuncrel} introduces the regularized probabilistic uncertainty relation underlying the proof of Theorem~\ref{thm:hoeldersepneu}, which is %Section~\ref{sec:concl} concludes the paper. 
provided in Section~\ref{sec:hoeldersep}.  In Section~\ref{sec:cs}, we particularize our results for mixed discrete-continuous source distributions 
and we derive a converse matching the corresponding achievability results.
Finally, in Section~\ref{sec:simple} we show how our mathematical  techniques  lead to a simplification of the proof of \cite[Thm.~18, 1)]{WV10} and to 
 significant strengthening  of the statement \cite[Thm.~18, 2)]{WV10}.

%relegated to the 
 %Appendix~\ref{app:hoeldersep}.
%Appendix along with additional auxiliary statements.

\section{Statement of the main results}
\label{sec:mainresults}

%\ensuremath{{[\am\:\: \bm]}}
We begin by introducing our information-theoretic framework for signal separation.
The recovery of  the vectors $\yv$ and $\zv$ from the noiseless observation $\wv$ in  \eqref{eq:problem1} can be rephrased  as the recovery of  $[\yv^{\operatorname{T}}\; \zv^{\operatorname{T}}]^{\operatorname{T}}$ from the linear measurements
\ban 		\wv = {[\am\:\: \bm]} \mleft[ \begin{matrix} \yv \\ \zv  \end{matrix} \mright].  \label{eq:model1} \ean 
%with measurement matrix $\hm:={[\am\:\: \bm]}$. 
An information-theoretic framework for analog compression, i.e., for the problem of recovering $\xrv$  from the linear measurements $\wrv= \hm\xrv$,
 was introduced in \cite{WV10}.
 The main achievability result in \cite{WV10} provides conditions on the compression rate $R$---in terms of the source vector's Minkowski dimension compression rate---for exact recovery to be possible at arbitrarily small probability of error  as the blocklength $n$ goes to infinity.
While the information-theoretic framework for  signal separation we develop here is inspired by the analog compression framework in \cite{WV10}, there are fundamental differences between these two problems. Specifically, the signal separation applications outlined in Section~\ref{sec:intro} (again, we refer to \cite[Sec.~1]{SKP12} for specifics) mandate taking specific structural properties of $\am$ and $\bm$ into account.
For example, for the recovery of signals corrupted by impulse noise or narrowband interference one of the matrices $\am$, $\bm$ equals the identity matrix or 
%one might be interested in the case where one of the matrices $\am$ and $\bm$ is the identity matrix or 
the DFT matrix. % In the deterministic setting, accounting for the stacked nature of the measurement matrix $\hm={[\am\:\: \bm]}$ has proven fruitful as it lead to a factor-of-$2$ improvement of  the recovery thresholds in \cite{SKP12}, \cite{EB02} relative to \cite{DE03}.
%To follow a similar approach,
This will be accounted for by taking
 $\bm$   to be deterministic and fixed throughout the paper.  
 %worknow
%  \rem{ As the results in \mbox{\cite{WV10}} lead to statements that apply to a.a.\ measurement matrices $\hm$ and fixing $\bm$ results in $\hm={[\am\:\: \bm]}$-matrices supported on a set of Lebesgue measure zero, it follows that \mbox{\cite{WV10}} is not applicable here.}
%In order to account for the stacked nature, we take one of the  matrices $\am$ and $\bm$, taken to be $\bm$ throughout, to be deterministic and fixed, while  $\am$ will be drawn from a random matrix $\arm$. %The resulting random measurement matrix in \eqref{eq:model1} is then given by $\hrm={[\arm\:\: \bm]}$, and a concrete realization is denoted by $\hm={[\am\:\: \bm]}$. 
%We note that \cite[Thm.~18]{WV10} provides a general achievability result that applies to almost all measurement matrices $\hm$.
%the main results in \cite{WV10} apply to almost all measurement matrices $\hm$.
%Now, we note that the main results in \cite{WV10} apply to almost all measurement matrices $\hm$.
% However, fixing $\bm$ and drawing $\am$ randomly   is not covered by this result, since the obtained realizations ${[\am \:\: \bm ]}$ form a class of $\hm$ of Lebesgue measure zero due to the fixed block $\bm$. In other words, \cite[Thm.~18]{WV10} does not tell us anything about the structure of \emph{pairs} $(\am, \bm)$ for which separation is possible. 
%The corresponding statements therefore do not imply anything for the signal separation problem where $\hm={[\am\:\: \bm]}$ where $\bm$ is fixed. 
As noted in the introduction, addressing this problem requires  new techniques, namely  probabilistic uncertainty relations akin to the (deterministic) Donoho-Stark \cite{DS89} and Elad-Bruckstein  \cite{EB02} uncertainty relations, extended to frames and undercomplete signal sets in \cite{KDB12}, and  underlying much of compressed sensing theory. These probabilistic uncertainty relations  will allow us to make statements that apply to a.a.\ $\am$ for a fixed $\bm$ (with the set of exceptions for $\am$ depending on the specific choice of $\bm$).

We next define the specifics of our setup.
%Next, we define the information-theoretic setup of the signal separation problem. First we define the 
 %random source vector  $\xrv=[\yrv^{\operatorname{T}}\; \zrv^{\operatorname{T}}]^{\operatorname{T}}$ of varying blocklength $n$ (where $\yrv$ and $\zrv$ are possibly dependent) such that the fraction of components in $\xrv$ corresponding to $\yrv$ and the fraction corresponding to $\zrv$ is approximately constant.

\begin{definition}\label{definitionX}
 Suppose that $(\IrY_i)_{i\in\IN}$ and $(\IrZ_i)_{ i\in\IN}$ are stochastic processes on $(\IR^\IN,\bs^{\otimes\IN})$. 
Then, for $n\in\IN$, we define the concatenated source vector $\xrv$ of dimension $n$ as $\xrv=[\IrX_1 \, \dots\, \IrX_n]^{\operatorname{T}}$ according to
\begin{align*}
\IrX_i &= \IrY_i,\quad \text{for}\ i\in\{1,\dots,n-\ell\}\\
\IrX_{n-\ell+i} &= \IrZ_i,\quad \text{for}\ i\in\{1,\dots,\ell\},
\end{align*}
where $\ell=\lfloor \lambda n \rfloor$ with the parameter $\lambda\in [0,1]$ representing the asymptotic fraction of components in $\xrv$ corresponding to the $\IrZ_i$'s.
\end{definition}

%\begin{remark}
%Throughout the paper, we assume that the measurement matrix $\hm={[\am\:\: \bm]}$ consists of a realization of a random matrix $\arm$  and a deterministic matrix $\bm$. When we consider the measurement matrix as a random matrix, we write $\hrm={[\arm\:\: \bm]}$.
%\end{remark}
We emphasize that the distributions of the components $\IrY_i$ and $\IrZ_i$ in the above definition are general in the sense that they can be a mixture of discrete,  continuous, and singular distributions, i.e., $\mu=\mu_{\text{d}}+\mu_{\text{c}}+\mu_{\text{s}}$.

%Second, we define achievable rates in terms of finding (a sequence of) pairs consisting of a measurement matrix $\hm={[\am\:\: \bm]}$ and a separator $g$. %Definitions~\ref{defsourcerate} and \ref{defminkowskirate} below are adopted from the corresponding definitions in \cite{WV10}.
%\begin{definition}\label{defsourcerate}
The encoding--decoding part comprises %an $(n,k)$ code consists of 
\begin{enumerate}[label=(\roman*)]
\item a measurement matrix $\hm={[\am\:\: \bm]} \colon\IR^{n-\ell}\times \IR^\ell \to \IR^k$, where $\am \in \mathbb R^{k\times (n-\ell)}$ and $\bm\in \mathbb R^{k\times \ell}$;
\item a separator $g:\IR^k\to \IR^{n-\ell}\times \IR^\ell$.
\end{enumerate} 
%\end{definition}
We  will deal with  separators $g$ that are measurable  and with $g$ that are, in addition, $\beta$-H\"older continuous, i.e., for a given $\beta>0$  they satisfy
\ba \| g(\xv_1) - g(\xv_2)\| \leqslant c \| \xv_1 - \xv_2\| ^\beta, \quad \text{ for all $\xv_1,\xv_2\in \mathbb R^k$}, \ea
where   $c>0$ is a constant.  H\"older continuous separators are relevant in the context of recovery from noisy, quantized, or otherwise perturbed measurements,
 but the class of H\"older continuous mappings is significantly smaller than that of measurable mappings.
\begin{definition}\label{defsourcerate2}
For $\xrv$ as in Definition \ref{definitionX} and a given measurement matrix $\hm={[\am\:\: \bm]}$,   we say that there exists a (measurable or $\beta$-H\"older continuous) separator that achieves rate $R\in [0,1]$ with error probability $\varepsilon\in (0,1)$ if there exists a sequence (w.r.t. $n$) of (measurable or $\beta$-H\"older continuous) maps $g$  such that $k=\lfloor Rn \rfloor$ and
\begin{align*}
\pr[g({[\am\:\: \bm]}\xrv)\neq \xrv]\overset{\textbf{.}}{\leqslant}\varepsilon.
\end{align*}
\end{definition}
%We define the optimal linear compression rate $R_\text{L}(\varepsilon)$ as the infimum over all $\varepsilon$-achievable rates. Here, the name ``linear'' reflects the restriction to linear measurements, employed throughout the paper.

Next, we quantify the description complexity of $\xrv$ with general  distribution (possibly containing a singular component) through the Minkowski dimension of approximate support sets for $\xrv$. The Minkowski dimension is sometimes also referred to as box-counting dimension, which explains the origin for the subscript $\text{B}$ in the notation $\MILU(\cdot)$ used below. 
We start with the definition of Minkowski dimension for general sets. %we define the Minkowski dimension, which plays a crucial role for characterizing the critical achievable rates.
%This quantity is sometimes also referred to as box-counting dimension, which is the origin for the subscript $\text{B}$ in the notation $\MILU(\cdot)$ used below.
\begin{definition}(Minkowski dimension, \cite{Fal04}).\label{defminwowskidim}
Let $\ss$ be a non-empty bounded set in  $\IR^n$.  Define the lower and upper Minkowski dimension of $\ss$ as 
\begin{align}
\MIL(\ss)&=\liminf_{\delta\to 0}\frac{\log N_\ss(\delta)}{\log\frac{1}{\delta}} \label{eq:defminkowski1}\\
\MIU(\ss)&=\limsup_{\delta\to 0}\frac{\log N_\ss(\delta)}{\log\frac{1}{\delta}}\label{eq:defminkowski2},
\end{align}
where $N_\ss(\delta)$ is the covering number of $\ss$ given by 
\begin{align}
N_\ss(\delta)&=\min\Big\{m \in\IN :\ss\subseteq\bigcup_{i\in\{1,\dots,m\}} B^n(\xv_i,\delta),\ \xv_i\in \IR^n\Big\}. \label{eq:origcovno}
\end{align}
If $\MIL(\ss)=\MIU(\ss)$, we define the Minkowski dimension of $\ss$ as $\MILU(\ss):=\MIL(\ss)=\MIU(\ss)$. 
\end{definition}
\begin{remark} 
%It is well-known that the Minkowski dimension can equivalently be  defined by replacing $N_\ss(\delta)$ in \eqref{eq:defminkowski1},  \eqref{eq:defminkowski2} with   modified covering numbers \cite[Equivalent Definitions~3.1]{Fal04}.
In Lemma~\ref{lem:equivalentdef} in Appendix~\ref{app:eqdef} we show that Minkowski dimension can be  defined equivalently by replacing $N_{\ss}(\delta)$ in \eqref{eq:defminkowski1},  \eqref{eq:defminkowski2} by the modified covering number%which requires that the covering ball centers lie within the set $\ss$:
\begin{align}
M_\ss(\delta)&=\min\Big\{m \in\IN :\ss\subseteq\bigcup_{i\in\{1,\dots,m\}} B^n(\xv_i,\delta),\ \xv_i\in \mathcal S\Big\} \label{eq:modcovno} ,
\end{align}
which is in terms of covering balls that have their  centers in the set $\ss$.
This equivalent definition is often convenient as the covering ball centers inherit structural properties from the set $\ss$.
%
% needed in the proofs of our main results.
%where the centers of the covering balls are restricted to lie within the set $\ss$.
\end{remark}
%\begin{remark} TO DO: EXPONENTIAL BOUNDS FOR COVERING NUMBER ERKLÄREN, VGL BEWEISE \end{remark}

As our framework  involves statements that are asymptotic in the blocklength $n$, we will  need a
description complexity measure that applies to  random processes. This leads to the notion of Minkowski dimension compression rate.
\begin{definition}(Minkowski dimension compression rate, \cite{WV10}).\label{defminkowskirate}
For $\xrv$ as in  Definition \ref{definitionX} and $\varepsilon >0$, we define the lower and upper Minkowski dimension compression rate as  
\begin{align}
\underline{R}_\text{B}(\varepsilon)&=\limsup_{n\to\infty} \underline{a}_n(\varepsilon),\quad\text{where}  \label{eq:deflowerminkrate} \\
\underline{a}_n(\varepsilon)&=\inf \lefto \{\frac{\MIL(\ss)}{n}   :  \ss \subseteq\IR^n,\ \pr[\xrv\in\ss]\ \geqslant 1-\varepsilon\right\}, \label{eq:appsupp}
\end{align} 
and 
\ban \overline{R}_\text{B}(\varepsilon)&=\limsup_{n\to\infty} \overline{a}_n(\varepsilon),\quad\text{where}  \label{eq:defupperminkrate} \\
\overline{a}_n(\varepsilon)&=\inf\lefto \{\frac{\MIU(\ss)}{n}  :  \ss \subseteq\IR^n,\ \pr[\xrv\in\ss]\ \geqslant 1-\varepsilon\right\}.
\ean
If $\underline{R}_\text{B}(\varepsilon)=\overline{R}_\text{B}(\varepsilon)$, we define the Minkowski dimension compression rate as $R_\text{B}(\varepsilon):=\underline{R}_\text{B}(\varepsilon)=\overline{R}_\text{B}(\varepsilon)$.
%simply write $R_\text{B}(\varepsilon)$.
\end{definition}

%\begin{remark}
%%Note that in \eqref{eq:appsupp} the infimum is taken with respect to the \emph{lower} Minkowski dimension, whereas the corresponding definition in \cite{WV10} is based on the \emph{upper} Minkowski dimension. 
%Our general achievability result, Theorem~\ref{th2} below, requires $R>\underline{R}_\text{B}(\varepsilon)$. The condition of the general achievability result in \cite{WV10} is $R>\overline{R}_\text{B}(\varepsilon)$. Therefore our result, when specialized to the setting of \cite{WV10}, i.e., $\lambda=0$,  constitutes a slight improvement of the corresponding result in \cite{WV10}. 
%\end{remark}  

The following theorem constitutes our first main result. % and states that for every full-rank matrix $\bm\in\IR^{k\times \ell}$, with $k\geqslant \ell$, and every rate $R$ with $R>\! \underline{R}_\text{B}(\varepsilon)$, for  a.a. $\am$ there exists a measurable separator achieving error probability $\varepsilon$. 
\begin{theorem}\label{th2}
Let $\xrv$ be as in Definition \ref{definitionX}. Take $\varepsilon >0$ and let  $R>\! \underline{R}_\text{B}(\varepsilon)$. Then, for every  full-rank matrix $\bm\in\IR^{k\times \ell}$, with $k\geqslant \ell$,  and for   a.a.  matrices $\am\in\IR^{k\times {(n-\ell)}}$, where $k=\lfloor Rn \rfloor$, there exists a measurable separator $g$ such that
\begin{align}\label{eq:resultmain}
\pr [g({[\am\:\: \bm]}\xrv)\neq\xrv] \overset{\textbf{.}}{\leqslant} \varepsilon .
\end{align}

\end{theorem}
\begin{IEEEproof}
See Section \ref{proofTh1}.
\end{IEEEproof}
\begin{remark}The set of exceptions for $\am$  depends on the specific choice of the full-rank matrix $\bm$.
The proof of Theorem \ref{th2} further reveals that the minimum $N\in\IN$ for \eqref{eq:resultmain} to hold for all $n\geqslant N$ depends  on the distribution of $\xrv$ only and is independent of the matrices $\am$ and $\bm$. 
\end{remark}
\begin{remark}\label{remgehtnicht}
%{\color{blue}
In \cite[Thm.~18, 1)]{WV10} it was shown---in the context of  analog compression---that every rate $R$ with $R>\! \overline{R}_\text{B}(\varepsilon)$ is achievable  for a.a.\ measurement matrices $\hm\in\IR^{k\times n}$. 
%As already mentioned, this  does imply that signal separation is possible for a fixed  submatrix $\bm$ in $\hm={[\am\:\: \bm]}$.
%and consider recovery of $\xv= [\yv^{\operatorname{T}}\; \zv^{\operatorname{T}}]^{\operatorname{T}}$ from $\wv = \am\yv + \bm\zv $.
% which is crucial for the signal separation problem. 
%In Theorem~\ref{th2} above this result is improved by replacing $\overline{R}_\text{B}(\varepsilon)$ with $\underline{R}_\text{B}(\varepsilon)$ and generalized  to hold for $\hm={[\am\:\: \bm]}$ for a given full-rank matrix $\bm$, with $k\geqslant \ell$, for a.a. matrices $\am\in\IR^{k\times (n-\ell)}$ (where the set of exceptions for $\am$ depends on  $\bm$). Since in a concrete separation problem we often encounter a particular structure for one of the matrices (assumed $\bm$ here, w.l.o.g.), %for example a certain dictionary under which the corresponding signal is sparse, 
%it is important to have the statement hold for a given $\bm$ for a.a.\ matrices $\am$, instead of only for a.a.\ $\hm={[\am\:\:  \bm]}$. 
The proof of \cite[Thm.~18, 1)]{WV10} relies on  intricate properties of invariant measures on Grassmannian manifolds   under the action of the orthogonal group. 
The  new proof technique we develop here is based on two key elements,  a probabilistic uncertainty relation formalized in Proposition~\ref{prop:inj2} and a concentration of measure result stated in Lemma~\ref{lemma2}. Specifically, the probabilistic uncertainty relation says that the $(n-k)$-dimensional null-space of $\hm={[\am\:\: \bm]}$ and the approximate support set $\ss$  of $\xrv$ do not intersect if the Minkowski dimension of $\ss$ is smaller than $k$.
%dimension counting argument provided in Corollary~\ref{prop:inj}.  The dimension counting argument says that the $(n-k)$-dimensional null-space of $\hm$ and the approximate support set $\ss$ in \eqref{eq:appsupp} of the source vector $\xrv$ will not intersect, if the Minkowski dimension of $\ss$ is smaller than $k$. 
Underlying this result  is the basic idea that two objects---in general relative position---whose dimensions do not add up to at least the dimension of their ambient space do not intersect. What is surprising  is that 
 Euclidean dimension (for the null-space of $\hm$) and Minkowski dimension (for the support set $\ss$) are compatible dimensionality notions in this context.
%}
\end{remark}

\begin{remark}\label{rem.rev}
As pointed out by an anonymous reviewer, it is possible to deduce a proof of Theorem \ref{th2} starting from the analog compression result 
\cite[Thm.~18, 1)]{WV10}, which applies  to  a.a.\ matrices $\hm={[\am\:\: \bm]}$, using a version of Fubini's theorem for complete measures \cite[Thm. 2.39]{Fo99}. 
The resulting overall proof is, however, more technical than our proof and does not  uncover the underlying probabilistic uncertainty relation.
%Our proof is less technical and more direct in the sense of applying  to a \emph{given} full-rank matrix $\bm$ and a.a. matrices $\am$, thereby    As mentioned in Remark \ref{remgehtnicht},  our proof technique also leads to a simplification of the proof of \cite[Thm.~18, 1)]{WV10} and a significant strengthening  of the statement  \cite[Thm.~18, 2)]{WV10}. 
Moreover, the  proof technique we develop also applies to the  analog compression problem \cite{WV10} and leads to a simplification of the proof of \cite[Thm.~18, 1)]{WV10} and 
 to significant strengthening of   the statement  \cite[Thm.~18, 2)]{WV10}, as detailed in Section~\ref{sec:simple}. 
\end{remark}

While Theorem~\ref{th2} provides guarantees for the  existence  of a \emph{measurable} separator, a natural follow-up question 
is whether we can make a similar statement under continuity/smoothness constraints imposed on the separator.
This question is relevant when separation is to be performed from quantized, noisy, or otherwise perturbed observations. It turns out that it is, indeed, possible for fixed $\bm$ and a.a.\ $\am$ to guarantee the existence of measurable separators that are, in addition,
H\"older continuous, even though
H\"older continuity is a much stronger property than  measurability alone.  It is therefore not surprising that the corresponding statement we obtain is  weaker, but actually only slightly so, than that for measurable separators. Specifically, we  establish the existence of a $\beta$-H\"older continuous separator with the threshold $R>\! \overline{R}_\text{B}(\varepsilon)$ instead of $R>\! \underline{R}_\text{B}(\varepsilon)$,   provided that   $ \beta < 1- \frac{\overline{R}_\text{B}(\varepsilon)}{R}$.
%with high probability with respect to a specific random model for $\am$ instead of for a.a.\ $\am$.

%For the case of Hölder continuous separators we do not have existence for a.a.\ blocks $\am$

\begin{theorem}\label{thm:hoeldersepneu}
Let $\xrv$ be as in Definition \ref{definitionX},  $R>\! \overline{R}_\text{B}(\varepsilon)$, for $\varepsilon >0$,  and  fix $\beta>0$ such that  %Suppose that the rows of $\arm\in\IR^{k\times (n-\ell)}$ are drawn i.i.d.\ uniformly from $B^{n-\ell}(\0v, r)$, for some $r>0$. Then, for every fixed full-rank matrix $\bm\in\IR^{k\times \ell}$, with $k=\lfloor Rn\rfloor \geqslant \ell$, and for all $\gamma,\varepsilon '\in (0,1)$ with $\gamma\geqslant \varepsilon ' >\varepsilon$ and all $\beta$ satisfying
\ba \beta < 1- \frac{\overline{R}_\text{B}(\varepsilon)}{R}. \ea
Then, for every fixed full-rank matrix $\bm\in\IR^{k\times \ell}$, with $k\geqslant \ell$,  and    a.a. matrices $\am\in\IR^{k\times {(n-\ell)}}$, where $k=\lfloor Rn \rfloor$, there exists a $\beta$-H\"older continuous separator $g$ such that
\begin{align}\label{eq:resulthoeldersepneu}
\pr [g({[\am\:\: \bm]}\xrv)\neq\xrv] \overset{\textbf{.}}{\leqslant} \varepsilon +\kappa,
\end{align}
where $\kappa>0$ is an arbitrarily small constant.
\end{theorem}

%\begin{theorem}\label{thm:hoeldersep}
%Let $\xrv$ be as in Definition \ref{definitionX}. Take $\varepsilon >0$  and   let  $R>\! \overline{R}_\text{B}(\varepsilon)$. Suppose that the rows of $\arm\in\IR^{k\times (n-\ell)}$ are drawn i.i.d.\ uniformly from $B^{n-\ell}(\0v, r)$, for some $r>0$. Then, for every fixed full-rank matrix $\bm\in\IR^{k\times \ell}$, with $k=\lfloor Rn\rfloor \geqslant \ell$, and for all $\gamma,\varepsilon '\in (0,1)$ with $\gamma\geqslant \varepsilon ' >\varepsilon$ and all $\beta$ satisfying
%\ba \beta < 1- \frac{\overline{R}_\text{B}(\varepsilon)}{R}, \ea
%the following holds: With probability (with respect to the distribution of $\arm$) strictly larger than $1-\varepsilon'/\gamma$  there exists a $\beta$-Hölder continuous separator that achieves rate $R$ with error probability $\gamma$.
%\end{theorem}
\begin{remark} As in Theorem~\ref{th2}, the set of exceptions for $\am$ %where there is no $\beta$-Hölder continuous separator 
depends on the specific choice of the full-rank matrix $\bm$. The constant $\kappa$ 
honors the fact that we have to excise a small set of concatenated source vectors on which the separator may fail to be H\"older-continuous.
% from having to excise an arbitrarily small set of points where the Hölder-continuity of the separator might fail.
The proof of Theorem~\ref{thm:hoeldersepneu} is based on a regularized probabilistic uncertainty relation reported in Section~\ref{sec:regprobuncrel}. 
The regularization accounts for the H\"older-continuity of the separator.
%For the proof of Theorem~\ref{thm:hoeldersepneu}, we again apply our alternative technique since the arguments of the corresponding result in \cite[Thm.~18]{WV10} can not be applied to our setting as the overall measurement matrix $\hm={[\am\:\: \bm]}$ has an arbitrary but fixed
%block $\bm$. Due to the regularity condition on the separator we introduce a regularized version of the probabilistic uncertainty relation in Proposition~\ref{prop:injneu}, which is the crucial tool for establishing Theorem~\ref{thm:hoeldersepneu}. 
\end{remark}

\section{Probabilistic Uncertainty Relation}\label{sec:probuncrel}

The central conceptual element in the proof of Theorem~\ref{th2} is a probabilistic uncertainty relation, which leads to  uniqueness   guarantees 
for recovery of $\yv$ and $\zv$ from $\wv$ in \eqref{eq:model1}. Formally, the  question of uniqueness boils down to asking whether  different concatenated  source vectors  $\xv=[\yv^{\operatorname{T}} \:\: \zv^{\operatorname{T}}]^{\operatorname{T}}$ and $\xv'=[{\yv'}^{\operatorname{T}} \:\: {\zv'}^{\operatorname{T}}]^{\operatorname{T}}$ exist such that 
\ban 	[\am\:\: \bm]\xv=[\am\:\: \bm]\xv'	 ,\ean
or, equivalently,
\ban 	\am\pv=\bm\qv,	\label{eq:sparse}\ean
%The conventional way in compressed sensing to achieve uniqueness is through uncertainty principles.
%Much of the theory for compressed sensing was sparked by the Donoho-Stark uncertainty principle, which states that there cannot exist vectors that are sparse in both time and frequency \cite{DS89}. More specifically, in terms of the signal separation problem \eqref{eq:model1}, for 
with difference vectors $\pv:=\yv-\yv '$ and $\qv:=\zv'-\zv$. 
In the context of compressed sensing where $\yv,\yv ', \zv,$ and $\zv'$ are sparse signals, $\pv$ and $\qv$ are  sparse  as well so that %(the number of non-zero entries in, e.g., $\yv-\yv '$  does not exceed sum of the numbers of non-zero entries in $\yv$ and $\yv'$), i.e.,
 \eqref{eq:sparse} would imply  the existence of a non-zero signal $\sv:=\am\pv=\bm\qv$ that can be sparsely represented in both dictionaries $\am$ and $\bm$. Uncertainty principles are at the heart of compressed sensing theory and state that no such $\sv$ can exist if the  signals $\yv,\yv ', \zv,$ and $\zv'$ and hence $\pv$ and $\qv$ are sufficiently sparse and the dictionaries $\am$ and $\bm$ are sufficiently incoherent, thereby guaranteeing that, for a given $\wv$, there is a unique pair $(\yv,\zv)$ such that $\wv=\am\yv+\bm\zv$. 
Specifically, the Donoho-Stark uncertainty principle \cite{DS89} applies to  the square matrices
$\am=\boldsymbol{I}_n$ and $\bm=\fm_n$, and states that there exists no pair of vectors
$(\pv,\qv) \neq \0v$ with $2n_{\pv}n_{\qv}< n$ satisfying \eqref{eq:sparse},  
%\begin{align} 		\am\pv=\bm\qv,	\label{eq:1} \end{align}
where $n_{\pv}$ and $n_{\qv}$ denote the number of non-zero entries in $\pv$ and $\qv$, respectively. 
Elad and Bruckstein  \cite{EB02} generalized the Donoho-Stark uncertainty principle to arbitrary orthonormal bases  $\am$ and $\bm$ and found that   no %(sufficiently) \emph{sparse}
  pair of  vectors
$(\pv,\qv)\neq \0v$  with $(n_{\pv}+n_{\qv})/2<1/\mu$ satisfying \eqref{eq:sparse} exists. Here,
%\begin{align} 		\am\pv=\bm\qv,	\label{eq:2} \end{align}
 \ba \mu:=\sup_{1\leqslant i,j \leqslant n} |\langle \av_i , \bv_j \rangle | \ea  is %the  maximum of the absolute values of the inner products between columns of $\am $ with columns of $\bm$, 
 the coherence of $\am=[\av_1  \dots   \av_n]$ and $\bm=[\bv_1 \dots  \bv_n]$. This uncertainty principle  was further extended to redundant and undercomplete dictionaries in \cite{KDB12}.
The essence of uncertainty relations is that uniqueness 
in   signal separation  or signal recovery 
can be enforced by demanding that the signals to be  separated or recovered, respectively, be sufficiently sparse, provided that the underlying dictionaries are incoherent enough.

% of the solution by considering only those signals that are sufficiently sparse.
The central tool in the proof of Theorem~\ref{th2} is a  probabilistic uncertainty relation obtained as follows.  
We first rewrite \eqref{eq:sparse} as $[\am \:\:\: \bm][\pv^{\operatorname{T}} \, -\!\qv^{\operatorname{T}}]^{\operatorname{T}}=\0v$ and then assume that 
$[\pv^{\operatorname{T}}\,   -\!\qv^{\operatorname{T}}]^{\operatorname{T}}$ lies 
in a set $\mathcal S$ of (sufficiently) small Minkowski dimension. 
The probabilistic uncertainty relation, stated formally in Proposition \ref{prop:inj2}, says that for fixed $\bm$ and for a.a.\ $\am$, there is no $[\pv^{\operatorname{T}}\,  -\!\qv^{\operatorname{T}}]^{\operatorname{T}}\in \mathcal S$\,$\setminus$$\{\0v\} $ such that % \eqref{eq:1} and hence
$[\am\:\: \bm][\pv^{\operatorname{T}}  \,\, -$\,$\qv^{\operatorname{T}}]^{\operatorname{T}}=\0v$ or equivalently \eqref{eq:sparse} 
holds. 
  % (the case $(\pv,\qv) = (\0v, \0v)$ corresponds to the case $(\yv,\zv)= (\yv ', \zv ')$).  
Minkowski dimension here replaces sparsity in terms of the number of non-zero entries as a measure of 
description complexity of the signals to be  separated. 
\vspace*{-2truemm}
%, quantified by the number of their non-zero entries in the classical sparse signal recovery or signal separation problem.
%Minkowski dimension allows to capture the description complexity of sparse signals as in the situation above, but also applies to more general settings where the signals in question have a low description complexity.
% In particular, it is the joint description complexity of $(\pv,\qv)$ that counts, which might be lower than the sum of the individual description complexities of $\pv$ and $\qv$ due to dependence of the signals.  %This approach is inspired by \cite{SYC91}. 
%The formal statement of our probabilistic uncertainty relation is as follows.
\begin{proposition}\label{prop:inj2}
Let $\mathcal S\subseteq\mathbb R^{n}$ be non-empty and bounded such that $\underline{\dim}_{\text{B}}(\mathcal S)<k$, and let $\bm\in\mathbb R^{k\times \ell}$, with $k\geqslant \ell$, be a  matrix with $\rank(\bm)=\ell$. Then, 
\ban 	\{ \xv\in \mathcal S\!\setminus\! \{\0v\} \;  :   \; {[\am\:\: \bm]} \xv= \0v  \} = \emptyset , \label{eq:empty2}\ean
for a.a.\ $\am \in \mathbb R^{k\times (n-\ell)}$.
\end{proposition}
\begin{IEEEproof}
%We first argue that it is enough to show a probabilistic version of the statement.
%The proof relies on a concentration of measure result presented in Lemma~\ref{lemma2} at the end of this section. 
We show that \eqref{eq:empty2} holds with probability (w.p.) $1$ for the random matrix $\arm=[\arv_1\, \dots \, \arv_k]^{\operatorname{T}}$, where the $\arv_i$ are i.i.d.\ uniform on   $B^{n-\ell}(\0v,r)$  and $r>0$ is arbitrary.
%We consider the random matrix $\arm=[\arv_1\, \dots \, \arv_k]^{\operatorname{T}}$, where the $\arv_i$ are i.i.d.\ uniform on   $B^{n-\ell}(\0v,r)$ with arbitrary $r>0$, and show that \eqref{eq:empty2} holds with probability $1$. %If the set of matrices $\am$ violating \eqref{eq:empty2} would have positive Lebesgue measure, then there also was a bounded set of matrices $\am$ with positive Lebesgue measure violating \eqref{eq:empty2}. This bounded set can be made to be contained in the set of realizations of $\arm$ by choosing $r$ large enough. 
Since  $r$ can, in particular,  be chosen arbitrarily large, this establishes that the Lebesgue measure of matrices $\am$ violating \eqref{eq:empty2}  is zero.
%We draw $\am$ from the random matrix $\arm=[\arv_1\, \dots \, \arv_k]^{\operatorname{T}}$ where the $\arv_i$ are i.i.d.\ uniform on   $B^{n-\ell}(\0v,r)$ with arbitrary $r>0$.
%%Let $\arm$ be distributed as specified in Lemma~\ref{lemma2}. 
%If the set of matrices $\am$ violating \eqref{eq:empty2} would have positive Lebesgue measure, then there also was a bounded set of matrices $\am$ with positive Lebesgue measure violating \eqref{eq:empty2}. 
%This bounded set can be made to be contained in the set of realizations of $\arm$ by choosing $r$ large enough. 
%We can therefore conclude that in order to show
%For sufficiently large $r$, such a bounded set  is contained in the set of realizations of $\arm$. Therefore, in order to show 
%that the Lebesgue measure of matrices $\am$ violating \eqref{eq:empty2}  is zero, it suffices to prove  that 
%In order to show that the Lebesgue measure of matrices $\am$ violating \eqref{eq:empty2}  is zero,
%The proof is therefore established by showing that
%\begin{align} \mathbb P\vast [\exists \mleft[ \begin{matrix} \yv \\ \zv  \end{matrix} \mright] \in \mathcal S\! \setminus\! \{\0v\} :  {[\arm\:\: \bm]} \mleft[ \begin{matrix} \yv \\ \zv  \end{matrix} \mright] = \0v\vast]= 0, \label{eq:suffices2}\end{align}
%for all $r>0$. %The idea is  to apply Lemm\mright]a~\ref{lemma2} with $\uv=\yv$ and $\vv=\bm \zv$, but for the bound in Lemma~\ref{lemma2} to make sense, we need to ensure that $\yv\neq \0v$. 
We split $\xv=[\yv^{\operatorname{T}} \:\: \zv^{\operatorname{T}}]^{\operatorname{T}}$, where $\yv\in\IR^{n-\ell}$ and $\zv\in\IR^{\ell}$, and note that,   
%We start by arguing that, 
 thanks to the full-rank assumption on $\bm$, it suffices to show that
\begin{align} \mathbb P\vast [\exists \mleft[ \begin{matrix} \yv \\ \zv  \end{matrix} \mright] \in \mathcal S\! \setminus\! \{\0v\} :  {[\arm\:\: \bm]} \mleft[ \begin{matrix} \yv \\ \zv  \end{matrix} \mright] = \0v\vast]= 0 \label{eq:suffices2}\end{align}
% \eqref{eq:suffices2} 
for sets $\ss$ that have the norm of the $\yv$-parts of their elements  bounded  away from zero. % from below by a positive constant.
To see this, we first note that $\bm$, by virtue of being full-rank,
 maps non-zero vectors to non-zero vectors. For $[\yv^{\operatorname{T}}\; \zv^{\operatorname{T}}]^{\operatorname{T}}\in \mathcal S\! \setminus\! \{\0v\}$, $\am\yv + \bm \zv=\0v$ is  therefore possible only  for $\yv\neq \0v$ as $\yv=\0v$ would lead to $\bm\zv=\0v$ which  in turn would result in  $[\yv^{\operatorname{T}}\; \zv^{\operatorname{T}}]^{\operatorname{T}}=\0v$.
%implies $\zv=\0v$, i.e., $[\yv^{\operatorname{T}}\; \zv^{\operatorname{T}}]^{\operatorname{T}}=\0v$.
We can hence rewrite \eqref{eq:suffices2} as
\ban 
&\mathbb P\vast [\exists \mleft[ \begin{matrix} \yv \\ \zv  \end{matrix} \mright] \in \mathcal S\! \setminus\! \{\0v\} :  {[\arm\:\: \bm]} \mleft[ \begin{matrix} \yv \\ \zv  \end{matrix} \mright] = \0v\vast]\\
&=\mathbb P\vast [\exists \mleft[ \begin{matrix} \yv \\ \zv  \end{matrix} \mright] \in \mathcal S\! \setminus\! \{\0v\} , \yv \neq \0v    :  {[\arm\:\: \bm]} \mleft[ \begin{matrix} \yv \\ \zv  \end{matrix} \mright] = \0v\vast]. \label{eq:thus} \ean 
%\ban \mathbb P\vast [\exists \mleft[ \begin{matrix} \yv \\ \zv  \end{matrix} \mright] \in \mathcal S\! \setminus\! \{\0v\} :  {[\arm\:\: \bm]} \mleft[ \begin{matrix} \yv \\ \zv  \end{matrix} \mright] = \0v\vast]=\mathbb P\vast [\exists \mleft[ \begin{matrix} \yv \\ \zv  \end{matrix} \mright] \in \mathcal S\! \setminus\! \lefto (\{\0v\}\times \mathbb R^\ell \right)   :  {[\arm\:\: \bm]} \mleft[ \begin{matrix} \yv \\ \zv  \end{matrix} \mright] = \0v\vast]. \label{eq:thus} \ean 
A union bound argument applied to 
%the right-hand side (RHS) of 
\eqref{eq:thus} then yields
\ban &\mathbb P\vast [\exists \mleft[ \begin{matrix} \yv \\ \zv  \end{matrix} \mright] \in \mathcal S\! \setminus\! \{\0v\} :  {[\arm\:\: \bm]} \mleft[ \begin{matrix} \yv \\ \zv  \end{matrix} \mright] = \0v\vast]\\ &\leqslant  \sum_{m=1}^\infty \mathbb P\vast [\exists \mleft[ \begin{matrix} \yv \\ \zv  \end{matrix} \mright] \in  \mathcal S\! \setminus\! \{\0v\}, \|\yv\| \geqslant \frac{1}{m} : {[\arm\:\: \bm]}  \mleft[ \begin{matrix} \yv \\ \zv  \end{matrix} \mright]  = \0v \vast ].\ \  \ \ \label{eq:inf2} \ean 
This allows us to conclude that \eqref{eq:suffices2} is established by showing that
\begin{align} \mathbb P\vast [\exists \mleft[ \begin{matrix} \yv \\ \zv  \end{matrix} \mright] \in \mathcal S' :  {[\arm\:\: \bm]} \mleft[ \begin{matrix} \yv \\ \zv  \end{matrix} \mright] = \0v\vast]= 0, \label{eq:suffices3}\end{align}
for all non-empty  bounded sets $\mathcal S'\subseteq\mathcal S$ with \ban \inf \vast \{ \| \yv  \| :  \mleft[ \begin{matrix} \yv \\ \zv  \end{matrix} \mright]\in \mathcal S' \vast\}  >0,  \label{eq:min} \ean  
as this implies that each term in  the series in \eqref{eq:inf2} equals zero. Note that we no longer need to excise $\0v$ from $\ss'$ in \eqref{eq:suffices3}, as $\ss'$ is  guaranteed not to contain $\0v$ by definition, cf.\  \eqref{eq:min}. 

We next employ a covering argument, which reduces the question of the existence of $[\yv^{\operatorname{T}}\; \zv^{\operatorname{T}}]^{\operatorname{T}}\in \mathcal S'$ such that %\\[-.8cm]
\ban \arm\yv +\bm \zv =\0v \label{eq:reduce} \ean to the question of the existence of covering ball centers  satisfying \eqref{eq:reduce}. % for a covering of $\mathcal S'$.
For reasons that will become clear towards the end of the proof, we 
%For convenience of exposition, we 
employ the modified covering number $M_{\ss '}(\delta)$ (defined in \eqref{eq:modcovno}), which requires  the covering ball centers  to lie in $\ss '$. This implies  that  the covering ball centers  $[\yv_i^{\operatorname{T}}\; \zv_i^{\operatorname{T}}]^{\operatorname{T}}$    satisfy %$\yv_i\neq \0v$.  
\ban \min_{i} \| \yv_i \| \geqslant  \inf \vast \{ \| \yv  \| : \mleft[ \begin{matrix} \yv \\ \zv  \end{matrix} \mright]\in \mathcal S' \vast\}>0  .  \label{eq:centersinf}\ean 
%which, after invoking the concentration of measure result in Lemma~\ref{lemma2} at the end of this section, will allow us to bound the resulting term $1/\|\yv_i\|^k$  uniformly over $i$.
%
%
%, since for each individual term in \eqref{eq:inf2}  $m$ is fixed, and hence \eqref{eq:suffices2} follows by \eqref{eq:thus}. Next, we reduce the problem of the existence of $[\yv^{\operatorname{T}}\; \zv^{\operatorname{T}}]^{\operatorname{T}}\in \mathcal S'$ such that 
%\ban \am\yv +\bm \zv =\0v \label{eq:reduce} \ean to the existence of ball center  satisfying \eqref{eq:reduce} for a covering of $\mathcal S'$.
%To ensure that the centers  $[\yv_i^{\operatorname{T}}\; \zv_i^{\operatorname{T}}]^{\operatorname{T}}$  of balls covering $\mathcal S'$ also satisfy $\yv_i\neq \0v$, we use the modified covering number $M_{\ss '}(\varepsilon)$  which requires that the covering ball centers lie within the set $\ss '$ (see \eqref{eq:modcovno}).
%\begin{align}
 %&:=\min\Big\{m \in\IN\mid \ss '\subseteq\bigcup_{i\in\{1,\dots,m\}} B^n(\xv_i,\varepsilon),\ \xv_i\in \mathcal S'\Big\}. %\label{eq:modcovno}
%\end{align}
%where the centers of the covering balls are restricted to lie within the set $\ss '$. 
%It is shown in Lemma~\ref{lem:equivalentdef} in Appendix~\ref{app:eqdef} that the  Minkowski dimension of a non-empty bounded set does not change when we restrict the ball centers lie within the set.
% Minkowski dimension can equivalently be  defined using this modified covering number.
By definition of  $\underline{\dim}_{\text{B}}(\cdot)$ in \eqref{eq:defminkowski1} there exists 
a sequence of covering ball radii  $\delta_j$ tending to zero  with corresponding covering ball  centers $\xv_1,\ldots , \xv_{M_{\mathcal S'}(\delta_j)}\in \ss'$    such that 
\ban 	
\lim_{j\to\infty}\frac{\log  M_{\mathcal S'}(\delta_j)}{\log \frac{1}{\delta_j}} =	\underline{\dim}_{\text{B}}(\mathcal S').	 \label{eq:liminfdim}	\ean
 %By \eqref{eq:min} we can assume $j$ to be sufficiently large for  $\min \{ \|\yv_i\|  \mid 1\leqslant i \leqslant  N_{\mathcal S'}(\varepsilon_j) \}  >0$ to hold. 
Next, we note that 
\ban \| {[\arm\:\: \bm]} \xv\|\leqslant c(k,r, \|\bm\| ) \|\xv\|, \quad \text{for all $\xv\in \mathbb R^n$}, \label{eq:asshown} \ean since i) $ \| {[\arm\:\: \bm]} \xv\|\leqslant \| {[\arm\:\: \bm]} \| \cdot \| \xv\|$, ii) $\| {[\arm\:\: \bm]} \|  \leqslant \| \arm\| + \|\bm\|$, and iii)
\ban \| \arm \yv \|	&= \sqrt{\langle \arv_1, \yv\rangle^2 + \ldots +\langle \arv_k, \yv\rangle^2 } \label{eq:lip1} \\ &\leqslant \sqrt{\|\arv_1\|^2\|\yv\|^2 + \ldots +\|\arv_k\|^2\|\yv\|^2} \label{eq:lip2} \\ &< r \sqrt{k} \|\yv\|	\label{eq:lip3}	,\ean
for all $\yv\in \mathbb R^{n-\ell}$, implying $\|\arm\|< r \sqrt{k}$, where %\eqref{eq:lip2} follows by the Cauchy-Schwarz inequality and
 in \eqref{eq:lip3} we used $\arv_i\in B^{n-\ell}(\0v,r)$, for $i=1,...,k$.
%As the norm of each row of $\arm$ is bounded, all realizations of ${[\arm\:\: \bm]}$ have a common Lipschitz constant $L$, i.e., $\| {[\am\:\: \bm]} \uv\|\leqslant L \|\uv\|$ for all $\uv\in \mathbb R^n$ and all realizations $\am$ of $\arm$.
 Putting things together, we find that 
\ban  	
&\mathbb P\vast[\exists \mleft[ \begin{matrix} \yv \\ \zv  \end{matrix} \mright] \in \mathcal S' :  {[\arm\:\: \bm]} \mleft[ \begin{matrix} \yv \\ \zv  \end{matrix} \mright]= \0v\vast]\\ 
&\leqslant \sum_{i=1}^{ M_{\mathcal S'}(\delta_j)} \mathbb P[\exists \xv \in B^n(\xv_i, \delta_j):  {[\arm\:\: \bm]}\xv = \0v]		\label{eq:formera2}\\  
& \leqslant \sum_{i=1}^{M_{\mathcal S'}(\delta_j)} \mathbb P[\exists \xv \in B^n(\xv_i, \delta_j): \| {[\arm\:\: \bm]} \xv \|<\delta_j ]\\ 
& \leqslant \sum_{i=1}^{M_{\mathcal S'}(\delta_j)} \mathbb P[\| \arm\yv_i + \bm \zv_i \|<(c(k,r, \|\bm\|)+1)\delta_j ] \label{eq:formerb2}\\ 
&\leqslant C(n,k,r, \|\bm\|) \,  M_{\mathcal S'}(\delta_j) \, \delta_j^k,  \label{eq:formerc2}\ean
where \eqref{eq:formera2} follows from a union bound over the covering balls  $B^n(\xv_i, \delta_j)$ of $\mathcal S'$ and  in \eqref{eq:formerb2} we set $\xv_i=[\yv_i^{\operatorname{T}}\; \zv_i^{\operatorname{T}}]^{\operatorname{T}}$ and used
\ba 
\|\arm\yv_i + \bm \zv_i\|
&=\| {[\arm\:\: \bm]}\xv_i \|\\ 
&\leqslant  \| {[\arm\:\: \bm]} (\xv_i-\xv )\| + \| {[\arm\:\: \bm]} \xv \|\\
&\leqslant c(k,r, \|\bm\| ) \delta_j+ \| {[\arm\:\: \bm]} \xv \| .\ea 
Finally,  \eqref{eq:formerc2} is by application of the concentration of measure result in Lemma~\ref{lemma2} below (for fixed, albeit arbitrarily large, $r$),
where we used \eqref{eq:centersinf} to deduce that $\yv_i\neq \0v$, which, in turn, allows us %combined with \eqref{eq:min} 
to absorb the  term $1/\|\yv_i\|^k$ into the constant $C(n,k,r, \|\bm\|)$.
%together with \eqref{eq:min}, \eqref{eq:centersinf}, which implies that the  term $1/\|\yv_i\|^k$ in Lemma~\ref{lemma2} can be upper-bounded by a constant, absorbed in $C(n,k,r, \|\bm\|)$.
%\ba \min_{i=1,...,M_{\mathcal S'}(\varepsilon_j)} \| \yv_i \| \geqslant  \inf \vast \{ \| \yv  \| \; \bigg \vert \; \mleft[ \begin{matrix} \yv \\ \zv  \end{matrix} \mright]\in \mathcal S' \vast\} >0,  \ea 
%which follows by $\xv_i\in \mathcal S'$  and implies that the  term $1/\|\yv_i\|^k$ in Lemma~\ref{lemma2} can be upper-bounded by a constant. 
Now, 
% \eqref{eq:formerc2} is a consequence of 
\ban 	
\lim_{j\to\infty}\frac{\log\lefto ( M_{\mathcal S'}(\delta_j)\delta_j^k\right)}{\log \frac{1}{\delta_j}}
&=\lim_{j\to\infty}\frac{\log  M_{\mathcal S'}(\delta_j)}{\log \frac{1}{\delta_j}}-k\label{eq:newinturnimplies}\\ 
&=\underline{\dim}_{\text{B}}(\mathcal S')-k\\
&<0	\label{eq:inturnimplies},\ean
where we used \eqref{eq:liminfdim} together with $\underline{\dim}_{\text{B}}(\mathcal S')\leqslant \underline{\dim}_{\text{B}}(\mathcal S)$ thanks to  $\mathcal S'\subseteq\mathcal S$. Since $\lim_{j\to\infty}\log (1/\delta_j) =\infty$, the convergence of the left-hand side (LHS) of \eqref{eq:newinturnimplies} to a finite negative number implies that  $\lim_{j\to\infty}\log\lefto ( M_{\mathcal S'}(\delta_j)\delta_j^k\right)=-\infty$ and hence $\lim_{j\to\infty}M_{\mathcal S'}(\delta_j)\delta_j^k =0$.  % implies   $M_{\mathcal S'}(\varepsilon)\leqslant M_{\mathcal S}(\varepsilon)$ for all $\varepsilon>0$ and 
Taking the limit $j\to\infty$ in  \eqref{eq:formera2}--\eqref{eq:formerc2} implies that the LHS in \eqref{eq:formera2} equals zero, which concludes the proof.
% We have therefore shown that $\mathbb P[\exists \uv \in \mathcal S':  {[\arm\:\: \bm]} \uv = 0 ]=0$.
\end{IEEEproof}

Proposition~\ref{prop:inj2} shows that we can enforce uniqueness of the solution  in the recovery of $\yv$ and $\zv$ from $\wv$ in \eqref{eq:model1} by requiring that $\xv=[\yv^{\operatorname{T}}\; \zv^{\operatorname{T}}]^{\operatorname{T}}$  lie in a set with small enough Minkowski dimension. 
We note that this condition is in terms of a general measure for the description complexity of $\xrv$, namely Minkowski dimension, and includes the case of traditional sparsity as measured in terms of the number of non-zero entries. Section~\ref{sec:cs} elaborates on this matter.

%For the proof of the achievability result for Hölder continuous separators, Theorem~\ref{thm:hoeldersepneu}, we will develop a regularized version of the probabilistic uncertainty relation in Proposition~\ref{prop:injneu} in Section~\ref{sec:regprobuncrel} below, which parallels the original probabilistic uncertainty relation in Proposition~\ref{prop:inj2}. The new  methodology developed in the probabilistic uncertainty relations  is the third main contribution of this paper.

It remains to establish the concentration of measure result employed in the proof of Proposition~\ref{prop:inj2}. % This concentration inequality will also turn out  instrumental in the proof of Theorem~\ref{thm:hoeldersepneu}. 
%, which is based on a regularized version of the probabilistic uncertainty relation  Proposition~\ref{prop:inj2} presented in Appendix~\ref{app:technical}.

%This result is instrumental for Proposition~\ref{prop:inj2} and consequently provides the basis for the proof of Theorem~\ref{th2}. Moreover, the concentration of measure result also plays an important role in the proof of our second main result Theorem~\ref{thm:hoeldersepneu} for which we present a regularized version of the probabilistic uncertainty relation  Proposition~\ref{prop:inj2} in Appendix~\ref{app:technical}.

%the proof of both our main results Theorem~\ref{th2} and Theorem~\ref{thm:hoeldersepneu}.

\begin{lemma}\label{lemma2}
Let $\arm=[\arv_1\, \dots \, \arv_k]^{\operatorname{T}}$ be a random matrix in $\mathbb R^{k\times p}$ where the $\arv_i$ are i.i.d.\ uniform on   $B^p(\0v,r)$, for $r>0$. Then, for each $ \uv\in\IR^{p}\! \setminus\! \{\0v\}$, each $\vv\in\IR^{k}$, and all $\delta>0$, we have
\begin{align*}
\pr[\|\arm\uv+\vv\|<\delta] \leqslant  C(p,k,r)\frac{\delta^k}{\|\uv\|^k}.
\end{align*}
%where $C(n,k,r)$ is a constant that depends on $n$, $k$, and $r$ only.
\end{lemma}
\begin{IEEEproof}
We start by noting that, by assumption, the random matrix $\arm$ is uniformly distributed in the $k$-fold product  set $B^p(\0v,r)\times \ldots \times B^p(\0v,r)$, which is of  Lebesgue measure $\alpha(p,r)^{k}$. We therefore have
\begin{align}
&\pr[\|\arm\uv+\vv\|<\delta]\\
&=\frac{1}{\alpha(p,r)^{k}}\leb^{kp}\{\am\in B^p(\0v,r)\times \ldots \times B^p(\0v,r):\\
&\phantom{=\frac{1}{\alpha(p,r)^{k}}\leb^{kp}\{} \|\am\uv+\vv\|<\delta\}\\
%&\leqslant \prod_{i=1^k}\alpha(n-l,M)^{-k}\leb^{n-l}\{\av_i\in B(0,M)\mid \|\av^{\operatorname{T}}\uv_1+v_i\|<\delta\} \\
&\leqslant \frac{1}{\alpha(p,r)^{k}} \prod_{i=1}^k\leb^{p}\{\av_i\in B^{p}(\0v,r): |\av^{\operatorname{T}}_i\uv+v_i|<\delta\}\ \ \label{eq:conc0} \\
%&= \prod_{i=1}^k\leb^{n-l}\Big\{\av_i\in B^{n-l}(0,M)\mid \Big|\av^{\operatorname{T}}_i\frac{\uv}{\|\uv\|}+\frac{v_i}{\|\uv\|}\Big|<\frac{\delta}{\|\uv\|}\Big\}\\
&=\frac{1}{\alpha(p,r)^{k}} \prod_{i=1}^k\leb^{p}\lefto\{\um \av_i\in B^{p}(\0v,r):\phantom{\frac{\uv}{\|\uv\|}}\right.\\
& \phantom{=\frac{1}{\alpha(p,r)^{k}} \prod_{i=1}^k\leb^{p}\{} \left.\Big|(\um\av_i )^{\operatorname{T}} \frac{\uv}{\|\uv\|}+\frac{v_i}{\|\uv\|}\Big|<\frac{\delta}{\|\uv\|}\right \} \label{eq:concz} \\ 
&=\frac{1}{\alpha(p,r)^{k}} \prod_{i=1}^k\leb^{p}\lefto\{\av_i\in B^{p}(\0v,r): \phantom{\frac{v_i}{\|\uv\|}}\right.\\
&\phantom{=\frac{1}{\alpha(p,r)^{k}} \prod_{i=1}^k\leb^{p}\{}\left.\Big|\av^{\operatorname{T}}_i\ev_1+\frac{v_i}{\|\uv\|}\Big|<\frac{\delta}{\|\uv\|}\right \} \label{eq:conca} \\ 
%&\overset{(b)}{\leqslant} (2M)^{k(n-l-1)}\prod_{i=1}^k\leb^{1}\Big\{a_i\in B^{1}(0,M)\mid \Big|a_i+\frac{v_i}{\|\uv\|}\Big |<\frac{\delta}{\|\uv\|}\Big\}\\
&\leqslant \frac{(2r)^{k(p-1)}}{\alpha(p,r)^{k}} \prod_{i=1}^k\leb^{1}\lefto\{a_i\in\IR : \Big|a_i +\frac{v_i}{\|\uv\|}\Big|<\frac{\delta}{\|\uv\|}\right \}\  \label{eq:concb}\ \ \\
&= \frac{(2r)^{k(p-1)}(2\delta)^k}{\alpha(p,r)^{k} \|\uv\|^k} , \label{eq:concc}
\end{align}
where 
\eqref{eq:conc0} holds by the multiplicativity of  Lebesgue measure and because $\|\am\uv+\vv\|<\delta$ implies $|\av^{\operatorname{T}}_i\uv+v_i|<\delta$, for all $i$,
\eqref{eq:concz} follows from $\uv\neq \0v$ and  the fact that $\leb^{p}$ is invariant under rotations, 
with the specific rotation $\um^{\operatorname{T}}$ considered here 
taking  $\uv/\|\uv \|$ into $\ev_1=[1\,\, 0 \,\,\dots \,\,0]^{\operatorname{T}}\in\IR^p$, in \eqref{eq:conca}  we relabel $\um\av_i \in B^{p}(\0v,r)$ as $\av_i \in B^{p}(\0v,r)$,
and  in 
\eqref{eq:concb} we denote  the first component of the vector $\av_i$  by $a_i$, we relax the condition on the magnitude of $a_i$ to $a_i\in \mathbb R$, 
and we use the monotonicity of  Lebesgue measure together with the fact that the magnitudes of the remaining components of $\av_i$ are less than or equal to $r$. Finally, \eqref{eq:concc} follows by noting that in \eqref{eq:concb} we take the product over the Lebesgue measures of intervals of length $2\delta/ \|\uv\|$.
\end{IEEEproof}
%\begin{remark}

The proofs of  Proposition~\ref{prop:inj2} and Lemma~\ref{lemma2} are  inspired by the proofs of \cite[Lem.~4.2, Lem.~4.3]{SYC91}, but use a new proof technique that is more direct.  
%A statement similar to  was proved in  \cite[Lem.~4.3]{SYC91}. Specifically, the result in  applies to general linear combinations of Lipschitz mappings, covers the case $\underline{\dim}_{\text{B}}(\mathcal S) \geqslant k$ as well, and %applies to linear combinations of Lipschitz mappings in place of the matrix ${[\arm\:\: \bm]}$, and 
% gives an upper bound on the lower Minkowski dimension of the set on the left hand side (LHS) in \eqref{eq:empty2}. In particular, for $\underline{\dim}_{\text{B}}(\mathcal S)<k$, the case considered here, the upper bound in \cite[Lem.~4.3]{SYC91} implies  \eqref{eq:empty2}. The proof of \cite[Lem.~4.3]{SYC91} is based on concentration properties of the  singular values of $\hm$ \cite[Lem.~4.2]{SYC91}. 
% Our proof is more direct, does not rely on specifics of the distribution of $\hrm={[\arm\:\: \bm]}$, but applies to $\underline{\dim}_{\text{B}}(\mathcal S)<k$ only, the case relevant here. %Finally, the authors of \cite{SYC91} do not interpret the result  \cite[Lem.~4.3]{SYC91} as a probabilistic uncertainty relation.
%\end{remark}
Finally, we note that the probabilistic uncertainty relation developed here is a quite general tool and has been  applied to %based on the probabilistic uncertainty relation in Proposition~\ref{prop:inj2},
%besides applying to the signal separation problem,
% turned out to also be suitable for  
establish information-theoretic limits of matrix completion \cite{RSB15} and of phase retrieval \cite{RT15}.

\section{Proof of Theorem \ref{th2}}\label{proofTh1}
Since $R>\underline{R}_{\text{B}}(\varepsilon)=\limsup_{n\to\infty} \underline{a}_n(\varepsilon)$ and $k=\lfloor Rn \rfloor$, both by assumption, we have
\ban \underline{a}_n(\varepsilon)\overset{\textbf{.}}{<} \frac{k}{n}, \label{eq:start}\ean
which, together with the definition of $\underline{a}_n(\varepsilon)$, implies that there exists a sequence\footnote{The symbol $\us$ actually denotes the sequence $\us_n$. We decided to  drop the index $n$  for simplicity of exposition.} of non-empty  compact sets\footnote{\label{fn}Since lower Minkowski dimension is invariant under set closure (\hspace*{-0.8truemm}\cite[Prop. 3.4]{Fal04}), we can assume, w.l.o.g., that $\us$ is compact.} $\us := \us_n\subseteq\IR^n$ such that 
\begin{align}
\MIL(\us) &\overset{\textbf{.}}{<} k \label{eq:suff1}\\ 
\text{and } \quad \pr[\xrv\in \us] &\geqslant 1-\varepsilon. \label{eq:suff2}
\end{align} 
In the remainder of the proof we take $n$ to be sufficiently large for \eqref{eq:suff1} %and \eqref{eq:suff2} 
to hold in the ${\#}$-sense and we drop the dot-notation. %For $\am\in\IR^{k\times (n-\ell)}$ and $\bm\in\IR^{k\times \ell}$  we 
%Furthermore, since (lower) Minkowski dimension is invariant under set closure (cf. \cite[Prop. 3.4]{Fal04}), we can assume without loss of generality that $\us$ is compact.  
Let $\bm\in\IR^{k\times \ell}$ be an arbitrary but fixed full-rank matrix with $k\geqslant \ell$. Now, consider the mapping $\IR^{k\times (n-\ell)}\times\IR^n\times \IR^k\to \IR$, $(\am,\uv,\vv)\mapsto \|[\am \:\: \bm]\uv-\vv\|$. Since this mapping is  continuous and $[\am \:\: \bm]\uv=\vv$ if and only if $\|[\am \:\: \bm]\uv-\vv\|= 0$, 
%Since $g$ is jointly continuous and $\us$ is compact, 
 \cite[Prop. 14.33 and Cor. 14.6]{rowe98} and the compactness of $\us$ imply that there exists a measurable mapping $f\colon\IR^{k\times (n-\ell)}\times\IR^k\to\IR^n$ satisfying\footnote{\label{foot:1}For the detailed arguments leading to this statement, we refer to \cite{albolekori17}.}  
\begin{align}\label{eq:decoder}
f(\am,\vv)\in 
%\begin{cases}
\{\uv\in\us : {[\am \:\: \bm]}\uv= \vv\},
%[\am,\bm]^{-1}(\vv) \cap \us,&\quad \text{if}\ [\am,\bm]^{-1}(\vv) \cap \us=\{ \x\}\\
%\{\ev\},&\quad \text{else},
%\end{cases}
\end{align}
if $\{\uv\in\us : {[\am \:\: \bm]}\uv= \vv\}\neq\emptyset$ 
and $f(\am,\vv)=\ev\in\IR^n\!\setminus\us$ else. 
The mapping $g=f(\am,\cdot)$ therefore constitutes a valid (i.e., measurable) separator. This separator is guaranteed to deliver a $\uv\in\us$ that is consistent with the observation $\vv$ (in the sense of $ {[\am \:\: \bm]}\uv= \vv$)  if at least one such consistent $\uv$ exists, otherwise an error is declared by delivering the ``error symbol'' $\ev$. 
We can now define the set 
\begin{align}
\as
&:=\{(\am,\xv)\in \IR^{k\times (n-\ell)}\times\us:f(\am, {[\am\:\: \bm]}\xv)\neq\xv\}\ \ \label{eq:as1}\\
&=\{(\am,\xv)\in \IR^{k\times (n-\ell)}\times\us:g({[\am\:\: \bm]}\xv)\neq\xv\}\label{eq:as2}
\end{align} 
%We formally define the separator\footnote{Taking ``error'' to be an arbitrary element of $\mathbb R^n$, we obtain a  measurable map $g\colon \IR^k\to \IR^{n-\ell}\times \IR^\ell$ as required in Definition~\ref{defsourcerate2}. Specifically,  the measurability of $g$ follows by noting that $g$ is a right-inverse of ${[\am \:\: \bm]}$ on $\us$ and hence measurable by \cite[Sec.~7.2]{Che01} and a constant (and hence measurable) mapping else.
%being the  measurable right inverse of  ${[\am\:\:\bm]}$ restricted to $\us$, see \cite[Sec.~7.2]{Che01}.
%the  set $\{\uv \mid {[\am \:\: \bm]}\uv= \vv\} \cap \us$ being measurable, i.e., $g$ is a piecewise measurable map defined by a measurable condition.
%Note that ``error'' has to be chosen to be an element in ${\mathbb R^n\! \setminus \! \us}$ as otherwise the step from \eqref{eq:deca} to  \eqref{eq:decc} could be invalidated.
%Note that, we cannot choose an element of $\mathcal U$ since in this case \eqref{eq:decb} would not hold true anymore as the separator  might reconstruct the correct $\xv$ by ``chance''.
%}  (sequence) based on the approximate support set sequence $\us$ according to
%. If there are no such  elements in $\us$, the separator declares an error and delivers $\ev$ as result. 
and upper-bound the probability of decoding error according to 
%which, if there is a unique element $\xv$ in $\us$ that is consistent with the
% observation $\vv$ in the sense of   $ {[\am \:\: \bm]}\xv= \vv$, returns this element $\xv$, %if there is a unique vector with these properties, 
% and else (i.e.,  if there is none or more than one consistent $\xv$ in $\us$)
% returns ``error''. With this definition,
%we have
\begin{align}
p_\mathrm{e}(\am)&:=\pr [g({[\am\:\: \bm]}\xrv)\neq\xrv]\label{eq:pe1}\\ 
&=\pr [g({[\am\:\: \bm]}\xrv)\neq\xrv, \xrv\in\us]\\ 
&\phantom{=}+\pr [g([\am\:\:\bm]\xrv)\neq\xrv, \xrv\notin\us]\label{eq:dec} \\  %&\leqslant \pr [g([\am,\bm]\xrv)\neq\xrv, \xrv\in\us]+\pr [\xrv\notin\us]\nonumber\\
&\leqslant \pr [g({[\am\:\: \bm]}\xrv)\neq\xrv, \xrv\in\us]+\varepsilon \label{eq:deca}\\
&= \pr [(\am,\xrv)\in\as]+\varepsilon,\quad \am\in \IR^{k\times (n-\ell)},\label{eq:pe2}
%&=\,\pr [g({[\am\:\: \bm]}\xrv)= \text{error}, \xrv\in\us]+\varepsilon ,  \label{eq:decb}\\
%&= \,\pr [\exists \uv \in \us_{\xrv}\!\setminus\! \{\0v\}: {[\am\:\: \bm]} \uv= \0v, \xrv\in\us]+\varepsilon ,  \label{eq:decc}
\end{align}
%where $\us_{\xrv}:=\{\vv - \xrv \mid \vv\in \us\}$, 
where \eqref{eq:deca} follows from \eqref{eq:suff2}. 
Since $\as$ is measurable${}^{\ref{foot:1}}$, %as a consequence of $f(\am,\vv)$ being measurable and the diagonal $\{\xv\times\xv:\xv\in\IR^n\}$  being measurable in $\IR^n\times\IR^n$, 
we can apply Fubini's theorem \cite[Thm. 1.14]{ma99} to the indicator function on $\as$ and get 
\begin{align}
&\int_{\IR^{k\times (n-\ell)}}\pr [(\am,\xrv)\in\as] \mathrm d\am\label{eq:newiszero}\\
%&=\int_{\IR^{k\times (n-\ell)}} \mu_{\xrv}\{\xv:(\am,\xv)\in\as\} \mathrm d\am\\
%&=\int_{\IR^n} \leb^{k(n-\ell)}\{\am:(\am,\xv)\in\as\} \mu_{\xrv}(\mathrm d\xv)\\
&=\int_{\us} \leb^{k(n-\ell)}\{\am:(\am,\xv)\in\as\} \mu_{\xrv}(\mathrm d\xv). \label{eq:iszero} 
\end{align}
Note that for $\vv={[\am \:\: \bm]}\xv$ with $\xv\in\us$, the separator $g$ can  make an error only if there exists a $\uv\in\us\!\setminus\!\{\xv\}$ that is consistent with $\vv$, i.e., if $\vv={[\am \:\: \bm]}\uv$ for some $\uv\in\us\setminus\!\{\xv\}$. We therefore have 
\begin{align}
\as&\subseteq \{(\am,\xv)\in \IR^{k\times (n-\ell)}\times\us: \ker({[\am \:\: \bm]})\cap \us_{\xv}\neq\{\0v\}\}, \label{eq:as3}
\end{align} 
where 
\begin{align}
\us_{\xv}=\{\uv-\xv:\uv\in\us\},\quad \xv\in\us, 
\end{align}
which implies   
\begin{align}
\{\am:(\am,\xv)\in\as\}\subseteq \{\am:\ker({[\am \:\: \bm]})\cap \us_{\xv}\neq\{\0v\}\}, 
\end{align}
for all $\xv\in\us$. 
The monotonicity of Lebesgue measure therefore yields  
%Furthermore, we have for all $\xv\in\us$
\begin{align}
&\leb^{k(n-\ell)}\{\am:(\am,\xv)\in\as\}\\
&\leq \leb^{k(n-\ell)}\{\am:\ker({[\am \:\: \bm]})\cap \us_{\xv}\neq\{\0v\}\}\label{eq:lebas1},%\\
%&=0,\label{eq:lebas2}
\end{align}
for all $\xv\in\us$.   
%where \eqref{eq:lebas1} follows from \eqref{eq:as3} and the monotonicity of Lebesgue measure and \eqref{eq:lebas2} is a consequence of 
The probabilistic uncertainty relation, Proposition~\ref{prop:inj2}, with $\ss=\us_{\xv}$ and $\MIL(\us_{\xv})<k$ ((lower) Minkowski dimension is invariant under translation, as seen by translating covering balls accordingly, and hence $\MIL(\us_{\xv})=\MIL(\us)$) implies that  \eqref{eq:lebas1} equals zero for all  $\xv\in\us$. 
Therefore,   \eqref{eq:newiszero} equals zero as well, which, by \eqref{eq:pe1}--\eqref{eq:pe2}, implies $p_\mathrm{e}(\am)\leq \varepsilon$ for a.a.  $\am$ and thereby completes the proof. 
 \endIEEEproof

%Therefore, the integral of \eqref{eq:pr2} with respect to $\mu_{\xrv}(\mathrm d\xv)$ equlas zero, and, 
%noting that \eqref{eq:pr2} can be written as an integral with respect to $\mathrm d\am$, we can apply Fubini's Theorem to interchange the two integrals and obtain
%\ban 	\int_{\mathbb R^{k\times (n-\ell)}}\pr [\exists \uv \in \us_{\xrv}\!\setminus\! \{\0v\}: {[\am\:\: \bm]} \uv= \0v, \xrv\in\us]\mathrm d\am=0.	\label{eq:fub}	\ean
%We therefore  have $\pr [\exists \uv \in \us_{\xrv}\!\setminus \! \{\0v\}: {[\am\:\: \bm]}\uv= \0v, \xrv\in\us]=0$ for a.a.\ $\am\in\mathbb R^{k\times (n-\ell)}$, which upon insertion into \eqref{eq:decb} yields
%\ba 	\pr [g({[\am\:\: \bm]}\xrv)\neq\xrv]\leqslant \varepsilon,		\ea
%for a.a.\ $\am$ and thereby completes the proof. 
%\endIEEEproof

%\appendices
\section{Regularized Probabilistic Uncertainty Relation}\label{sec:regprobuncrel}
In this section, we develop the regularized  probabilistic uncertainty relation the proof of Theorem~\ref{thm:hoeldersepneu} is based on. 
%In this section, we collect the  technical results needed in the proof of Theorem~\ref{thm:hoeldersepneu}, which we present in Section~\ref{sec:hoeldersep}. The main  result in this section is 
%a regularized version of the probabilistic uncertainty relation presented in Proposition~\ref{prop:injneu} which parallels the original probabilistic uncertainty relation in Proposition~\ref{prop:inj2}.
We start with results on the existence  of H\"older continuous separators.

\begin{definition}
For $\mathcal A\subseteq \mathbb R^n$, $\mathcal B\subseteq \mathbb R^m$, and $\beta >0$, a map $f\colon \mathcal A \to \mathcal B$ is $\beta$-H\"older continuous if there exists a constant $c>0$ such that for all $\xv_1,\xv_2\in \mathcal A$ we have
\ba \|f(\xv_1)-f(\xv_2)\| \leqslant c \|\xv_1 -\xv_2\|^\beta  .\ea
\end{definition}

\begin{lemma}\label{lem:hoelder}
For a map $f\colon \mathcal A \to \mathcal B$, where $\mathcal A\subseteq \mathbb R^n$ and $\mathcal B\subseteq \mathbb R^m$, there exist $c>0$ and $\beta>0$ such that
\ban c \|\xv_1 -\xv_2\|^{1/\beta} \leqslant \|f(\xv_1)-f(\xv_2)\|,\label{eq:shift}
\ean
for all $\xv_1,\xv_2\in \mathcal A$, 
if and only if $f$ is injective and $f^{-1}\colon f(\mathcal A)\to \mathcal A$ is $\beta$-H\"older continuous.
\end{lemma}
\begin{IEEEproof}
If \eqref{eq:shift} holds, then $f$ is injective as for all $\xv_1,\xv_2\in \as$ with $\xv_1\neq \xv_2$, we have
\ba 	\|f(\xv_1)-f(\xv_2)\| \geqslant c \|\xv_1 -\xv_2\|^{1/\beta} >0 , \ea
and hence  $f(\xv_1)\neq f(\xv_2)$. Therefore, $f^{-1}\colon f(\mathcal A)\to \mathcal A$ is well-defined. Moreover,  for all $\yv_1,\yv_2\in  f(\mathcal A)$ we can find $\xv_1,\xv_2\in \as$ such that $f(\xv_i)=\yv_i$ and hence  $\beta$-H\"older continuity of $f^{-1}$ follows from 
\ba 
\|f^{-1}(\yv_1)-f^{-1}(\yv_2)\|&= \|\xv_1-\xv_2\|\\ &\leqslant \frac{1}{c^\beta}\| f(\xv_1)-f(\xv_2)\|^\beta\\ &= \frac{1}{c^\beta} \| \yv_1-\yv_2\|^\beta, \ea
where the inequality is by \eqref{eq:shift}.

Conversely, suppose that $f$ is injective and $f^{-1}\colon f(\mathcal A)\to \mathcal A$ is $\beta$-H\"older continuous. Then,  for all $\xv_1,\xv_2\in \as$, by  $\beta$-H\"older continuity of $f^{-1}$ there exists  a constant $C$ such that
\ban 	\| f^{-1}(f(\xv_1)) - f^{-1}(f(\xv_2))\| \leqslant C \| f(\xv_1) - f(\xv_2)\| ^\beta .	 \label{eq:settingbeta}\ean
Since $f^{-1}(f(\xv_i))=\xv_i$ by injectivity of $f$, this implies \eqref{eq:shift} with $c:=1/C^{1/\beta}$. % in \eqref{eq:settingbeta}.
%
%
%, then there exists a constant $C$ such that for all $\yv_1,\yv_2\in f(\as)$ 
%\ba  \|	f^{-1}(\yv_1) - f^{-1}(\yv_2)\| \leqslant C \| \yv_1-\yv_2\|^\beta . 	\ea
%
%
%By replacing $\xv_i$  in \eqref{eq:shift}   by $f^{-1}(f(\xv_i))$ for $i=1,2$ %and shifting the constants $c$ and $\beta$  to the other side 
%the statement follows directly from the definition of  $\beta$-Hölder continuity.
\end{IEEEproof}

For a linear map $f$, e.g., the map  induced by a realization of the random matrix ${[\arm \:\: \bm]}$, verifying \eqref{eq:shift}  reduces to checking the condition
\ban 	\inf_{\xv\,\in\,(\as\ominus\as)\setminus\{\0v\}}  \frac{\|f(\xv)\|}{\|\xv\|^{1/\beta}}>0.\label{eq:simpl}	\ean
%\ban 	c \|\xv \|^{1/\beta} \leqslant \|f(\xv)\|, \quad \text{for all $\xv\in \as\ominus\as$}	.	\label{eq:simpl}\ean
%Note that for $c>0$, \eqref{eq:simpl} is equivalent to 
%\ban 	c \|\xv \|^{1/\beta} \leqslant \|f(\xv)\|, \quad \text{for all $\xv\in \as\ominus\as$}	.	\label{eq:simpl2}\ean

We next provide a sufficient condition---that is convenient to check---for %the existence of a constant $c$ such that
  \eqref{eq:simpl}  to hold. For expositional simplicity, we 
 formulate the condition  for general sets $\ss$ in place of $\as\ominus\as$. 
 %In fact, in the proof of Theorem~\ref{thm:hoeldersepneu}, we will  verify \eqref{eq:simpl} for $\ss=\as\ominus \{\av\}$  in place of $\as\ominus\as$, for  fixed elements $\av\in \as$, and employ Fubini's theorem  similar to the one in the proof of Theorem~\ref{th2}. 
% In the proof of Theorem~\ref{thm:hoeldersepneu}, this will allow us to verify \eqref{eq:simpl} with $\ss=\as\ominus \{\xv\}$  in place for $\as\ominus\as$, for a fixed $\xv\in \as$, and employ a Fubini-maneuver similar to the one in the proof of Theorem~\ref{th2}.
%This condition essentially consists of ensuring that the images of elements in $\as\ominus\as$ with a ball cut out map to vectors outside a ball that is at most exponentially in $\beta$ smaller
The  condition we establish essentially consists of checking whether the elements in the set  obtained upon excision of a ball of radius $2^{-j\beta}$ from $\ss$ map to points outside a ball of radius $2^{-j}$.
%checking lower bounds on the norm of images under $f$ after cutting out balls from the domain $\as\ominus\as$. The lower bound to be met is exponentially---in $\beta$---smaller than the radius of the balls cut out.
%The radius of the balls cut out scales exponentially in $\beta$, which accounts for the $\beta$-Hölder continuity. Intuitively, we can decompose $\as\ominus\as$ in \eqref{eq:shift} into different levels and verify \eqref{eq:shift} on these levels individually. 
A related approach was used in \cite[p.~3736]{WV10}. 
\begin{lemma}\label{lem:memberneu}
Let  $\ss$ be a nonempty and bounded set in $\mathbb R^n$, $\ss\neq\{\0v\}$, $f\colon \ss \to \mathbb R^k$, $\beta\in (0,1)$, and  
$\delta_j:=2^{-j}$.
%$(\delta_j)_{j\in\mathbb N}$ be a decreasing sequence tending to $0$ for $j\to \infty$ with $\alpha:=\inf_{j\in\mathbb N}\delta_{j+1}/\delta_j>0$.
 If there is a $J\in \mathbb N$ such that for  all $j\geqslant J$ we have
\ban 	\text{ 	$\|f(\xv)\|\geqslant \delta_j $, \quad for all $\xv\in \ss \! \setminus \!B^n(\0v, \delta_j^{\beta} )$},	\label{eq:assumption0} \ean
then 
\ban 	\inf_{\xv\,\in\,\ss\setminus\{\0v\}}  \frac{\|f(\xv)\|}{\|\xv\|^{1/\beta}}>0.\label{eq:assumption1}	\ean
\end{lemma}
\begin{IEEEproof}
We show that there exists a constant $c(\ss)>0$ such that 
\ban c(\mathcal S) \|\xv\|^{1/\beta}\leqslant \|f( \xv)\|, \quad \text{for all $\xv \in \ss$},  \ean
which is equivalent to \eqref{eq:assumption1}. 
Let $\xv\in\mathcal S$, $\mathcal S_j:= \ss \! \setminus \!B^n(\0v, \delta_j^{\beta} ) $,  and $i_{\xv} :=\min \{ i\in \mathbb N : \xv\in \mathcal S_i \}$  (see Figure~\ref{fig:2} for an illustration).  

%\vspace*{-25truemm}
\begin{figure}[h]\centering
\vspace*{-20truemm}
\begin{tikzpicture}[scale=3]
 \draw[thick] (-1,0) .. controls (-1,1.3) and (0.2,-.2) .. (.2,.6)
               .. controls (0.2,2) and (1,0.555) .. (1,0)
		.. controls (1,-.3)  and (.9,-.5) .. (.7,-.5)
		.. controls (.5,-.5) and (-.3,-.4) .. (-.3,-.9)
		.. controls (-.3,-1.2) and (-.5,-1.3) .. (-.6,-1.3)
		.. controls (-1,-1.3) and (-1,-.4) .. (-1,0);

\filldraw[fill=darkblue!20, thick] (-.3,-.1) circle (14pt);
\filldraw[fill=darkblue!10, thick] (-.3,-.1) circle (9pt);
\filldraw[fill=darkblue!5, thick] (-.3,-.1) circle (5pt);

%\filldraw[pattern=horizontal lines light gray] (-.3,-.1) circle (11pt);
%\filldraw[pattern=horizontal lines light blue] (-.3,-.1) circle (8pt);
%\filldraw[pattern=horizontal lines gray] (-.3,-.1) circle (5pt);
\node at (-.3,-.1) {\tiny $\times$};
\node at (-.51,-.2)  {\tiny $\times$};
\node at (-.47,-.28)  {$\xv$};
\node at (-.27,-.18)  {$\mathbf 0$};
\node at (1,-.5)  {$\mathcal S$};

%\node at (.3,-.8)  {$B^n(\mathbf 0, 2^{-(j-1)\beta})$};
\draw[->] (.3,-.8) node[right] {$B^n(\mathbf 0, 2^{-(i_{\xv}-2)\beta})$} .. controls (.2,-.8)  and (0.35,-.25) .. (0.2,-.2);
\draw[->] (.2,-1) node[right] {$B^n(\mathbf 0, 2^{-(i_{\xv}-1)\beta})$} .. controls (.1,-1)  and (0.2,-.25) .. (0.01,-.22);
\draw[->] (.1,-1.2) node[right] {$B^n(\mathbf 0, 2^{-i_{\xv}\beta})$} .. controls (0,-1.2)  and (0.1,-.35) .. (-0.15,-.22);
\end{tikzpicture}
\caption{\small Illustration of the set $\ss_j$ for $j=i_{\xv}-2,i_{\xv}-1 ,i_{\xv}$.}
\label{fig:2}
\end{figure}
%By assumption,   there is a $J\in \mathbb N$ such that  \eqref{eq:assumption0} holds for all $j\geqslant J$.
 We then find that
\ban
	\|f( \xv )\| &\geqslant \begin{cases}\delta_{i_{\xv}}, &\text{if $i_{\xv}\geqslant J$}\\ 	\delta_{J}, &\text{if $i_{\xv}< J$}	\end{cases}\label{eq:membneu}  \\ &\geqslant \begin{cases}\|\xv\|^{1/\beta}/2, &\text{if $i_{\xv}\geqslant J$} \\ 	2^{-J} \frac{\|\xv\|^{1/\beta}}{\sup_{\uv\in\mathcal S}\|\uv\|^{1/\beta} },&\text{if $i_{\xv}< J$}	\end{cases}  \label{eq:memcneu} \\ %&\label{eq:mema}\geqslant \begin{cases}\|\xv\|^{1/\beta}/2, &\text{if $i_{\xv}\geqslant J$}\\ 	2^{-J} \frac{\|\xv\|^{1/\beta}}{2^{1/\beta}\sup_{\uv\in\mathcal S}\|\uv\|^{1/\beta}},&\text{if $i_{\xv}< J$}	\end{cases}\\ 
	&\geqslant c(\mathcal S) \|\xv\|^{1/\beta}, \label{eq:memdneu}
\ean
where \eqref{eq:membneu} follows, for ${i_{\xv}}\geqslant J$, from $\xv\in \mathcal S_{i_{\xv}}$ together with \eqref{eq:assumption0}; and  for $i_{\xv}<J$, from $\xv\in \mathcal S_{J}$ and \eqref{eq:assumption0} with $j=J$. In \eqref{eq:memcneu} we used %that $i_{\uv}$ is the minimum $i$  such that $ \uv\in \mathcal T_i(\xv )$ which implies that 
$\|\xv\|<2^{-(i_{\xv}-1)\beta}$, for $i_{\xv}\geqslant J$, 
%$\delta_{i_{\xv}}> (\delta_{i_{\xv}}/\delta_{i_{\xv}-1})\|\xv\|^{1/\beta} \geqslant  \alpha \|\xv\|^{1/\beta}$, for $i_{\xv}\geqslant J$, 
and for $i_{\xv}<J$ we apply the trivial  bound $\|\xv\| \leqslant \sup_{\uv\in\mathcal S}\|\uv\|$. Finally, in \eqref{eq:memdneu} we set  $c(\mathcal S)=  \min \lefto\{1/2 , \frac{2^{-J}}{\sup_{\uv\in\mathcal S}\|\uv\|^{1/\beta}}\right \}$, and we note that $c(\mathcal S)>0$  by virtue of  $\mathcal S$ being bounded and $J<\infty$.
\end{IEEEproof}

We are now ready to present the announced regularized probabilistic uncertainty relation. In the original probabilistic uncertainty relation, stated  in Proposition~\ref{prop:inj2}, we showed that for fixed $\bm$, for  a.a.\ $\am$ there are no non-zero vectors in $\ss$ that map to zero under ${[\am\:\:\bm]}$ provided that the lower Minkowski dimension of $\ss$ is sufficiently small. 
The regularized version of this result  states that the norm of the image of  a vector
%\footnote{It is more convenient to formulate the regularized probabilistic uncertainty relation for vectors $\xv\in\mathbb R^n$}  
$\xv\in\ss$ under ${[\am\:\:\bm]}$   does not become too small relative to $\|\xv\|$.
This will then allow us to deduce the existence of a H\"older continuous separator in Theorem~\ref{thm:hoeldersepneu} by applying Lemma~\ref{lem:hoelder}.

%bounded away from zero do not get too small.

%To prove our general achievability result for Hölder continuous separators, we need a regularized version of this statement providing a guarantee that the norms of images under ${[\am\:\:\bm]}$  of vectors  in $\ss$ bounded away from zero do not get too small. 
\begin{proposition}\label{prop:injneu}
Let $\bm\in\mathbb R^{k\times \ell}$, with $k\geqslant \ell$, have $\rank(\bm)=\ell$, let $\mathcal S\subseteq\mathbb R^{n}$ be non-empty and bounded with $\ss\neq\{\0v\}$ and $\overline{\dim}_{\text{B}}(\mathcal S)<k$, and fix $\beta\in\IR$ such that 
\ban 0<\beta < 1-\frac{ \overline{\dim}_{\text{B}}(\mathcal S)}{k}. \label{eq:betaa} \ean
Then, for a.a.\ $\am\in\mathbb R^{k\times (n-\ell)}$, we have 
\ban 	\inf_{\xv\,\in\,\ss\setminus\{\0v\}}  \frac{\|[\am \:\: \bm]\xv\|}{\|\xv\|^{1/\beta}}>0.\label{eq:properties}	\ean
\end{proposition}
\begin{IEEEproof}
As in the proof of Proposition~\ref{prop:inj2}, 
we show that   for the random matrix $\arm=[\arv_1\, \dots \, \arv_k]^{\operatorname{T}}$, with the $\arv_i$ i.i.d.\ uniform on   $B^{n-\ell}(\0v,r)$ with arbitrary $r>0$, \eqref{eq:properties} holds w.p. $1$. 
Since $r$ can be taken arbitrarily large, this establishes that the Lebesgue measure of matrices $\am$ for which  \eqref{eq:properties} does not hold is zero. 
As \eqref{eq:properties} is an inequality of the form \eqref{eq:assumption1} we can apply Lemma~\ref{lem:memberneu} with $f(\xv)={[\arm \:\: \bm]}\xv$  to conclude that showing
%w.p. $1$ (w.r.t. $\arm$) there exists a $J\in \mathbb N$ such that
%Let 
%$\delta_j:=2^{-j}$
%%$(\delta_j)_{j\in\mathbb N}$ be a sequence tending to such that 
%%\ban 	\frac{\log  M_{\mathcal S}(\delta_j)}{\log \frac{1}{\delta_j}} \; \xrightarrow{j\to\infty}\;	\underline{\dim}_{\text{B}}(\mathcal S).	 \label{eq:liminfdimneu}	\ean
%%Without loss of generality we may assume that $(\delta_j)_{j\in\mathbb N}$ is decreasing and $\inf_{j\in\mathbb N}\delta_{j+1}/\delta_j>0$. We
%and  $\mathcal S_j:= {\ss \! \setminus \!B^n(\0v, \delta_j^{\beta} )} $.
%Using Lemma~\ref{lem:memberneu}  for $f={[\arm \:\: \bm]}$ and $\xv=[\yv^{\operatorname{T}}\; \zv^{\operatorname{T}}]^{\operatorname{T}}$
% it suffices to show that with probability $1$ there is a $J\in \mathbb N$ such that for all $j\geqslant J$
\ban & \mathbb P[\exists J\in \mathbb N :\| [\arm \:\: \bm] \xv\|\geqslant \delta_j ,\\ 
&\phantom{\mathbb P[\exists J\in \mathbb N :} \text{ for all  $j\geqslant J$ and all $\xv\in \mathcal S_j$}]=1, \ean
with $\delta_j:=2^{-j}$ and  $\mathcal S_j:= {\ss \! \setminus \!B^n(\0v, \delta_j^{\beta} )} $, establishes the proof.  
%and all $j\geqslant J$. 
Applying the Borel-Cantelli  Lemma  \cite[Thm.~2.3.1]{Dur13} to the complementary events it follows that it  suffices to show that
\ban  \sum_{j=0}^\infty \mathbb P[ \exists \xv\in \mathcal S_j : \| [\arm \:\: \bm] \xv\|<\delta_j ] < \infty. \label{eq:accomplishedneu} \ean
The basic idea for establishing \eqref{eq:accomplishedneu} is to cover $\ss_j$ with balls of radius $\delta_j$ and to upper-bound
 the probabilities in \eqref{eq:accomplishedneu} by probabilities that are  in terms of
  the corresponding covering ball centers. 
  Specifically, with the  minimum number of balls of radius $\delta_j$ needed to cover $\ss_j$ denoted by $M_j:=M_{\mathcal S_j}(\delta_j)$   
  and the corresponding ball centers $\xv^{(j)}_1,\ldots, \xv^{(j)}_{M_j}\in \mathcal S_j$, we establish that 
 \ban  
 &\mathbb P[ \exists \xv\in \mathcal S_j : \| [\arm \:\: \bm] \xv\|<\delta_j ]\label{eq:newprobboundneu}\\ 
 &\leqslant \sum_{i=1}^{M_j}\mathbb P[\|{[\arm\:\: \bm]}  \xv_i^{(j)}\|<(L +1)\delta_j]. \label{eq:probboundneu}  \ean
  Here, $L:= c(k,r,\|\bm\|)$ is the constant in \eqref{eq:asshown}. %  and we set  ${[\arm\:\: \bm]} := {[\arm\:\: \bm]}$.
To prove \eqref{eq:newprobboundneu}--\eqref{eq:probboundneu}, first note that the existence of an $\xv\in \mathcal S_j$ such that $\|{[\arm\:\: \bm]} \xv\|<\delta_j$ implies 
$\xv\in B^n(\xv_{i_0}^{(j)},\delta_j)$ for some $i_0\in\{1,...,M_j\}$, since the balls $ B^n(\xv_i^{(j)},\delta_j)$, $i=1,...,M_j$, cover $\mathcal S_j$. It then follows that
\ban \|{[\arm\:\: \bm]} \xv_{i_0}^{(j)}\| &\leqslant  \|{[\arm\:\: \bm]} \xv\|+ \|{[\arm\:\: \bm]}( \xv_{i_0}^{(j)}-{\xv} )\| \label{eq:est1} \\ &<   \delta_j + L\delta_j=(L+1)\delta_j,  \label{eq:est3} \ean
where we used $\|{[\arm\:\: \bm]}  \xv\| <\delta_j$ and \eqref{eq:asshown}. From \eqref{eq:est1}, \eqref{eq:est3}, and a union bound argument we then get  \eqref{eq:newprobboundneu}--\eqref{eq:probboundneu}.
%\ban \|{[\arm\:\: \bm]} \xv\| &\geqslant \|{[\arm\:\: \bm]} \xv_{i_0}^{(j)}\|- \|{[\arm\:\: \bm]}( \xv_{i_0}^{(j)}-{\xv} )\|\\ &\geqslant  (L +1)\delta_j - L\delta_j\\ &=\delta_j, \ean
%where we used $\|{[\arm\:\: \bm]}  \xv_{i_0}^{(j)}\|\geqslant (L+1)\delta_j$ and \eqref{eq:conse}.
% $\|{[\arm\:\: \bm]}  \xv_i^{(j)}\|\geqslant (L+1)\delta_j$ implies that for all $\xv\in B^n(\xv_i^{(j)},\delta_j)$, we have
%\ban \|{[\arm\:\: \bm]} \xv\| &\geqslant \|{[\arm\:\: \bm]} \xv_i^{(j)}\|- \|{[\arm\:\: \bm]}( \xv_i^{(j)}-{\xv} )\|\\ &\geqslant  (L +1)\delta_j - L\delta_j\\ &=\delta_j, \ean
%together with the fact that %by definition of $\mathcal Q_j(\xv_0)$, and since
% the balls $ B^n(\xv_i^{(j)},\delta_j)$, $i=1,...,M_j$, cover $\mathcal S_j$.
%We can therefore conclude that the existence of an $\xv=[\yv^{\operatorname{T}}\; \zv^{\operatorname{T}}]^{\operatorname{T}}\in \mathcal S_j$ such that $\|{[\arm\:\: \bm]} \xv\|<\delta_j$ implies  $\|{[\arm\:\: \bm]}  \xv_i^{(j)}\| < (L+1)\delta_j$ for at least one $i\in\{1,...,M_j\}$. A union bound argument  leads to
%\ban  \mathbb P[ \exists [\yv^{\operatorname{T}}\; \zv^{\operatorname{T}}]^{\operatorname{T}}\in \mathcal S_j : \| \arm \yv +\bm \zv\|<\delta_j ] \leqslant \sum_{i=1}^{M_j}\mathbb P[\|{[\arm\:\: \bm]}  \xv_i^{(j)}\|<(L +1)\delta_j]. \label{eq:probboundneu}  \ean 
We  now  turn to bounding the terms in the sum  of \eqref{eq:probboundneu} and will then use these bounds in  \eqref{eq:accomplishedneu} to establish the final result.
Let us start by writing the covering ball centers as \ba \xv_i^{(j)} =\mleft[ \begin{matrix}\yv_i^{(j)} \\ \zv_i^{(j)}  \end{matrix} \mright], \ea 
with $\yv_i^{(j)}\in\mathbb R^{n-\ell}$ and  $\zv_i^{(j)}\in \mathbb R^{\ell}$, and splitting them into two groups according to 
\ban
\{ \xv^{(j)}_1,\ldots, \xv^{(j)}_{M_j} \}= \mathcal X_1^{(j)}\cup \mathcal X_2^{(j)},
\ean
where 
\ban \mathcal X_1^{(j)}&:= \{\xv_i^{(j)} : \|\yv_i^{(j)}\| < c \|\zv_i^{(j)}\| \} \label{eq:defX1neu}\\  \mathcal X_2^{(j)}&:= \{\xv_i^{(j)} : \|\yv_i^{(j)}\|  \geqslant  c \|\zv_i^{(j)}\| \}  , \label{eq:defX2neu}\ean
with the constant $c>0$  chosen below. The reasoning behind this splitting is as follows. For ball centers in $\mathcal X_1^{(j)}$, 
we  establish that the corresponding probabilities  of \eqref{eq:probboundneu} equal zero for sufficiently large $j$, whereas for ball centers in $\mathcal X_2^{(j)}$, we use the concentration inequality in Lemma~\ref{lemma2} to establish that the corresponding terms in the sum   in \eqref{eq:probboundneu} are sufficiently small to result in a finite upper bound as required in  \eqref{eq:accomplishedneu}.
We first note that for all ball centers  
\ban \|\xv_i^{(j)}\|^2=\|\yv_i^{(j)}\|^2+\|\zv_i^{(j)}\|^2\geqslant \delta_j^{2\beta},  \ean
by virtue  of  $\xv_i^{(j)}\in \mathcal S_j$.
We now turn to the set  $ \mathcal X_1^{(j)}$. From \eqref{eq:defX1neu} we  get
\ban  (c^2+1)\|\zv_i^{(j)}\|^2 \geqslant \|\yv_i^{(j)}\|^2+\|\zv_i^{(j)}\|^2\geqslant   \delta_j^{2\beta}. \label{eq:zlowneu} \ean
%Since $\bm$ has full-rank, by assumption, there exists a constant $C_{\bm}>0$ such that $\|\bm \zv\|\geqslant C_{\bm} \|\zv\|$ for all $\zv\in\mathbb R^\ell$.  
This allows us to deduce that
\ban  \| {[\arm\:\: \bm]} \xv_i^{(j)} \| 	 &\geqslant \| \bm \zv_i^{(j)}\| - \|\arm \yv_i^{(j)}\|		\label{eq:low1neu} \\ 
						& 	\geqslant  C_{\bm} \|\zv_i^{(j)}\| - \| \arm \| \| \yv_i^{(j)}\| 		\label{eq:low2neu} \\ 
						& 	\geqslant  C_{\bm} \|\zv_i^{(j)}\| -c \| \arm \| \| \zv_i^{(j)}\| 	\label{eq:low3neu}	\\ 
						&\geqslant \frac{C_{\bm}-c L}{\sqrt{1+c^2}}\delta_j^\beta ,		\label{eq:low4neu}	
	\ean
where in \eqref{eq:low1neu} we applied the reverse triangle inequality, for \eqref{eq:low2neu} we note that there exists a constant $C_{\bm}>0$ such that $\|\bm \zv\|\geqslant C_{\bm} \|\zv\|$, for all $\zv\in\mathbb R^\ell$, as a consequence of $\bm$ being full-rank,  
 in \eqref{eq:low3neu} we used $\xv_i^{(j)}\in \mathcal X_1^{(j)}$,  and  in \eqref{eq:low4neu} we employed \eqref{eq:zlowneu} and $\|\arm\|\leqslant \|{[\arm\:\: \bm]} \|\leqslant L$, where $L$ was defined right after \eqref{eq:probboundneu}. 
Since $\beta\in(0,1)$, $\delta_j^\beta$ can be made arbitrarily large relative to $\delta_j$ (i.e., $\delta_j^\beta/\delta_j=2^{j(1-\beta)}$ can be made arbitrarily large) by taking $j$ sufficiently large. Specifically, choosing\footnote{This is possible since $C_{\bm}/L >0$.} $c>0$ to ensure $C_{\bm}-c L>0$, we can find a $J_1\in \mathbb N$ such that 
	\ban 				\frac{C_{\bm}-c L}{\sqrt{1+c^2}}\delta_j^\beta&\geqslant  (L+1)\delta_j, \quad \text{for all $j\geqslant J_1$.}	\label{eq:J1}	\ean
By \eqref{eq:low4neu} this implies 
\ban \| {[\arm\:\: \bm]} \xv_i^{(j)}\| \geqslant  (L+1)\delta_j, \label{eq:low5neu} \ean
 for all $j\geqslant J_1$, and hence establishes that 
 \ban \mathbb P[\|{[\arm\:\: \bm]}  \xv_i^{(j)}\|<(L +1)\delta_j]=0  ,\label{eq:verw1} \ean
 % for all $\xv_i^{(j)}\in \mathcal T_j(\xv_0)$ the norm of their images under ${[\arm\:\: \bm]}$ is lower-bounded, showing that 
% the terms on the RHS of \eqref{eq:probboundneu} equal zero 
for $\xv_i^{(j)}\in \mathcal X_1^{(j)}$ and $j\geqslant J_1$. %Each term on the RHS of \eqref{eq:probboundneu} for which  $j<J_1$ can  be bounded by $1$ resulting in a total bound of $J_1$.

Next, consider $\xv_i^{(j)}\in \mathcal X_2^{(j)}$. From \eqref{eq:defX2neu} we get
\ban  \left (1+\frac{1}{c^2} \right ) \|\yv_i^{(j)}\|^2\geqslant  \|\yv_i^{(j)}\|^2+\|\zv_i^{(j)}\|^2 \geqslant \delta_j^{2\beta},  \label{eq:lowX2neu}\ean
which, using the  concentration inequality in Lemma~\ref{lemma2}, allows us to conclude that 
\ban &\mathbb P[\|\arm  \yv_i^{(j)}+\bm  \zv_i^{(j)}\| < (L +1)\delta_j]\\
 &\leqslant C(n,k,r)\frac{(L+1)^k\delta_j^k}{\|\yv_i^{(j)}\|^k}\label{eq:finalize1neu}  \\ &\leqslant C(n,k,r,L)\left (\sqrt{1+\frac{1}{c^2}}\right )^k \frac{2^{-jk}}{2^{-\beta j k}}  . \label{eq:finalize2neu} \ean
%for $\xv_i^{(j)}\in \mathcal X_2^{(j)}$, where in \eqref{eq:finalize1neu} we used the  concentration inequality Lemma~\ref{lemma2} noting that, by \eqref{eq:lowX2neu}, $\|\yv_i^{(j)}\|\neq 0$ and \eqref{eq:finalize2neu} follows from \eqref{eq:lowX2neu}.
Putting things together, we obtain
\ban	 &\sum_{j=0}^\infty \mathbb P[ \exists \xv\in \mathcal S_j : \| [\arm \:\: \bm] \xv\|<\delta_j ]\\ 
&\leqslant  J_1+ \sum_{j=J_1}^\infty\! \mathbb P[ \exists \xv\in \mathcal S_j : \| [\arm \:\: \bm] \xv\|<\delta_j ]  \label{eq:chain1neu}\\
&=J_1+\sum_{j=J_1}^\infty \sum_{i=1}^{M_j} \mathbb P[\|{[\arm\:\: \bm]}  \xv_i^{(j)}\|<(L +1)\delta_j] \label{eq:chain2neu}\\
&= J_1+\sum_{j=J_1}^\infty \sum_{\xv_i^{(j)}\in \mathcal X_2^{(j)}}\!\!\!\!\!\mathbb P[\|{[\arm\:\: \bm]}  \xv_i^{(j)}\|<(L +1)\delta_j] \label{eq:chain3neu}\\
%\sum_{j=0}^{\infty}	\mathbb P[{[\arm\:\: \bm]} \notin \mathcal Q_j(\xv_0)] &\leqslant  J_1+\sum_{j=J_1}^{\infty}\mathbb P[{[\arm\:\: \bm]} \notin \mathcal Q_j(\xv_0)] \label{eq:chain1neu}\\
%		 &=J_1	+\sum_{j=J_1}^{\infty}\sum_{i=1}^{M_j}\mathbb P[\|{[\arm\:\: \bm]}  \xv_i^{(j)}\| < (L +1)\delta_j] \label{eq:chain2neu}\\ 
%		&=J_1	+\sum_{j=J_1}^{\infty}\sum_{\xv_i^{(j)}\in \mathcal X_2^{(j)}}\mathbb P[\|\arm  \yv_i^{(j)}+\bm  \zv_i^{(j)}\| < (L +1)\delta_j]	\label{eq:chain3neu}\\
		&\leqslant J_1+\sum_{j=J_1}^{\infty}\sum_{\xv_i^{(j)}\in \mathcal X_2^{(j)}}\!\!\!\! C(n,k,r,L)\left (\sqrt{1+\frac{1}{c^2}}\right )^k \frac{2^{-jk}}{2^{-\beta j k}}\ \ \ \ \ \ \  \label{eq:chain5neu}\\
		 &\leqslant J_1+C(n,k,r,L,c) \sum_{j=J_1}^{\infty} M_j 2^{-jk(1-\beta)}		\label{eq:chain6neu}\\
		 &\leqslant J_1+C(n,k,r,L,c,\mathcal S) \sum_{j=J_1}^{\infty} 2^{jd'}2^{-jk(1-\beta)} 	\label{eq:chain7neu}\\ 
		&= J_1+C(n,k,r,L,c,\mathcal S)  \sum_{j=J_1}^{\infty}2^{-jk \left (1-\frac{d'}{k}-\beta \right )}	\label{eq:chain8neu}\\ 
		&<\infty.  	\label{eq:chain9neu} \ean
Here, in  \eqref{eq:chain1neu} we upper-bounded the probability of the terms for
 $j<J_1$ by $1$, where
$J_1$  was defined in \eqref{eq:J1},
 \eqref{eq:chain2neu} is by \eqref{eq:newprobboundneu}--\eqref{eq:probboundneu}, \eqref{eq:chain3neu} follows %since  for $\xv_i^{(j)}\in \mathcal X_1^{(j)}$ and $j\geqslant J_1$ we have $\mathbb P[\|{[\arm\:\: \bm]}  \xv_i^{(j)}\| < (L +1)\delta_j]=0$ by  
from  \eqref{eq:verw1},
  in \eqref{eq:chain5neu} we invoked \eqref{eq:finalize2neu}, %in \eqref{eq:chain5} we used \eqref{eq:lowX2}, 
  and  \eqref{eq:chain6neu} holds since $|\mathcal X_2^{(j)}|\leqslant M_j$. For \eqref{eq:chain7neu}, we 
  set $d'= \overline{\dim}_{\text{B}}(\mathcal S)+\alpha$ with $\alpha>0$ small enough so that $1-\frac{d'}{k}>\beta$, which is possible by \eqref{eq:betaa},  and we used 
  \ban	M_j=M_{\ss_j}(\delta_j)&\leqslant N_{\ss_j}(\delta_j/2) \label{eq:triangleargument}  \\ &\leqslant N_{\ss}(\delta_j/2) \label{eq:subneu} \\ &\leqslant C(\ss) \delta_j^{-d'}, 	\label{eq:Mepsneu}  \ean
  %
  %used
 % \ban M_j\leqslant C(\mathcal U) \delta_j^{-k'}=C(\mathcal U) 2^{jk'} \label{eq:Mepsneu} \ean
  for all $j\in\mathbb N$,  where \eqref{eq:triangleargument} follows from a triangle inequality argument (cf.\ \eqref{eq:absetzen}), \eqref{eq:subneu} holds as\footnote{Note that we resort to the original covering number $N_\delta(\as)$ in this argument, as for the modified 
   covering number $M_\delta(\as)$ the relation $\as\subseteq\bs$ does not imply, in general, that $M_\delta(\as)\leqslant M_\delta(\bs)$. % , which is why we use the original covering number $N_\delta(\as)$ in this argument.
   } $\ss_j\subseteq \ss$, for all $j$, and \eqref{eq:Mepsneu}
 is a consequence of \begin{enumerate}[label=\roman*)] \item  $ \overline{\dim}_{\text{B}}(\mathcal S)<d'$ and thus $N_{\ss}(\delta_j/2)\leqslant (\delta_j/2)^{-d'}=2^{d'}\delta_j^{-d'}$ for sufficiently large $j$ by definition of $\limsup$, and % by \eqref{eq:dimUneu} together with
%Lemma~\ref{lem:equivalentdef} in Appendix~\ref{app:eqdef} (translation of $\us$ does not change the covering number), and  
  \item $C(\ss)$ taken sufficiently large  so that \eqref{eq:Mepsneu} also holds for
   the (finite number of) $j$'s for which  $N_{\ss}(\delta_j/2)\leqslant (\delta_j/2)^{-d'}$ does not hold. Note that $C(\ss)$ is guaranteed to be finite as  the set $\ss$ is bounded and therefore the covering numbers $N_{\ss}(\delta_j/2)$ are finite for all $j$. %  in the constant $C(\mathcal S) $.
   \end{enumerate}
Finally, \eqref{eq:chain9neu} follows from 
\ban 1-\frac{d'}{k} >\beta  ,\ean
which is by choice of $d'$.
%By  \eqref{eq:chain1neu}--\eqref{eq:chain9neu} we established \eqref{eq:accomplishedneu} thereby completing 
This completes the proof.
\end{IEEEproof}

\section{Proof of Theorem~\ref{thm:hoeldersepneu}}\label{sec:hoeldersep}
 We start with preparatory material.
Since, by assumption, $\beta>0$ is fixed and satisfies $1- \frac{\overline{R}_\text{B}(\varepsilon)}{R}>\beta $, we can find an $\alpha>0$  such that
\ban 1 - \frac{\overline{R}_\text{B}(\varepsilon)+\alpha}{R}>\beta \label{eq:delta} .\ean
Let $k':= (\overline{R}_\text{B}(\varepsilon)+\alpha)n$. By definition of $\overline{R}_\text{B}(\varepsilon)$, we can find a sequence of  non-empty compact\footnote{Since upper Minkowski dimension is invariant under set closure (\hspace*{-0.85truemm}\cite[Prop. 3.4]{Fal04}), we can assume, w.l.o.g., that $\us$ is compact.} sets $\mathcal U\subseteq \mathbb R^n$ such that
\begin{align}
\MIU(\us) &\overset{\textbf{.}}{<} k'  \label{eq:dimU}\\ 
\text{and } \quad \pr[\xrv\in \us] &\geqslant 1-\varepsilon.  \label{eq:probU}
\end{align} 
Moreover, %by definition of $k$ (in the statement of the theorem) and $k'$, 
we have
\ban
1-\frac{k'}{k}	&=\frac{\lfloor Rn \rfloor -  (\overline{R}_\text{B}(\varepsilon)+\alpha)n}{\lfloor Rn \rfloor} \label{eq:beta1}\\ 
			&\overset{\textbf{.}}{\geqslant } \frac{\lfloor Rn \rfloor - (\overline{R}_\text{B}(\varepsilon)+\alpha)n}{ Rn } \label{eq:beta0}\\
			&> \frac{R-\frac{1}{n}-\overline{R}_\text{B}(\varepsilon)-\alpha}{R} \label{eq:beta2}\\ 
			&\overset{\textbf{.}}{>} \beta , \label{eq:beta3} \ean
where  \eqref{eq:beta0} follows from $\lfloor Rn \rfloor \leqslant Rn$ and $\lfloor Rn \rfloor -  (\overline{R}_\text{B}(\varepsilon)+\alpha)n > (R-\frac{1}{n})n-  (\overline{R}_\text{B}(\varepsilon)+\alpha)n \overset{\textbf{.}}{>} 0$ (since $R>\overline{R}_\text{B}(\varepsilon)+\alpha$, by choice of $\alpha$), in \eqref{eq:beta2} we used $Rn-1<\lfloor Rn \rfloor$, and in \eqref{eq:beta3} we invoked \eqref{eq:delta}. In the remainder of the proof, we take $n$ sufficiently large for \eqref{eq:dimU}--\eqref{eq:beta3} to hold in the ${\#}$-sense and drop the dot-notation. 
%BRUCH (motivation...)
%Next,  we employ the regularized probabilistic uncertainty relation  to find a  constant $c(\us_{\xv}, \am,\bm)>0$, where $\us_{\xv}:=\{\vv - \xv : \vv\in \us\}$ for fixed $\xv\in\mathbb R^n$, such that 
%\begin{align} c(\us_{\xv}, \am,\bm) \| \vv - \xv \|^{1/\beta}\leqslant \| \hm \vv-\hm \xv\|	, \quad \text{for all $\vv\in\us$}.\label{eq:c}
%\end{align}
%which by Lemma~\ref{lem:hoelder} implies that the inverse to the restricted map $\hm \colon \us_{\xv} \to \hm(\us_{\xv})$ is $\beta$-Hölder continuous. The main technical difficulty that remains is to extend this ``localized'' argument to all of $\us$ universally with respect to $\xv$. To overcome this difficulty we will employ a similar F ubini-maneuver as in the proof of Theorem~\ref{th2}.
%
%
%to establish the existence of a $\beta$-Hölder continuous separator. 
For $\bm$  as in the statement of the theorem, $\am\in\IR^{k\times {(n-\ell)}}$, and 
 $\xv\in\mathbb R^n$,  we set\footnote{We use the convention  $\inf(\emptyset)=\infty$.}
 %\footnote{To avoid the case where the infimum is not well-defined, we assume that the set $\us$ is not a singleton, i.e., it does not consist of a single point $\us=\{\uv\}$ only. If $\us$ is a singleton, finding a  $\beta$-Hölder continuous separator $g$ satisfying \eqref{eq:resulthoeldersepneu}  is trivial as every $\beta$-Hölder continuous map $g \colon \mathbb R^k \to \mathbb R^n$ with $g(\hm\uv)=\uv$ is a valid separator.}
\begin{align}\label{eq:As}
\as:=\Bigg\{(\am,\xv):\inf_{\uv\,\in\,\us_{\xv}\setminus\{\0v\}}  \frac{\|[\am\:\:\bm] \uv\|}{\|\uv\|^{1/\beta}}=0\Bigg\}, 
\end{align}
with 
\begin{align}
\us_{\xv}=\{\uv-\xv:\uv\in\us\}. 
\end{align}
Since (upper) Minkowski dimension is invariant under translation (as seen by translating covering balls accordingly), we have $\MIU(\us_{\xv})=\MIU(\us)$ which, together with  \eqref{eq:dimU} and \eqref{eq:beta1}--\eqref{eq:beta3}, implies  $1-\frac{\MIU(\us_{\xv})}{k}>1-\frac{k'}{k} >\beta$ for all $\xv\in\IR^n$. 
We can therefore apply the regularized probabilistic uncertainty relation, Proposition~\ref{prop:injneu}, to each $\us_{\xv}$ with  $\xv\in\IR^n$  and get    %(recall the definition of $\as$ in \eqref{eq:as})
\begin{align}\label{eq:pr2neu}
\leb^{k(n-\ell)}\{\am : (\am,\xv)\in\as \}=0,
\end{align}
for all $\xv\in \mathbb R^n$. Integrating \eqref{eq:pr2neu} w.r.t. $\mu_{\xrv}(\mathrm d\xv)$ yields
\ban 	
\int_{\mathbb R^{n}}\leb^{k(n-\ell)}\{\am : (\am,\xv)\in\as \} \mu_{\xrv}(\mathrm d\xv) =0 .	\label{eq:combining} 
\ean
We next show that $\as$ is measurable, which will allow us to change the order of integration in \eqref{eq:combining} by applying Fubini's theorem  \cite[Thm. 1.14]{ma99} to the indicator function on $\as$. This will be accomplished by showing that the sets 
\begin{align}\label{eq:ajs}
\as_j:=\Bigg\{(\am,\xv):\inf_{\uv\,\in\,\us_{\xv}\setminus\{\0v\}}  \frac{\|[\am\:\:\bm] \uv\|}{\|\uv\|^{1/\beta}}>\frac{1}{j}\Bigg\}
\end{align}
are measurable for all $j\in\mathbb N$ and using\footnote{Complements and countable unions of measurable sets are again measurable.} 
\begin{align}\label{eq:Ascomp}
\as^\mathrm{c}
&=\bigcup_{j\in\mathbb N}\as_j,  
\end{align}
where $\as^\mathrm{c}$ denotes the complement of $\as$ in $\mathbb R^{k\times (n-\ell)}\times \mathbb R^{n}$. 
Indeed, for each $j\in\mathbb N$, we can write 
\begin{align}
\as_j
%&=\{(\am,\xv):c(\us_{\xv}, \am,\bm)\geq\alpha\}\\
%&=\Big\{(\am,\xv):\inf_{\uv\,\in\,\us_{\xv}\setminus\{\0v\}}  \frac{\|[\am\:\:\bm] \uv\|}{\|\uv\|^{1/\beta}}\geq\alpha\Big\}\\
%&=\big\{(\am,\xv):  \|\hm \uv\|\geq\alpha\|\uv\|^{1/\beta},\ \text{for all}\ \uv\in\us_{\xv}\setminus\{\0v\}\big\}\\
&=\big\{(\am,\xv):  j\|[\am\:\:\bm] \uv\|\geq\|\uv\|^{1/\beta},\ \text{for all}\ \uv\in\us_{\xv}\big\}\\
&=\big\{(\am,\xv):  j\|[\am\:\:\bm](\uv-\xv)\|\geq\|\uv-\xv\|^{1/\beta},\\
&\phantom{=\big\{}\, \text{for all}\ \uv\in\us\big\}\\
&=\big\{(\am,\xv):\inf_{\uv\in\us}\big(j\|[\am\:\:\bm] (\uv-\xv)\|-\|\uv-\xv\|^{1/\beta}\big)\\
&\phantom{=\big\{}\,\geq0\big\},\label{eq:Aa}
\end{align} 
where \eqref{eq:Aa} is measurable as a consequence of \cite[Prop. 14.40]{rowe98}, upon noting that  $\us$ is compact and 
\begin{align}
h(\am,\xv,\uv):=j\|[\am\:\:\bm](\uv-\xv)\|-\|\uv-\xv\|^{1/\beta}
\end{align} 
is a continuous mapping. Fubini's theorem therefore yields 
\ban  
&\int_{\mathbb R^{n}}\leb^{k(n-\ell)}\{\am : (\am,\xv)\in\as \} \mu_{\xrv}(\mathrm d\xv) \label{eq:newfub2neu}\\%\label{eq:fub1neu}\\
%&=\int_{\mathbb R^{k\times (n-\ell)}} \mu_{\xrv}\{\xv : (\am,\xv)\in\as \} \mathrm d\am\label{eq:fubneu}\\
&=\int_{\mathbb R^{k\times (n-\ell)}} \pr [(\am,\xrv)\in\as] \mathrm d\am. \label{eq:fub2neu} 
%&\int_{\mathbb R^{n}}\leb^{k(n-\ell)}\{\am \mid  c(\us_{\xv}, \am, \bm)=0 \} \mu_{\xrv}(\mathrm d\xv) \label{eq:fub1neu} \\  &= \int_{\mathbb R^{n}} \left [ \int_{\mathbb R^{k\times (n-\ell)}} \mathds{1}_{\lefto \{(\widetilde \am, \, \widetilde \xv) \,  \mid  \,  c(\us_{\widetilde  \xv},\,\widetilde  \am,\, \bm)=0 \right \} }(\am,\xv) \, \mathrm d\am \right ]  \mu_{\xrv}(\mathrm d\xv)\\ &=\int_{\mathbb R^{k\times (n-\ell)}} \left [\int_{\mathbb R^{n}}  \mathds{1}_{\lefto \{(\widetilde \am, \, \widetilde \xv) \,  \mid  \,   c(\us_{\widetilde \xv},\, \widetilde  \am,\, \bm)=0 \right \}}(\am,\xv)  \,  \mu_{\xrv}(\mathrm d\xv)  \right ] \mathrm d\am \label{eq:fubneu}\\
%&=\int_{\mathbb R^{k\times (n-\ell)}}\pr [c(\us_{\xrv}, \am, \bm)=0]\mathrm d\am ,   \label{eq:fub2neu}
 \ean
%where in \eqref{eq:fub2neu} we applied .  
Combining \eqref{eq:combining} with \eqref{eq:newfub2neu}--\eqref{eq:fub2neu}, we  conclude that 
\ban \pr [(\am,\xrv)\in\as] =0,\quad\text{for a.a.} \ \am,\label{eq:obtainneu}\ean
which is equivalent to 
\ban \pr [(\am,\xrv)\in\as^\mathrm{c}] =1,\quad\text{for a.a.} \ \am.\label{eq:obtainneu2}\ean
Using \eqref{eq:Ascomp}, \eqref{eq:obtainneu2}, and $\as_{j}\subseteq\as_{j+1}$, for all $j\in\mathbb N$, \cite[Lem. 3.4, Part (a)]{ba95} implies that 
\ban 
\lim_{j\to\infty}\pr \lefto  [(\am,\xrv)\in\as_j\right ] = 1,\quad\text{for a.a.} \ \am.  \ean
Therefore, for every $\kappa>0$, there exists a $J(\am)\in \mathbb N$ such that 
\ban 
\pr \lefto  [(\am,\xrv)\in\as_{J(\am)}\right ] \geqslant 1-\kappa,\quad\text{for a.a.} \ \am.  \ean
Moreover, since $\pr [\xrv\in\us]\geq 1-\varepsilon$, a union bound argument yields 
\ban 
\pr \lefto  [\xrv\in  \us_{\am} \right ] \geqslant 1-\kappa-\varepsilon,\quad\text{for a.a.} \ \am,  \ean
where 
\ban
\us_{\am}:= \Big\{\xv : (\am,\xv)\in\as_{J(\am)}\Big\}\cap\us. 
\ean
Since $\us_{\am}\subseteq \as_{J(\am)}$, we can conclude that for a.a.\ $\am$ the following holds: 
\begin{align}
 \| \xv_1 - \xv_2 \|^{1/\beta}\leqslant J(\am)\|[\am\:\:\bm](\xv_1-\xv_2)\|, 
\end{align}
for all $\xv_1,\xv_2\in\us_{\am}$.
By Lemma~\ref{lem:hoelder}, $\hm={[\am\:\:\bm]}$ is therefore injective on $\mathcal U_{\am}$ and its inverse  $\hm^{-1} \colon \hm(\mathcal U_{\am}) \to \mathcal U_{\am}$ is $\beta$-H\"older continuous, and by \cite[Thm.~1, ii)]{Min69} (restated below for completeness)  with $\vs=\mathbb R^k$, $\ws=\mathbb R^n$, and $\bs(\am)=\hm(\mathcal U_{\am})$ the inverse $\hm^{-1}$ can be extended to a $\beta$-H\"older continuous mapping $g_{\hm} \colon \mathbb R^k \to \mathbb R^n$. Again, this statement holds for  a.a. $\am$. Finally, thanks to  injectivity of $\hm$ on $\us_{\am}$, for a.a. $\am$, we have $g_{\hm}(\hm \xv)=\xv$ for all $\xv \in \us_{\am}$ and for a.a. $\am$, and therefore
\ba \mathbb P[g_{\hm}(\hm \xrv)\neq \xrv] \leqslant \mathbb P[\xrv \notin \mathcal U_{\am}] \leq\varepsilon +\kappa, \ea
for a.a. $\am$. This completes the proof.
\endIEEEproof

Finally, for the reader's convenience, we provide the following (reformulated) version of the statement \cite[Thm.~1, ii)]{Min69}.
\begin{theorem}
Let $\mathcal V,\mathcal W$ be Euclidean spaces and let $g \colon \bs \to \mathcal W$ be $\beta$-H\"older continuous with  $0<\beta < 1$ and $\bs\subseteq \mathcal V$.  Then, $g$ can be extended to a  $\beta$-H\"older continuous mapping on all of $\mathcal V$.
\end{theorem}

\section{To sparse signal separation}\label{sec:cs}

% Theorems~\ref{th2} and~\ref{thm:hoeldersepneu} provide achievability statements.  
 
Converses  for the achievability statements in Theorems~\ref{th2} and~\ref{thm:hoeldersepneu}  seem difficult to obtain for 
general sources. We can, however, build on  \cite[Thm.~15]{WV10}, which establishes a converse  for the  analog compression problem for sources  of mixed  discrete-continuous distribution, and derive a   converse to   Theorems~\ref{th2} and~\ref{thm:hoeldersepneu} for mixed  discrete-continuous sources. % distributions. 
%In the context of analog compression, \cite[Thm.~15]{WV10} establishes a converse for  sources of mixed  discrete-continuous distribution.
% It turns out,  that these results can be extended to the signal separation problem and we find a   converse statement for both  Theorem~\ref{th2} and~\ref{thm:hoeldersepneu} in the case of concatenated source vectors 
% of mixed  discrete-continuous distribution. 
%In this section, we establish a converse for sources of mixed  discrete-continuous distribution.
Mixed  discrete-continuous sources  are of particular interest as their Minkowski dimension effectively quantifies the number of non-zero entries and hence reflects  the traditional  sparsity notion used, e.g., in \cite{SKP12, DK10,DS89,CRT06,Don06}.
%establish the connection to the traditional sparse signal separation problem considered, e.g., in \cite{Li12, WM10, DK10, CT12, PBS13, SKP12, DH01, DS89}. %, 
Specifically, we  consider concatenated source vectors $\xrv$ with independent entries of mixed discrete-continuous  distribution and  possibly different mixture parameters for  the constituent processes $(\IrY_i)_{i\in \mathbb N}$ and  $(\IrZ_i)_{i\in \mathbb N}$.

\begin{definition}\label{definitionM}
We say that $\xrv$ in Definition \ref{definitionX} has a mixed discrete-continuous distribution if 
for each $n\in\IN$ the random variables $\IrX_i$ for $i\in\{1,\dots,n\}$ are independent and distributed according to  
\begin{align}\label{eq:auchverweisen}
\mu_{\IrX_i}=
\begin{cases}
(1-\rho_1)\mu_{\text{d}_1}+\rho_1\mu_{\text{c}_1},\quad i\in\{1,\dots, n-\ell\}\\
(1-\rho_2)\mu_{\text{d}_2}+\rho_2\mu_{\text{c}_2},\quad i\in\{n-\ell+1,\dots, n\} ,
\end{cases}
\end{align}
where $0\leqslant \rho_i\leqslant 1$ are mixture parameters, $\ell=\lfloor \lambda n \rfloor$, the $\mu_{\text{d}_i}$ are discrete distributions, and the $\mu_{\text{c}_i}$ are absolutely continuous (w.r.t. Lebesgue measure) distributions.
\end{definition}

%The model of mixed discrete-continuous sources incorporates the traditional sparse signal model. 
Before  stating the converse, we  extend---to concatenated source vectors---\!\!  \cite[Thm.~6]{WV10}, which shows that, indeed, the Minkowski dimension compression rate of mixed discrete-continuous sources reflects the traditional notion of sparsity.
Specifically, if the discrete parts $\mu_{\text{d}_1}$, $\mu_{\text{d}_2}$ are Dirac measures at $0$, i.e.,  $\mu_{\text{d}_1} =\mu_{\text{d}_2}= \delta_0$, then the non-zero entries of $\xrv$ can be generated only by the continuous parts $\mu_{\text{c}_1}$, $\mu_{\text{c}_2}$.
%discrete parts always yield a zero entry in the source vector. 
%In order to quantify the number of non-zero entries in the concatenated source vector $\xrv$, we set
With
\ba \widetilde{\IrY}_i&:=\mathds{1}_{\mathbb R\setminus\{0\}}(\IrY_i), \quad i=1,...,n-\ell , \\
 \widetilde{\IrZ}_i&:=\mathds{1}_{\mathbb R\setminus\{0\}}(\IrZ_i), \quad i=n-\ell+1,...,n,
  \ea
the fraction of non-zero entries in  $\xrv$ is given by 
\ban  \frac{1}{n}\left (\sum_{i=1}^{n-\ell} \widetilde{\IrY}_i + \sum_{i=n-\ell+1}^n  \widetilde{\IrZ}_i \right).		\label{eq:fractionofnonzeros}\ean
Letting $n\to \infty$ in \eqref{eq:fractionofnonzeros}, we  obtain
\ba &\frac{1}{n}\left (\sum_{i=1}^{n-\ell} \widetilde{\IrY}_i + \sum_{i=n-\ell+1}^n  \widetilde{\IrZ}_i \right) \\&= \frac{n-\ell}{n} \frac{1}{n-\ell}\sum_{i=1}^{n-\ell} \widetilde{\IrY}_i  + \frac{\ell}{n}\frac{1}{\ell} \sum_{i=n-\ell+1}^n \widetilde{\IrZ}_i\\
&  \xrightarrow{\mathmakebox[.6cm]{\pr}} (1-\lambda) \rho_1 + \lambda\rho_2, \ea
where we used  
\begin{align}
\lim_{n\to\infty}(n-\ell)/n
&=\lim_{n\to\infty}(n-\lfloor \lambda n \rfloor)/n\\
&=(1-\lambda),
\end{align}
as $(1-\lambda)n\leqslant n-\lfloor \lambda n\rfloor <(1-\lambda)n+1$.  Similarly,  
\begin{align}
\lim_{n\to\infty}\ell/n
&=\lim_{n\to\infty}\lfloor \lambda n \rfloor /n\\
&=\lambda,
\end{align}
as $\lambda n -1<\lfloor \lambda n\rfloor \leqslant \lambda n$, and 
\begin{align}
\frac{1}{n-\ell}\sum_{j=1}^{n-\ell}  \widetilde{\IrY}_i&\xrightarrow{\mathmakebox[.6cm]{\pr}} \rho_1\\
\frac{1}{\ell}\sum_{j=n-\ell+1}^n \widetilde{\IrZ}_i&\xrightarrow{\mathmakebox[.6cm]{\pr}} \rho_2 , 
\end{align}
by the weak law of large numbers and  $\mathbb E[\widetilde{\IrY}_i]=\mathbb P[\widetilde{\IrY}_i=1] =\rho_1$, $\mathbb E[\widetilde{\IrZ}_i]=\mathbb P[\widetilde{\IrZ}_i=1] =\rho_2$.
%
% we have
%\ban 		\frac{1}{n} \lefto | \lefto \{ i \in \{1,...,n\} \mid \IrX_i \neq 0 \right \}  \right | \xrightarrow{\mathmakebox[.6cm]{\pr}} (1-\lambda) \rho_1 + \lambda\rho_2 ,	\label{eq:fract}\ean 
This shows that the fraction of non-zero entries in  $\xrv$ converges---in probability---to $ (1-\lambda) \rho_1 + \lambda\rho_2$. The next result establishes that the Minkowski dimension compression rate $R_{\text{B}}(\varepsilon)$  of mixed discrete-continuous sources equals, for all $\varepsilon\in(0,1)$,  the asymptotic  fraction of non-zero entries given by $ (1-\lambda) \rho_1 + \lambda\rho_2$. %, i.e., $R_{\text{B}}(\varepsilon)$ plays the role of a sparsity threshold.

\begin{proposition}\label{lemmax}
Suppose that $\xrv$ is distributed according to Definition \ref{definitionM}. Then, we have
\begin{align}\label{eq:result1}
R_{\text{B}}(\varepsilon) =(1-\lambda) \rho_1 + \lambda\rho_2  ,
\end{align}
for all $\varepsilon\in (0,1)$.
\end{proposition}
\begin{IEEEproof} The proof follows closely \cite[Thm. 15]{WV10}  and is therefore not detailed here. Interested readers can, however, consult the Online Addendum to this paper \cite[Sec.~II]{SRAB15extended} for the proof of \cite[Thm. 15]{WV10} adapted to our setting.
%adapted to our setting and is provided in the Online Addendum \cite[Sect.~II]{SRAB15extended}  for completeness. 
\end{IEEEproof}

%similar to the proof of \cite[Thm. 15]{WV10} and is provided in Appendix~\ref{prooflemmax}  for completeness. \end{IEEEproof}

%Theorem~\ref{th2} shows that the optimal compression rate   is lower-bounded by the (lower) Minkowski dimension compression rate $\underline{R}_{\text{B}}(\varepsilon)$.

%Lemma~\ref{lemmax} allows to interpret the general achievability result  Theorem~\ref{th2} for mixed discrete-continuous sources in a quantitative fashion.
%As mentioned above, general converses are difficult to get, but in  the mixed discrete-continuous case we can strengthen the achievability statement through the following converse.

We are now ready to state the converse for measurable separators.

\begin{proposition}\label{lemma3}
Suppose that $\xrv$ is distributed according to Definition \ref{definitionM} and let $\varepsilon\in (0,1)$. Then, the existence of
 % $R\geqslant {R}_{\text{B}}(\varepsilon)$ is necessary  for the existence of 
  a measurement matrix $\hm={[\am\:\: \bm]}:\IR^{n-\ell}\times \IR^\ell \to \IR^k$ and a corresponding measurable separator $g\colon \IR^k\to \IR^{n-\ell}\times \IR^\ell$, with $k=\lfloor Rn \rfloor$, such that  
\begin{align}\label{eq:errorsepa}
\pr [g({[\am\:\: \bm]}\xrv)\neq\xrv] \overset{\textbf{.}}{\leqslant} \varepsilon,
\end{align}
%
%
% and  $R>\! {R}_{\text{B}}(\varepsilon)$. Then, for each full-rank matrix $\bm\in\IR^{k\times \ell}$, with $k\geqslant \ell$, and for all matrices $\am\in\IR^{k\times {(n-\ell)}}$,  where $k=\lfloor Rn \rfloor$, the exists
%
%Moreover, for every $\varepsilon$ with $0<\varepsilon<1$, $R\geqslant {R}_{\text{B}}(\varepsilon)$ is also a necessary condition for \eqref{eq:errorsepa} to hold. %, i.e., $R_{\text{L}}(\varepsilon)=R_{\text{B}}(\varepsilon)$.
imply  $R\geqslant {R}_{\text{B}}(\varepsilon)$.
\end{proposition}
\begin{IEEEproof} The proof does not have to account for the fact that $\hm={[\am \:\: \bm]}$ contains a fixed
block $\bm$ and follows closely the converse part of \cite[Thm. 6]{WV10}. We therefore do not include the details here, but, again,  refer the interested reader to the Online Addendum \cite[Sec.~III]{SRAB15extended}.
%The proof therefore follows by adapting the converse part of \cite[Thm. 6]{WV10}  to our setting. For completeness we provide the proof  in the Online Addendum \cite[Sect.~III]{SRAB15extended}.
\end{IEEEproof}

%Finally, we combine Lemmata~\ref{lemmax} and \ref{lemma3} to get an analytical expression for the optimal compression rate.
%
%
%\begin{corollary}\label{sep}
%Suppose that $\xrv$ has a mixed discrete-continuous distribution according to Definition \ref{definitionM} and let $\varepsilon\in (0,1)$.  
%Then, the critical rate is given by
%\begin{align}\label{eq:rate}
%(1-\lambda)\rho_1+\lambda\rho_2, \end{align}
%i.e., for every $R$ greater than \eqref{eq:rate} there are ${[\am\:\: \bm]}$  and $g$ achieving \eqref{eq:errorsepa}, and for every $R$ smaller than \eqref{eq:rate} there are no such ${[\am\:\: \bm]}$  and $g$.
%\end{corollary}

Combining the achievability statements in Theorems~\ref{th2} and \ref{thm:hoeldersepneu}, and Propositions~\ref{lemmax} and \ref{lemma3}, we can conclude that, for mixed discrete-continuous sources, 
\begin{align}\label{eq:rate}
R_{\text{B}}(\varepsilon) =(1-\lambda)\rho_1+\lambda\rho_2, \end{align}
is the critical rate
in the following sense: 
\begin{itemize}
\item For  $R>R_{\text{B}}(\varepsilon) $, for  every fixed full-rank matrix $\bm\in\IR^{k\times \ell}$, with $k\geqslant \ell$,  and for a.a.\ $\am$  (where the set of exceptions for $\am$ depends on the specific choice of $\bm$), there exists a measurable separator $g$ satisfying \eqref{eq:errorsepa}, as well as a $\beta$-H\"older continuous separator  $g$ for fixed $\beta$ with  $ \beta < 1- \frac{\overline{R}_\text{B}(\varepsilon)}{R}$ satisfying \eqref{eq:errorsepa} with $\varepsilon$ replaced by $\varepsilon+\kappa$ for arbitrarily small $\kappa>0$, %there are ${[\am\:\: 
\item for $R=R_{\text{B}}(\varepsilon) $, we cannot make a general statement on the existence of a separator, 
\item and for  $R<R_{\text{B}}(\varepsilon) $  there does not exist a single pair   $(g, {[\am\:\: \bm]})$, with $g$ measurable,  satisfying \eqref{eq:errorsepa}.
\end{itemize}
%This result essentially states that the critical rate is determined by the fraction of continuously distributed components in the concatenated source vector.
As $R\approx k/n$, where $n$ is the ambient dimension and $k$ the number of measurements, the threshold $R_{\text{B}}(\varepsilon) =(1-\lambda)\rho_1+\lambda\rho_2$ identifies the critical number of measurements relative to the ambient dimension as the number of non-zero entries in  $\xrv$. % approximately given by $n ((1-\lambda)\rho_1+\lambda\rho_2)$.
 
{\it Relation to classical uncertainty relations in compressed sensing.} Comparing the threshold obtained from our probabilistic uncertainty relations to the thresholds available in the compressed sensing literature, we note the following. The Donoho-Stark \cite{DS89} and Elad-Bruckstein \cite{EB02} uncertainty principles hold for \emph{all}  $\yv$ and $\zv$, but suffer from the ``square-root bottleneck'' \cite{Tro08}. It is well known that the inequalities leading to the square-root bottleneck are saturated by very special combinations of signals and dictionaries, e.g., a Dirac comb for $\am=\boldsymbol{I}_n$ and $\bm=\fm_n$ \cite{DS89}. Relaxing these deterministic thresholds by considering random models for  the signals and dictionaries  \cite{PBS13,CRT06,Don06,Tro08b} leads to thresholds that exhibit a ``$\log n$-factor''. 
 The thresholds that follow from our probabilistic uncertainty relations exclude an arbitrarily small set of signals, a set of $\am$-matrices of Lebesgue measure zero, are asymptotic in $n$, and suffer neither from the square-root bottleneck nor from the $\log n$-factor.
% The threshold that follows from our probabilistic uncertainty relations is the information-theoretic limit for random signals and excludes a set of $\am$-matrices of Lebesgue measure zero, but suffers neither from the square-root bottleneck nor from the $\log n$-factor.
% Our  probabilistic uncertainty relations therefore yield significantly stronger statements than the Donoho-Stark and the Elad-Bruckstein uncertainty principles \cite{DS89, EB02}, which hold for \emph{all}  $\yv$ and $\zv$, but suffer from the ``square-root bottleneck'' \cite{Tro08}.
% The threshold we obtain also does not suffer from the ``$\log n$''-factor prevalent in the compressed sensing literature.
 Moreover, they are best possible as the same threshold would be obtained if the support sets of $\yv$ and $\zv$ were known  a priori and only the values of the non-zero entries in the concatenated source vector  were to be recovered. 
 % our threshold is as good as knowing the support sets of $\yv$ and $\zv$, which would also mandate the number of measurements to be at least equal to the number of non-zero entries in the concatenated source vector. 
The set of exceptions for $\am$ in the ``a.a.-statement'' in Theorems~\ref{th2} and \ref{thm:hoeldersepneu} depending on the specific choice of $\bm$ can be interpreted as a mild incoherence condition between $\am$ and $\bm$ akin to those in  \cite{DS89, EB02}.
%Moreover,
 %the critical  rate we identify does not depend on 
 %coherence quantities of the measurement matrix $\hm={[\am \:\: \bm]}$, as is the case for classical sparsity thresholds \cite{SKP12, PBS13}.
%which usually arise in recovery thresholds in the sparse signal separation problem, see, e.g., 
In fact, we have a phase transition phenomenon, which states that above the critical rate $R_\text{B}(\varepsilon)$ a.a.\ matrices $\am$ are ``incoherent'' to a given matrix $\bm$, whereas below the critical rate there is not a single pair of matrices $\am$ and $\bm$ that admits separation via a measurable separator $g$. 
Finally, 
% (where the set of exceptions for $\am$ depends on  $\bm$, which can be seen as a mild form of an incoherence condition), provided that the rate exceeds $R_\text{B}(\varepsilon)$ and, conversely, if the rate is strictly smaller than $R_\text{B}(\varepsilon)$, no pair of matrices $\am$ and $\bm$ admits recovery. 
% In this respect, under the rate constraint $R>\! R_\text{B}(\varepsilon)$, 
as already noted, our regularized probabilistic uncertainty relation does not seem to have a counterpart in classical compressed sensing theory.
%We emphasize, however, that  the set of exceptions for $\am$ depending on the choice of  $\bm$, can be interpreted as a mild form of an incoherence condition.
%When the distribution of one of the  constituents is purely discrete, i.e., either $\rho_1=0$ or $\rho_2=0$, the critical separation rate is determined solely by the   mixture parameter of the other constituent. Finally, if the dimension of one of the constituents is much larger than the dimension of the other, i.e., $\lambda \approx 0$ or $\lambda\approx 1$, then the characteristics of the higher-dimensional constituent dominate the threshold in \eqref{eq:rate}.

\begin{remark}\label{rem:declipping}
The results above show % in Proposition~\ref{lemmax}
 that for mixed discrete-continuous sources $\xrv$ the Minkowski dimension compression rate is small if the asymptotic fraction $(1-\lambda) \rho_1 + \lambda\rho_2 $ of non-zero entries in $\xrv$  is small, i.e., if the source vectors  are sparse in the classical sense. Another factor that can lead to  small Minkowski dimension compression rate is statistical  dependence between the constituents $\yrv$ and $\zrv$. For example, consider the declipping problem \cite{SKP12}
where a signal that is sparse in the dictionary $\am$ is to be recovered from its clipped version. Specifically, we observe 
 $\am \yrv+\zrv$ with $\zrv=g_a(\am \yrv)-\am \yrv$, where $g_a$ denotes entry-wise clipping to the values  $\pm a$.  %This models clipping of a signal that is sparse in the dictionary $\am$.
%This models the measurement setting where the original signal  is sparse in a random basis $\am$ and, 
If  clipping is not too aggressive, the signal $\zrv$ will be  sparse in the identity basis $\bm=\boldsymbol{I}_\ell$ (see Fig.~\ref{fig:1}).
%\begin{center}
\begin{figure*}[h]
\begin{center}
\begin{tikzpicture}[scale=.8]
\begin{axis}[xmin=0, xmax=256,xlabel=sample index, 
			ymin=-2.2,ymax=2.2, ylabel=amplitude, xtick={0,32,64,96,128,160,192,224,256}, grid=major]
\addplot+[no markers, sharp plot, darkgreen, thick, densely dotted] coordinates {(0,1)  (256,1)};
\addplot+[no markers, sharp plot, darkgreen, thick, densely dotted] coordinates {(0,-1)  (256,-1)};
\addplot[no markers, mark options={scale=0.001}, darkblue, dashed]	file {data/file1.dat};
\addplot[no markers, smooth, firebrick3, thick]	file {data/file2.dat};
\end{axis}
\begin{axis}[xshift=9cm, yshift=3.4cm, xmin=0, xmax=256,	ymin=-2.2,ymax=2.2, xtick={0,32,64,96,128,160,192,224,256}, grid=major]
\addplot[no markers, smooth, darkblue, thick]	file {data/file1.dat};
\end{axis}
\begin{axis}[xshift=9cm, yshift=-3.4cm, xmin=0, xmax=256,	ymin=-2.2,ymax=2.2,    xtick={0,32,64,96,128,160,192,224,256}, grid=major]
\addplot[no markers, smooth, darkgreen, thick] file {data/file3.dat};
\end{axis}
\draw[->, thick] (8.5,6) -- (7.3,3.1);
\draw[->, thick] (8.5,-.3) -- (7.3,2.5);
\node at (7.2,2.8) {\large $+$};

\node at (16.6,6.2) {$\boldsymbol{A}\boldsymbol{\mathsf{y}}$};
\node at (16.5,-.6) {$\zrv$};

\end{tikzpicture}
\caption{\small Declipping of signals as a sparse signal separation problem.}
\label{fig:1}
\end{center}
\end{figure*}
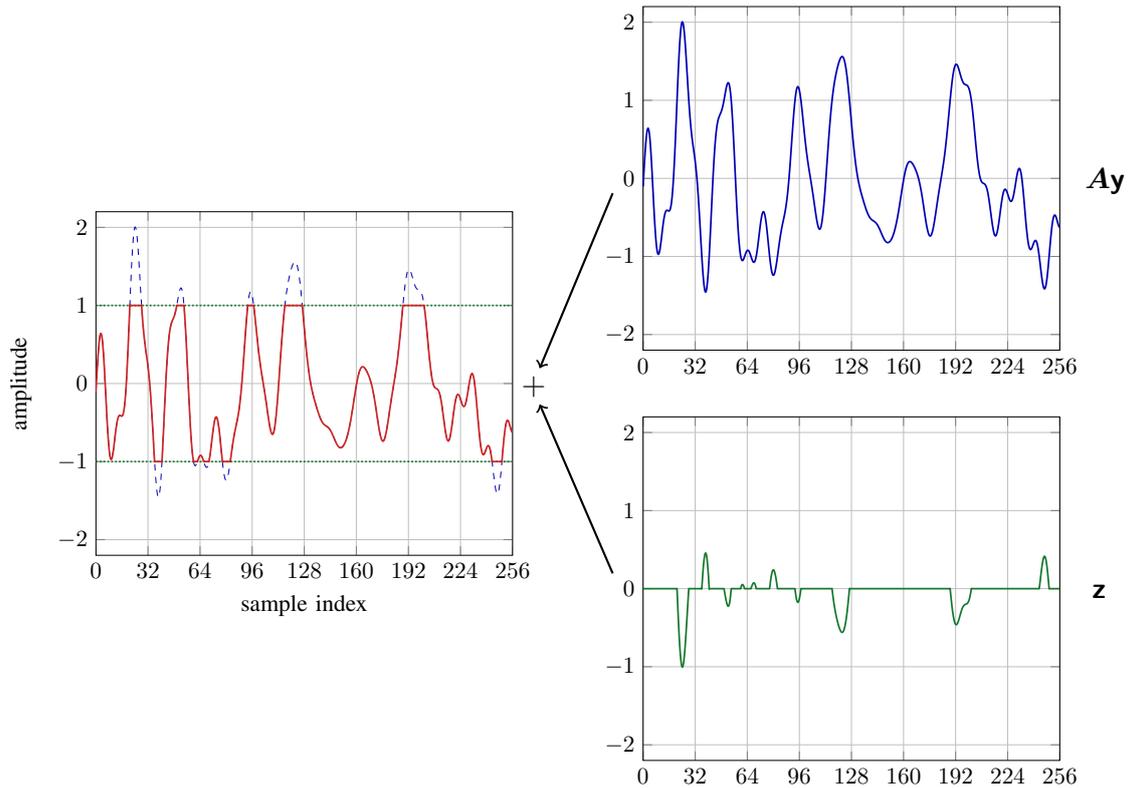
%\end{center}
 Here $\yrv$ and $\zrv$ are of the same dimension, i.e., $\lambda=1/2$. Moreover,  $\zrv$ is completely determined by $\yrv$, which, as 
proved in Lemma~\ref{lem:example} in Appendix~\ref{app:example}, implies that 
\ba R_\text{B}^{\xrv} (\varepsilon) = \frac{1}{2}R_\text{B}^{\yrv} (\varepsilon),  \ea 
where $R_\text{B}^{\xrv} (\varepsilon)$ is
the Minkowski dimension compression rate of  $\xrv= [\yrv^{\operatorname{T}}\; \zrv^{\operatorname{T}}]^{\operatorname{T}}$ and $R_\text{B}^{\yrv} (\varepsilon)$ is
 the  Minkowski dimension compression rate of $\yrv$ only.
%, which means that $[\yrv^{\operatorname{T}}\; \zrv^{\operatorname{T}}]^{\operatorname{T}}$ has half of the same Minkowski dimension compression rate of $\yrv$ alone, which is shown in Lemma~\ref{lem:example} in Appendix~\ref{app:example}. 
If the components of $\yrv$ are i.i.d.\ and of discrete-continuous mixture $(1-\rho)\mu_{\text{d}}+\rho\mu_{\text{c}}$, it  follows from Proposition~\ref{lemmax} %for $\lambda=0$
 that $R_\text{B}^{\yrv} (\varepsilon)=\rho$
% the
%Minkowski dimension compression rate of  $\xrv = [\yrv^{\operatorname{T}}\; \zrv^{\operatorname{T}}]^{\operatorname{T}}$ is given by
and consequently
\begin{align}\label{eq:deduce}
R_\text{B}^{\xrv} (\varepsilon)  =\frac{1}{2}\rho  .
\end{align}
The description complexity of  $\xrv$ is therefore determined by the fraction of components in $\yrv$ that are continuously distributed. % and by taking into account that $\yrv$ occupies half of $\xrv$.
As expected, we find that the critical rate here is half  the critical rate for a mixed discrete-continuous source with independent $\yrv$ and $\zrv$ and $\rho_1=\rho_2=\rho$.
% In this example, we have knowledge of the support set of  $\zrv$ as the set of components where  $\am \yrv+\zrv$ has value $\pm a$. However, we can deduce that this knowledge  does not reduce the description complexity, since \eqref{eq:deduce} would be unchanged if $\zrv$ would be deterministically $\0v$. This contrasts results from \cite{SKP12} where knowledge of the support set has been found to improve recovery thresholds.
%For all $R>\frac{1}{2}\rho$, we can thus find a separator that achieves $R$ with error probability $\varepsilon$ by Theorem~\ref{th2}.
\end{remark}

\section[]{Strengthening of \cite[Thm.~18]{WV10} and simplifying its proof}
\label{sec:simple}

In this section, we sketch how the probabilistic uncertainty relation,  Proposition~\ref{prop:inj2}, and the regularized probabilistic uncertainty relation,  Proposition~\ref{prop:injneu}, 
can be applied to devise a  simplification of the proof of \cite[Thm.~18, 1)]{WV10} and a 
 significant strengthening  of the statement \cite[Thm.~18, 2)]{WV10}.
% simplified  proof of a strengthened version of \cite[Thm.~18]{WV10}. The result \cite[Thm.~18]{WV10} has two parts.
We begin by restating   \cite[Thm.~18, 1)]{WV10} in our notation and terminology.
 %$ which is geometrically appealing as it says that a generic $(n-k)$-dimensional subspace will intersect a $d$-dimensional object with $d<k$ at most trivially. % we obtain the following special case. % which intuitively says that a set of Minkowski dimension smaller than $k$ and the (generally) $(n-k)$-dimensional kernel of a $k\times n$ matrix intersect at most trivially, since the two dimensions do not add up to the full dimension $n$ of the ambient space.
%It is surprising  that Euclidean dimensionality and Minkowski dimensionality behave compatibly in this context.

\begin{theorem}  [{\!\! \cite[Thm.~18, 1)]{WV10}}]\label{thm:wv10}
Let $\xrv=[\IrX_1 \, \dots\, \IrX_n]^{\operatorname{T}}$ be a source vector of dimension $n$  with underlying  stochastic source process $(\IrX_i)_{i\in\IN}$ on $(\IR^\IN,\bs^{\otimes\IN})$. Take $\varepsilon >0$ and let    $R>\! \overline{R}_\text{B}(\varepsilon)$. Then, for a.a.\ $\hm\in\mathbb R^{k\times n}$, there exists a measurable decoder $g$ such that
\ba \pr[g(\hm \xrv)\neq \xrv]\overset{\textbf{.}}{\leqslant}\varepsilon, 	\ea
where $k=\lfloor Rn\rfloor$.
\end{theorem}

This statement can be recovered
%in Theorem~\ref{thm:wv10} 
 from Theorem~\ref{th2} by setting $\lambda=0$, which, in fact, yields a slight improvement upon  \cite[Thm.~18, 1)]{WV10}, namely  the condition $R>\! \overline{R}_\text{B}(\varepsilon)$ in  \cite[Thm.~18, 1)]{WV10}  is replaced  by $R>\! \underline{R}_\text{B}(\varepsilon)$.
 % strengthened version of , where the assumption $R>\! \overline{R}_\text{B}(\varepsilon)$ in  \cite[Thm.~18, 1)]{WV10}  is replaced  by $R>\! \underline{R}_\text{B}(\varepsilon)$. This version improves the result in \cite{WV10} whenever the source is such that $\overline{R}_\text{B}(\varepsilon)>\underline{R}_\text{B}(\varepsilon)$.
For our simplified proof, %of a strengthened version of this result,
 we start by particularizing the probabilistic uncertainty relation,  Proposition~\ref{prop:inj2}, to  $\ell=0$. 
\begin{corollary}\label{prop:inj}
Let $\mathcal S\subseteq\mathbb R^n$ be  non-empty and bounded such that $\underline{\dim}_{\text{B}}(\mathcal S)<k$. Then, we have
\ban 	\{ \xv \in \mathcal S\!\setminus\! \{\0v\} : \hm \xv = \0v\} = \emptyset , 	\label{eq:empty} \ean
for  a.a.\ $\hm \in \mathbb R^{k\times n}$.
\end{corollary}

We refer to this result as a probabilistic null-space property as it is a statement on the intersection
 of the null-space of  $\hm$ with the set $\ss$. %, where no separation aspect is involved, we call this result a probabilistic null-space property.
%As introduced in Section~\ref{sec:mainresults}, the framework of almost lossless analog compression in \cite{WV10} for the case of  linear measurements and  a measurable decoder considers a general stochastic source process $\xrv$. The problem is to reconstruct $\xrv$ from linear measurements $\hm \xrv$, where $\hm$ is the measurement matrix.
% The result in \cite[Thm.~18, 1)]{WV10} says that for $R>\! \overline{R}_\text{B}(\varepsilon)$, for a.a.\ $\hm\in\mathbb R^{k\times n}$, there exists a measurable decoder $g$ such that
%\ba \pr[g(\hm \xrv)\neq \xrv]\overset{\textbf{.}}{\leqslant}\varepsilon, 	\ea
%where $k=\lfloor Rn\rfloor$.
%The  probabilistic null-space property in  Corollary~\ref{prop:inj}, allows us to  give an  
Our alternative, simplified proof of \cite[Thm.~18, 1)]{WV10} goes as follows. As in the proof of  Theorem~\ref{th2} we choose a sequence of compact sets $\us\subseteq\IR^n$ satisfying \eqref{eq:suff1} and \eqref{eq:suff2}. % hold, and  w
Let $\ev\in\IR^n\!\setminus\!\us$. Again, it follows from \cite[Prop. 14.33 and Cor. 14.6]{rowe98}  and the compactness of $\us$  that there exists a measurable mapping $f\colon\IR^{k\times n}\times\IR^k\to\IR^n$ satisfying  
\begin{align}\label{eq:decoder2}
f(\hm,\vv)\in 
\{\uv\in\us : \hm\uv= \vv\},&\quad \text{if}\ \{\uv\in\us : \hm\uv= \vv\}\neq\emptyset,
%[\am,\bm]^{-1}(\vv) \cap \us,&\quad \text{if}\ [\am,\bm]^{-1}(\vv) \cap \us=\{ \x\}\\
\end{align}
and $f(\am,\vv)=\ev$ else.
The mapping $g=f(\hm,\cdot)$ therefore constitutes a valid (i.e., measurable) decoder. 
Let 
\begin{align}
p_\mathrm{e}(\hm)&:=\pr [g(\hm\xrv)\neq\xrv],\quad \hm\in\IR^{k\times n}.
\end{align}
Repeating the steps in \eqref{eq:as1}--\eqref{eq:lebas1} with $\hm$ in place of ${[\am \:\: \bm]}$, it follows that 
$p_\mathrm{e}(\hm)\leq\varepsilon$ for a.a. $\hm\in\IR^{k\times n}$ provided  we can show that 
\begin{align}
\leb^{kn}\{\hm:\ker(\hm)\cap \us_{\xv}\neq\{\0v\}\}&=0, \label{eq:laststep2}
\end{align}
for all $\xv\in\us$, where 
\begin{align}
\us_{\xv}=\{\uv-\xv:\uv\in\us\},\quad \xv\in\us. 
\end{align}
Applying  Corollary~\ref{prop:inj} with $\ss=\us_{\xv}$,  \eqref{eq:laststep2} holds as a consequence of   $\MIL(\us_{\xv})<k$, thereby 
completing the proof.
The application of the probabilistic null-space property, Corollary~\ref{prop:inj},  replaces the arguments in \cite[Thm.~18, 1)]{WV10} that are based  on properties of invariant measures on Grassmannian manifolds. % thereby allowing for a simplification of the proof. %Moreover, we note that these arguments are not applicable  to the signal separation problem.  
Finally, we note that, instead of  particularizing Proposition~\ref{prop:inj2} to $\ell=0$, the  probabilistic null-space property in Corollary~\ref{prop:inj} can also be proved directly with considerably less logistic effort, as  done in \cite{SRB13}. 
%Thanks to the condition on the dimension of $\ss$ in Corollary~\ref{prop:inj} being with respect to \emph{lower} Minkowski dimension, the argument above continues to hold when we replace the assumption $R>\! \overline{R}_\text{B}(\varepsilon)$ in  \cite[Thm.~18, 1)]{WV10}  by $R>\! \underline{R}_\text{B}(\varepsilon)$, thereby improving the result in \cite{WV10} whenever the source is such that $\overline{R}_\text{B}(\varepsilon)>\underline{R}_\text{B}(\varepsilon)$.

%The proof idea in \cite[Thm.~18, 1)]{WV10} is to first restrict to random orthogonal projections $\hrm$ and then apply a concentration result  for the invariant measure on the Grassmannian manifold  under the action of the orthogonal group to bound the probability of ``bad'' $\hrm$ in terms of the invariant measure applied to the set of spaces onto which $\hrm$ projects.
%which  can be studied using tools from the theory on Grassmanian manifolds.

Next, we restate \cite[Thm.~18, 2)]{WV10} in our notation and terminology.
\begin{WVV} %[{\!\! \cite[Thm.~18, 2)]{WV10}}] \label{thm:wv10_2}
Let $\xrv=[\IrX_1 \, \dots\, \IrX_n]^{\operatorname{T}}$ be a source vector of dimension $n$  with underlying  stochastic  process $(\IrX_i)_{i\in\IN}$ on $(\IR^\IN,\bs^{\otimes\IN})$. 
 Take $\varepsilon >0$,  and   let  $R>\! \overline{R}_\text{B}(\varepsilon)$ and $\beta>0$ be fixed such that  
\ban \beta < 1- \frac{\overline{R}_\text{B}(\varepsilon)}{R}. \label{eq:betahoeld} \ean
Then, there exist 
$\hm\in\mathbb R^{k\times n}$ and a corresponding $\beta$-H\"older continuous decoder $g$ such that
\ban \pr[g(\hm \xrv)\neq \xrv]\overset{\textbf{.}}{\leqslant}\varepsilon+ \kappa, \label{eq:achhoeld}	\ean
where $k=\lfloor Rn\rfloor$ and $\kappa>0$ can be chosen arbitrarily small.
\end{WVV}

Particularizing Theorem~\ref{thm:hoeldersepneu} to  $\lambda=0$, we obtain a substantial strengthening of  \cite[Thm.~18, 2)]{WV10}, as  \cite[Thm.~18, 2)]{WV10} states the existence of an $\hm$ with a corresponding $g$ satisfying \eqref{eq:achhoeld}, whereas our result says that for \emph{a.a.}  $\hm$ there is a corresponding $g$ satisfying \eqref{eq:achhoeld}.
% stating that for a.a.\ $\hm$ there exists a $\beta$-Hölder continuous decoder $g$ satisfying \eqref{eq:achhoeld}. % provided that \eqref{eq:betahoeld} holds. This improves the statement in Theorem~\ref{thm:wv10_2} which only asserts the existence of such a matrix $\hm$. 
The  crucial element in  accomplishing  this strengthening is  the regularized probabilistic uncertainty relation in Proposition~\ref{prop:injneu}.

\appendices

\section{Minkowski dimension compression rate for declipping example}
\label{app:example}

We consider the declipping problem where $\lambda=1/2$ and   we observe 
% $\wrv=\am \yrv+\zrv$ where the source is given by $\xrv =  [\yrv^{\operatorname{T}}\; \zrv^{\operatorname{T}}]^{\operatorname{T}}$
$\zrv=g_a(\am\yrv)- \am\yrv$, with  $g_a$ denoting entry-wise clipping to the values  $\pm a$ for some $a>0$. We introduce the notation
\ba &\underline{R}_\text{B}^{\xrv} (\varepsilon), \underline{a}_n^{\xrv}(\varepsilon),\overline{R}_\text{B}^{\xrv} (\varepsilon),\overline{a}_n^{\xrv}(\varepsilon)\\ &\underline{R}_\text{B}^{\yrv} (\varepsilon), \underline{a}_n^{\yrv}(\varepsilon),\overline{R}_\text{B}^{\yrv} (\varepsilon),\overline{a}_n^{\yrv}(\varepsilon)\ea
for the quantities in Definition~\ref{defminkowskirate} corresponding to the processes $\xrv=  [\yrv^{\operatorname{T}}\; \zrv^{\operatorname{T}}]^{\operatorname{T}}$ and $\yrv$, respectively.

\begin{lemma}\label{lem:example}
For  $\varepsilon>0$, we have%\footnote{The intuitive reason for the factors $2$ in \eqref{eq:intuitive} is that for $\yrv$ the corresponding  $\xrv$ with the same amount of information has twice the length.} 
\ban \underline{R}_\text{B}^{\xrv} (\varepsilon) = \frac{1}{2}\underline{R}_\text{B}^{\yrv} (\varepsilon) \quad \text{and} \quad \overline{R}_\text{B}^{\xrv} (\varepsilon) = \frac{1}{2}\overline{R}_\text{B}^{\yrv} (\varepsilon). \label{eq:intuitive}\ean
\end{lemma}
\IEEEproof
We only prove the first identity and note that  the second is obtained by simply replacing  $\MIL(\cdot)$ by $\MIU(\cdot )$ in the arguments below. 
Let us begin by showing that 
\ban  \underline{R}_\text{B}^{\xrv} (\varepsilon) \leqslant \frac{1}{2} \underline{R}_\text{B}^{\yrv} (\varepsilon). \label{eq:begin} \ean
Recall that $\ell=\lfloor \frac{n}{2}  \rfloor$, and suppose that we are given a set $\ss\subseteq\mathbb R^{n-\ell}$ such that $\pr[\yrv \in \ss]\geqslant 1-\varepsilon$. Set 
\ban \ts:=\{ [\yv^{\operatorname{T}}\;\, (g_a(\am\yv)- \am\yv)^{\operatorname{T}}]^{\operatorname{T}} : \yv \in \ss \}\subseteq \mathbb R^n,  \ean
and note that  for all $\yv_1,\yv_2\in \mathbb R^{n-\ell}$ we have
\ban 
\|\yv_1 - \yv_2\|&\leqslant \| [\yv_1^{\operatorname{T}}\;\,\,\, (g_a(\am\yv_1)- \am\yv_1)^{\operatorname{T}}]^{\operatorname{T}}\label{eq:newcontractive1}\\
&\phantom{=\|}  - [\yv_2^{\operatorname{T}}\;\,\,\, (g_a(\am\yv_2) -\am\yv_2)^{\operatorname{T}}]^{\operatorname{T}} \| \label{eq:contractive1} \ean 
and 
\ban &\| [\yv_1^{\operatorname{T}}\;\,\,\, (g_a(\am\yv_1)- \am\yv_1)^{\operatorname{T}}]^{\operatorname{T}}  - [\yv_2^{\operatorname{T}}\;\,\,\, (g_a(\am \yv_2) - \am\yv_2)^{\operatorname{T}}]^{\operatorname{T}} \|\nonumber\\ 
&\leqslant \|\yv_1 - \yv_2\|  +\|g_a(\am\yv_1)-\am\yv_1 -  (g_a(\am\yv_2)- \am\yv_2)\|\ \  \label{eq:contractive2}\\ &\leqslant  (1+\|\am\|) \|\yv_1 - \yv_2\|  +\|g_a(\am \yv_1)- g_a(\am \yv_2)\| \label{eq:contractive2a}\\ &\leqslant (1+2\|\am\|)\|\yv_1 - \yv_2\|, \label{eq:contractive3}	\ean
where \eqref{eq:contractive1}, \eqref{eq:contractive2}, and \eqref{eq:contractive2a} follow from  the triangle inequality, and \eqref{eq:contractive3} holds as $|g_a(y_1)-g_a(y_2)|\leqslant |y_1-y_2|$, for all $y_1,y_2\in \mathbb R$.
%$g_a$ decreases the distance between each pair of entries in $\yv_1$ and $\yv_2$.
Combining \eqref{eq:contractive1} and \eqref{eq:contractive2}--\eqref{eq:contractive3}, it  follows that for  $\delta>0$
\ban 	N_\ss(\delta) \leqslant N_\ts(\delta) \leqslant N_\ss((1+2\|\am\|)\delta)	,	\label{eq:same} \ean
which implies  $\MIL(\ss)=\MIL(\ts)$. Since $2(n-\ell) =2(n-\lfloor n/2 \rfloor) \leqslant n+2$, we obtain
\ba \frac{1}{2} \frac{\MIL(\ss)}{n-\ell} = \frac{1}{2} \frac{\MIL(\ts)}{n-\ell} \geqslant  \frac{\MIL(\ts)}{n+2}   , \ea 
and therefore $ \frac{1}{2}\underline{a}_{n-\ell}^{\yrv}(\varepsilon) \geqslant \frac{n}{n+2} \underline{a}_n^{\xrv}(\varepsilon) $. As 
$\lim_{n\to\infty}n/(n+2)=1$, we get \eqref{eq:begin}.

To prove
\ban  \underline{R}_\text{B}^{\xrv} (\varepsilon) \geqslant \frac{1}{2} \underline{R}_\text{B}^{\yrv} (\varepsilon), \label{eq:end} \ean
we consider a set $\us\subseteq \mathbb R^n$ such that $\pr[\xrv \in \us]\geqslant 1-\varepsilon$. Setting $\vs:=\{\yv \in \mathbb R^{n-\ell} : [\yv^{\operatorname{T}}\;\, (g_a(\am\yv) -\am\yv)^{\operatorname{T}}]^{\operatorname{T}} \in \us\}$, we have $\pr[\yrv \in \vs]=\pr[\xrv \in \us]$. For the set $\widetilde \us :=\{  [\yv^{\operatorname{T}}\;\, (g_a(\am\yv)-\am\yv)^{\operatorname{T}} ]^{\operatorname{T}} : \yv\in \vs\}$ we have $\MIL(\widetilde \us)=\MIL(\vs)$ by the same arguments as in \eqref{eq:contractive1}--\eqref{eq:same}. Moreover, by definition of $\vs$ we have $\widetilde \us \subseteq \us$ and therefore $\MIL(\widetilde \us)\leqslant \MIL(\us)$, which implies 
\ba  \frac{1}{2} \frac{\MIL(\vs)}{n-\ell}= \frac{1}{2} \frac{\MIL(\widetilde \us)}{n-\ell} \leqslant   \frac{1}{2} \frac{\MIL(\us)}{n-\ell}  \leqslant  \frac{\MIL(\us)}{n},\ea
where in the last step we used $2(n-\ell) \geqslant n$. This shows  that $\underline{a}_n^{\xrv}(\varepsilon) \geqslant  \frac{1}{2} \underline{a}_{n-\ell}^{\yrv}(\varepsilon)  $ which  establishes \eqref{eq:end} and
%\ban  2\underline{R}_\text{B}^{\xrv} (\varepsilon) \geqslant \underline{R}_\text{B}^{\yrv} (\varepsilon),  \ean
thereby  completes the proof.
\endIEEEproof

\section{Alternative definition of Minkowski dimension}
\label{app:eqdef}
In this section, we prove that Minkowski dimension can  equivalently be defined through the modified covering number \eqref{eq:modcovno}. 
Similar arguments for different modifications of the covering number \eqref{eq:origcovno} can be found  in  \cite[Equivalent Definitions~3.1]{Fal04}.
\begin{lemma}\label{lem:equivalentdef}The Minkowski dimension of a non-empty bounded set $\ss \subseteq\IR^n$
does not change when  the covering  balls
 in the definitions \eqref{eq:defminkowski1}, \eqref{eq:defminkowski2} 
are restricted to  have their centers inside the set $\ss$, that is, we have
\begin{align}
\liminf_{\delta\to 0}\frac{\log N_\ss(\delta)}{\log\frac{1}{\delta}}&=\liminf_{\delta\to 0}\frac{\log M_\ss(\delta)}{\log\frac{1}{\delta}}	\label{eq:center1}\\
\limsup_{\delta\to 0}\frac{\log N_\ss(\delta)}{\log\frac{1}{\delta}}&=\limsup_{\delta\to 0}\frac{\log M_\ss(\delta)}{\log\frac{1}{\delta}} \label{eq:center2} ,
\end{align}
where $N_\ss(\delta)$ is the covering number of $\ss$ given by 
\begin{align*}
N_\ss(\delta)&=\min\Big\{m \in\IN : \ss\subseteq\!\bigcup_{i\in\{1,\dots,m\}}\! B^n(\xv_i,\delta),\ \xv_i\in \IR^n\Big\} ,
\end{align*}
and $M_\ss(\delta)$ is the covering number of $\ss$ with the covering balls centered in $\ss$, i.e.,
\begin{align*}
M_\ss(\delta)&=\min\Big\{m \in\IN : \ss\subseteq\bigcup_{i\in\{1,\dots,m\}} B^n(\xv_i,\delta),\ \xv_i\in \mathcal S\Big\}.
\end{align*}
\end{lemma}
\IEEEproof
Since $N_\ss(\delta)\leqslant M_\ss(\delta)$, the ``$\leqslant$''-part in \eqref{eq:center1} and \eqref{eq:center2} is immediate. To establish  the ``$\geqslant$''-part, we consider a set of covering balls of $\ss$ of radius $\delta/2$ and corresponding centers $\xv_1, ..., \xv_{N_{\ss}(\delta/2)}\in \mathbb R^{n}$. Note that these centers do not necessarily lie in $\ss$.
% let $\xv_1, ..., \xv_{N_{\ss}(\varepsilon/2)}\in \mathbb R^{n}$ be the centers of balls with radius $\varepsilon/2$ that cover $\ss$ but where the centers do not necessarily lie in $\ss$. 
Since $N_\ss(\delta/2)$ is the minimum number of balls with radius $\delta/2$ needed to cover $\ss$, the intersection $B^n(\xv_i,\delta/2)\cap \ss$ must be non-empty for all $i=1,...,N_\ss(\delta/2)$. We now choose an arbitrary point $\yv_i \in (B^n(\xv_i,\delta/2)\cap \ss)$ and note that  % implies that  for all $\uv\in\mathbb R^n$
\ban 	\|\uv-\yv_i\| \leqslant \|\uv-\xv_i\|+\|\xv_i-\yv_i\| \leqslant \|\uv-\xv_i\| +\delta/2,\label{eq:triangleargument2}\ean
for all $\uv\in\mathbb R^n$, 
which implies $B^n(\xv_i,\delta/2)\subseteq B^n(\yv_i,\delta)$, $i=1,..., N_\ss(\delta/2)$. It therefore follows that
 $B^n(\yv_i,\delta)$, $i=1,..., N_\ss(\delta/2)$, is a covering of $\ss$ with balls of radius $\delta$ all centered in $\ss$. This implies  \ban M_\ss(\delta)\leqslant N_\ss(\delta/2), \label{eq:absetzen} \ean and hence 
\ban 	\frac{\log M_\ss(\delta)}{\log\frac{1}{\delta}}\leqslant \frac{\log  N_\ss(\delta/2)}{\log\frac{1}{\delta}}=\frac{\log  N_\ss(\delta/2)}{\log\frac{2}{\delta}} \underbrace{\frac{\log\frac{2}{\delta}}{\log\frac{1}{\delta}}}_{\xrightarrow{\delta\to 0}1}.	\label{eq:taking}\ean
Taking $\liminf_{\delta\to 0}$ and $\limsup_{\delta\to 0}$ on both sides of \eqref{eq:taking} yields the ``$\geqslant$''-part in \eqref{eq:center1} and \eqref{eq:center2}, respectively, according to  
\ba \liminf_{\delta\to 0}\frac{\log  N_\ss(\delta/2)}{\log\frac{2}{\delta}} = \liminf_{\delta\to 0}\frac{\log  N_\ss(\delta)}{\log\frac{1}{\delta}},\\ 
 \limsup_{\delta\to 0}\frac{\log  N_\ss(\delta/2)}{\log\frac{2}{\delta}} = \limsup_{\delta\to 0}\frac{\log  N_\ss(\delta)}{\log\frac{1}{\delta}}.
\ea
%as replacing $\delta$ by $\delta/2$  simply amounts to a reparametrization. 
\endIEEEproof

\section*{Acknowledgments}
The authors are thankful to the anonymous reviewers for very insightful comments. 
In particular, we would like to acknowledge one of the  reviewers for pointing out an alternative proof of Theorem \ref{th2} (cf. Remark \ref{rem.rev}).  

%\balance
\bibliographystyle{IEEEtran}
\bibliography{IEEEabrv,refs}

\begin{thebibliography}{10}
\providecommand{\url}[1]{#1}
\csname url@samestyle\endcsname
\providecommand{\newblock}{\relax}
\providecommand{\bibinfo}[2]{#2}
\providecommand{\BIBentrySTDinterwordspacing}{\spaceskip=0pt\relax}
\providecommand{\BIBentryALTinterwordstretchfactor}{4}
\providecommand{\BIBentryALTinterwordspacing}{\spaceskip=\fontdimen2\font plus
\BIBentryALTinterwordstretchfactor\fontdimen3\font minus
  \fontdimen4\font\relax}
\providecommand{\BIBforeignlanguage}[2]{{%
\expandafter\ifx\csname l@#1\endcsname\relax
\typeout{** WARNING: IEEEtran.bst: No hyphenation pattern has been}%
\typeout{** loaded for the language `#1'. Using the pattern for}%
\typeout{** the default language instead.}%
\else
\language=\csname l@#1\endcsname
\fi
#2}}
\providecommand{\BIBdecl}{\relax}
\BIBdecl

\bibitem{SRB13}
D.~Stotz, E.~Riegler, and H.~B\"olcskei, ``Almost lossless analog signal
  separation,'' \emph{Proc. IEEE Int. Symp. Inf. Theory}, pp. 106--110, Jul.
  2013.

\bibitem{SRAB15extended}
D.~Stotz, E.~Riegler, E.~Agustsson, and H.~B{\"o}lcskei, ``Addendum to
  ``{A}lmost lossless analog signal separation and probabilistic uncertainty
  relations'','' \emph{available online:
  \url{http://www.nari.ee.ethz.ch/commth/research/downloads/sigsep_addendum.pdf}},
  {May} 2017.

\bibitem{SKP12}
C.~Studer, P.~Kuppinger, G.~Pope, and H.~B{\"o}lcskei, ``Recovery of sparsely
  corrupted signals,'' \emph{IEEE Trans. Inf. Theory}, vol.~58, no.~5, pp.
  3115--3130, May 2012.

\bibitem{Li12}
X.~Li, ``Compressed sensing and matrix completion with constant proportion of
  corruptions,'' \emph{Constructive Approximation}, vol.~37, no.~1, pp. 79--99,
  Feb. 2013.

\bibitem{WM10}
J.~Wright and Y.~Ma, ``Dense error correction via $\ell^1$-minimization,''
  \emph{IEEE Trans. Inf. Theory}, vol.~56, no.~7, pp. 3540--3560, Jul. 2010.

\bibitem{DMM09}
D.~L. Donoho, A.~Maleki, and A.~Montanari, ``Message-passing algorithms for
  compressed sensing,'' \emph{Proc. Natl. Acad. Sci.}, vol. 106, no.~45, pp.
  18\,914--18\,919, 2009.

\bibitem{DK10}
D.~L. Donoho and G.~Kutyniok, ``Microlocal analysis of the geometric separation
  problem,'' \emph{Comm. Pure and Appl. Math.}, vol.~66, no.~1, pp. 1--47, Jan.
  2013.

\bibitem{Tro08}
J.~A. Tropp, ``{On the conditioning of random subdictionaries},'' \emph{Applied
  and Computational Harmonic Analysis}, vol.~25, pp. 1--24, 2008.

\bibitem{CT12}
M.~B. McCoy and J.~A. Tropp, ``{Sharp recovery bounds for convex demixing, with
  applications},'' \emph{Found. Comp. Math.}, vol.~14, no.~3, pp. 503--567,
  {Jun.} 2014.

\bibitem{PBS13}
G.~Pope, A.~Bracher, and C.~Studer, ``Probabilistic recovery guarantees for
  sparsely corrupted signals,'' \emph{IEEE Trans. Inf. Theory}, vol.~59, no.~5,
  pp. 3104--3116, May 2013.

\bibitem{DH01}
D.~L. Donoho and X.~Huo, ``Uncertainty principles and ideal atomic
  decomposition,'' \emph{IEEE Trans. Inf. Theory}, vol.~47, no.~7, pp.
  2845--2862, Nov. 2001.

\bibitem{DS89}
D.~L. Donoho and P.~B. Stark, ``Uncertainty principles and signal recovery,''
  \emph{SIAM Journal on Applied Mathematics}, vol.~49, no.~3, pp. 906--931,
  {Jun.} 1989.

\bibitem{WV10}
Y.~Wu and S.~Verd\'u, ``R\'enyi information dimension: Fundamental limits of
  almost lossless analog compression,'' \emph{IEEE Trans. Inf. Theory},
  vol.~56, no.~8, pp. 3721--3748, Aug. 2010.

\bibitem{SYC91}
T.~D. Sauer, J.~A. Yorke, and M.~Casdagli, ``{Embedology},'' \emph{Journal of
  Statistical Physics}, vol.~65, no. 3-4, pp. 579--616, Nov. 1991.

\bibitem{EB02}
M.~Elad and A.~M. Bruckstein, ``A generalized uncertainty principle and sparse
  representation in pairs of bases,'' \emph{IEEE Trans. Inf. Theory}, vol.~48,
  no.~9, pp. 2558--2567, Sep. 2002.

\bibitem{KDB12}
P.~Kuppinger, G.~Durisi, and H.~B{\"o}lcskei, ``Uncertainty relations and
  sparse signal recovery for pairs of general signal sets,'' \emph{IEEE Trans.
  Inf. Theory}, vol.~58, no.~1, pp. 263--277, Jan. 2012.

\bibitem{Fal04}
K.~Falconer, \emph{Fractal Geometry: Mathematical Foundations and
  Applications}, {2nd}~ed.\hskip 1em plus 0.5em minus 0.4em\relax John Wiley \&
  Sons, 2004.

\bibitem{Fo99}
G.~B. Folland, \emph{Real {A}nalysis}, 2nd~ed., ser. Pure and Applied
  Mathematics.\hskip 1em plus 0.5em minus 0.4em\relax New York, NY: Wiley,
  1999.

\bibitem{RSB15}
E.~Riegler, D.~Stotz, and H.~B\"olcskei, ``Information-theoretic limits of
  matrix completion,'' \emph{Proc. IEEE Int. Symp. Inf. Theory}, pp.
  1836--1840, {Jun.} 2015.

\bibitem{RT15}
E.~Riegler and G.~Taub{\"o}ck, ``Almost lossless analog compression without
  phase information,'' \emph{Proc. IEEE Int. Symp. Inf. Theory}, pp. 999--1003,
  {Jun.} 2015.

\bibitem{rowe98}
R.~T. Rockafellar and R.~J.-B. Wets, \emph{Variational Analysis}, 3rd~ed.\hskip
  1em plus 0.5em minus 0.4em\relax Heidelberg, Germany: Springer, 2009.

\bibitem{albolekori17}
G.~Alberti, H.~B\"olcskei, C.~De~Lellis, G.~Koliander, and E.~Riegler,
  ``Lossless analog compression,'' \emph{IEEE Trans. Inf. Theory, \emph{in
  preparation}}.

\bibitem{ma99}
P.~Mattila, \emph{Geometry of Sets and Measures in Euclidean Space: Fractals
  and Rectifiability}.\hskip 1em plus 0.5em minus 0.4em\relax Cambridge, UK:
  Cambridge Univ. Press, 1999.

\bibitem{Dur13}
R.~Durrett, \emph{Probability: Theory and Examples}, 4th~ed.\hskip 1em plus
  0.5em minus 0.4em\relax Cambridge University Press, 2010.

\bibitem{ba95}
R.~G. Bartle, \emph{{T}he {E}lements of {I}ntegration and {L}ebesgue
  {Measure}}.\hskip 1em plus 0.5em minus 0.4em\relax New York, NY: Wiley, 1995.

\bibitem{Min69}
G.~J. Minty, ``On the extension of {Lipschitz}, {Lipschitz-H{\"o}lder}
  continuous, and monotone functions,'' \emph{Bulletin of the American
  Mathematical Society}, vol.~76, no.~2, pp. 334--339, Mar. 1970.

\bibitem{CRT06}
E.~Cand\`es, J.~Romberg, and T.~Tao, ``Robust uncertainty principles: {E}xact
  signal reconstruction from highly incomplete frequency information,''
  \emph{IEEE Trans. Inf. Theory}, vol.~52, no.~2, pp. 489--509, Feb. 2006.

\bibitem{Don06}
D.~L. Donoho, ``Compressed sensing,'' \emph{IEEE Trans. Inf. Theory}, vol.~52,
  no.~4, pp. 1289--1306, Apr. 2006.

\bibitem{Tro08b}
J.~A. Tropp, ``{On the linear independence of spikes and sines},''
  \emph{Journal of Fourier Analysis and Applications}, vol.~14, no.~5, pp.
  838--858, 2008.

\end{thebibliography}

\begin{IEEEbiographynophoto}{David Stotz}
received the MASt in Mathematics in 2009 from the University of Cambridge, United Kingdom, and the Dipl.-Math. degree in 2010 from the University of Freiburg, Germany. In 2011 he joined the Communication Technology Laboratory at ETH Zurich, where he graduated with the Dr. sc. degree in 2015 followed by postdoctoral research. He teaches Mathematics at Kantonsschule Schaffhausen, Schaffhausen, Switzerland.

\end{IEEEbiographynophoto}
\begin{IEEEbiographynophoto}{Erwin Riegler}
(M'07) received the Dipl-Ing. degree in Technical Physics (with distinction) in 2001 and the Dr. techn. degree in Technical Physics (with distinction) in 2004, both from Vienna University of Technology. From 2005 to 2006 he was a post-doctoral researcher at the Institute for Analysis and Scientific Computing, Vienna University of Technology. From 2007 to 2010 he was a senior researcher at the Telecommunications Research Center Vienna (FTW). From 2010 to 2014 he was a post-doctoral researcher at the Institute of Telecommunications, Vienna University of Technology. Since 2014 he has been a senior researcher with the Communication Theory Group at ETH Zurich, Switzerland.

Dr. Riegler was a visiting researcher at the Max Planck Institute for Mathematics in the Sciences in Leipzig, Germany (Sep. 2004 to Feb. 2005), the Communication Theory Group at ETH Zurich, Switzerland (Sep. 2010 to Feb. 2011 and June 2012 to Nov. 2012), the Department of Electrical and Computer Engineering at The Ohio State University in Columbus, Ohio (Mar. 2012), and the Department of Signals and Systems at Chalmers University of Technology in Gothenburg, Sweden (Nov. 2013). He is a co-author of a paper that won a Student Paper Award at the 2012 International Symposium on Information Theory.

His research interests include noncoherent communications, machine learning, interference management, large system analysis, and transceiver design. 
\end{IEEEbiographynophoto}
\begin{IEEEbiographynophoto}{Eirikur Agustsson}
received a MSc degree in Electrical Engineering and Information Technology in 2014 from ETH Zurich, a BSc degree in Mathematics in 2012 and a BSc degree in Electrical Engineering in 2011, both from the University of Iceland. He is currently a PhD candidate in the Computer Vision Lab at ETH Zurich, under the supervision of Prof. Luc Van Gool. His main research interests include machine learning, information theory, and data compression.
\end{IEEEbiographynophoto}
\begin{IEEEbiographynophoto}{Helmut B\"olcskei}
(S'94--M'98--SM'02--F'09) was born in M\"odling, Austria on May 29, 1970, and received the Dipl.-Ing. and Dr. techn. degrees in electrical engineering from Vienna University of Technology, Vienna, Austria, in 1994 and 1997, respectively. In 1998 he was with Vienna University of Technology. From 1999 to 2001 he was a postdoctoral researcher in the Information Systems Laboratory, Department of Electrical Engineering, and in the Department of Statistics, Stanford University, Stanford, CA. He was in the founding team of Iospan Wireless Inc., a Silicon Valley-based startup company (acquired by Intel Corporation in 2002) specialized in multiple-input multiple-output (MIMO) wireless systems for high-speed Internet access, and was a co-founder of Celestrius AG, Zurich, Switzerland. From 2001 to 2002 he was an Assistant Professor of Electrical Engineering at the University of Illinois at Urbana-Champaign. He has been with ETH Zurich since 2002, where he is a Professor of Electrical Engineering. He was a visiting researcher at Philips Research Laboratories Eindhoven, The Netherlands, ENST Paris, France, and the Heinrich Hertz Institute Berlin, Germany. His research interests are in information theory, mathematical signal processing, machine learning, and statistics.

He received the 2001 IEEE Signal Processing Society Young Author Best Paper Award, the 2006 IEEE Communications Society Leonard G. Abraham Best Paper Award, the 2010 Vodafone Innovations Award, the ETH "Golden Owl" Teaching Award, is a Fellow of the IEEE, a 2011 EURASIP Fellow, was a Distinguished Lecturer (2013-2014) of the IEEE Information Theory Society, an Erwin Schr\"odinger Fellow (1999-2001) of the Austrian National Science Foundation (FWF), was included in the 2014 Thomson Reuters List of Highly Cited Researchers in Computer Science, and is the 2016 Padovani Lecturer of the IEEE Information Theory Society. He served as an associate editor of the IEEE Transactions on Information Theory, the IEEE Transactions on Signal Processing, the IEEE Transactions on Wireless Communications, and the EURASIP Journal on Applied Signal Processing. He was editor-in-chief of the IEEE Transactions on Information Theory during the period 2010-2013. He served on the editorial board of the IEEE Signal Processing Magazine and is currently on the editorial boards of "Foundations and Trends in Networking? and "Foundations and Trends in Communications and Information Theory?. He was TPC co-chair of the 2008 IEEE International Symposium on Information Theory and the 2016 IEEE Information Theory Workshop and serves on the Board of Governors of the IEEE Information Theory Society. He has been a delegate of the president of ETH Zurich for faculty appointments since 2008.
\end{IEEEbiographynophoto}

\end{document}